\documentclass[longauth]{aa} 

\usepackage{graphicx}
\usepackage{txfonts}
\usepackage{xcolor}
\usepackage{natbib}

\usepackage[colorlinks=true,linkcolor=bluea,allcolors=blue]{hyperref}
\begin{document} 
	
	\title{The Galaxy Activity, Torus, and Outflow Survey (GATOS): N. Unveiling physical processes in local active galaxies}
	\subtitle{Unsupervised hierarchical clustering of JWST MIRI/MRS observations}
    \titlerunning{Unsupervised hierarchical clustering of JWST MIRI/MRS observations}
	
	\author{L. Hermosa Mu{\~n}oz \inst{1}
		\and
		J. R. Gonz{\'a}lez Fern{\'a}ndez\inst{2}
		\and
		A. Alonso-Herrero \inst{1}
        \and
		I. Garc{\'i}a-Bernete\inst{1}
        \and
        O. Gonz{\'a}lez-Mart{\'i}n\inst{3}
        \and
        M. Pereira-Santaella\inst{4}
        \and
        E. L{\'o}pez-Rodr{\'i}guez\inst{5}
        \and 
        C. Ramos Almeida\inst{6,7}
        \and
        S. Garc{\'i}a-Burillo\inst{8}
        \and
        L. Zhang\inst{9}
        \and
        A. Audibert\inst{6,7}
        \and
        E. Bellocchi\inst{10,11}
        \and 
        F. Combes\inst{12,13}
        \and
        T. D{\'i}az-Santos\inst{14,15}
        \and
        D. Esparza-Arredondo\inst{3} 
        \and
        B. Garc{\'i}a-Lorenzo\inst{6,7}
        \and 
        M. Garc{\'i}a-Mar{\'i}n\inst{16}
        \and
        E. K. S. Hicks\inst{17,9,18}
        \and
        {\'A}. Labiano\inst{19}
        \and
        N. A. Levenson\inst{20}
        \and 
        M. Mart{\'i}nez-Paredes\inst{21}
        \and
        C. Packham\inst{9}
        \and 
        R. A. Riffel\inst{22,23}
        \and
        D. Rigopoulou\inst{24}
        \and 
        J. Schneider\inst{9}
        \and 
        M. Villar-Mart{\'i}n\inst{23}
	}
	
	\institute{
		1. Centro de Astrobiolog{\'i}a (CAB) CSIC-INTA, Camino Bajo del Castillo s/n, 28692 Villanueva de la Ca{\~n}ada, Madrid, Spain \\
		\email{lhermosa@cab.inta-csic.es}
		\\
        2. Universidad Internacional de la Rioja (UNIR), Av. de la Paz 137, 26006 Logro{\~n}o, La Rioja, Spain \\
        3. Instituto de Radioastrononom{\'i}a y Astrof{\'i}sica (IRyA), Universidad Nacional
Aut{\'o}noma de M{\'e}xico, Antigua Carretera a P{\'a}tzcuaro 8701 Ex-Hda. San
Jos{\'e} de la Huerta, Morelia, Michoac{\'a}n, 58089, Mexico \\
		4. Instituto de F{\'i}sica Fundamental, CSIC, Calle Serrano 123, 28006 Madrid, Spain \\
        5. Kavli Institute for Particle Astrophysics \& Cosmology (KIPAC), Stanford University, Stanford, CA 94305, USA \\
        6. Instituto de Astrof{\'i}sica de Canarias, C/ V{\'i}a L{\'a}ctea s/n, 38205 La Laguna, Tenerife, Spain \\
        7. Departamento de Astrof{\'i}sica, Universidad de La Laguna, 38205 La Laguna, Tenerife, Spain \\
        8. Observatorio Astron{\'o}mico Nacional (OAN-IGN) - Observatorio de Madrid, Alfonso XII, 3, 28014, Madrid, Spain \\
        9. Department of Physics and Astronomy, The University of Texas at San Antonio, 1 UTSA Circle, San Antonio, Texas, 78249, USA \\
		10. Departmento de F{\'i}sica de la Tierra y Astrof{\'i}sica, Fac. de CC F{\'i}sicas, Universidad Complutense de Madrid, E-28040 Madrid, Spain \\
		11. Instituto de F{\'i}sica de Partículas y del Cosmos IPARCOS, Fac. de CC F{\'i}sicas, Universidad Complutense de Madrid, E-28040 Madrid, Spain \\
		12. Observatoire de Paris, LUX, PSL University, Sorbonne Universit{\'e}, CNRS, F-75014 Paris, France \\
		13. Coll{\`e}ge de France, 11 Place Marcelin Berthelot, 75231 Paris, France \\
		14. Institute of Astrophysics, Foundation for Research and Technology - Hellas (FORTH), Heraklion 70013, Greece \\
        15. School of Sciences, European University Cyprus, Diogenes Street, Engomi 1516, Nicosia, Cyprus \\
		16. European Space Agency, c/o Space Telescope Science Institute, 3700 San Martin Drive, Baltimore MD 21218, USA \\
		17.  Department of Physics and Astronomy, University of Alaska Anchorage, Anchorage, AK 99508-4664, USA \\
		18. Department of Physics, University of Alaska, Fairbanks, Alaska 99775-5920, USA \\
		19. Telespazio UK for the European Space Agency (ESA), ESAC, Camino Bajo del Castillo s/n, 28692 Villanueva de la Ca{\~n}ada, Spain \\
		20. Space Telescope Science Institute, 3700 San Martin Drive, Baltimore, MD 21218, USA \\
		21. 1142 Sunset Point Rd, Clearwater, Florida 33755, USA\\
		22. Departamento de Física, CCNE, Universidade Federal de Santa Maria, Av. Roraima 1000, 97105-900, Santa Maria, RS, Brazil \\
		23. Centro de Astrobiolog{\'i}a (CAB) CSIC-INTA,  Ctra. de Ajalvir km 4, Torrej{\'o}n de Ard{\'o}z, 28850, Madrid, Spain \\
		24. Department of Physics, University of Oxford, Keble Road, Oxford, OX1 3RH, UK \\
	}
	
	\date{Received M DD, YYYY; accepted M DD, YYYY}
	
	
	\abstract{With the rise of the integral field spectroscopy (IFS), we are currently dealing with large amounts of spatially resolved data, whose analysis has become challenging, especially when observing complex objects such as nearby galaxies.
	}
	{We aim to develop a method to automatically separate regions with different physical properties (ionisation, kinematics, etc.) within the central parts ($1" \sim 160$\,pc, on average) of galaxies. This can allow us to better understand the systems, and provide an initial characterisation of the main ionisation sources affecting its evolution.}
	{We have developed an unsupervised hierarchical clustering algorithm to analyse data cubes based on spectral similarity. It clusters together spaxels with similar spectra, which is useful to disentangle between regions affected by different processes, such as ionisation sources. We have applied this method to a sample of 15 nearby (distances $<$100\,Mpc) galaxies, 7 from the Galaxy Activity, Torus, and Outflow Survey (GATOS) and 8 archival sources, all observed with the medium resolution spectrometer (MRS) of the Mid-Infrared Instrument (MIRI) on board of the James Webb Space Telescope (JWST). The sample spans sources with various morphologies, AGN types, and/or starbursts. From the clusters, we computed their median spectrum and measured the line and continuum properties. We used these measurements to train random forest models and create several empirical mid-IR diagnostic diagrams for the MRS channel 3 wavelength range, ranging from 11.5 to 18$\mu$m, that includes among others the bright [Ne\,II], [Ne\,III], and [Ne\,V] lines, several H$_{2}$ transitions, and PAH features.}
	{The clustering technique allows to differentiate emission coming from an active galactic nucleus (AGN), a nuclear starburst, the disc and star forming (SF) regions in the galaxies, and other composite regions, potentially ionised by several sources simultaneously. This is supported by the results from the empirical diagnostic diagrams, that are indeed able to separate physically distinct regions. This innovative method serves as a tool to identify regions of interest in any data cube prior to an in-depth analysis of the sources. In a future work we will explore other wavelength ranges and a larger sample, that would help us to obtain statistically significant conclusions. }
	{}
    
	
	\maketitle
	%
	
	\section{Introduction}
	\label{Sect1:Introduction}
	
	With the rise of integral field spectroscopy (IFS) data available in the scientific archives, astronomers can now study in great detail different objects and physical processes (e.g. kinematics, ionisation, density, temperature, etc.) both in a spatially and spectrally resolved way in galaxies. However, the complexity of data analysis and interpretation has equally increased. This is particularly true for the study of the central parts of nearby galaxies, where multiple physical processes are occurring simultaneously \citep[e.g.][]{Bacon2001,Emsellem2004,Cappellari2011,Sanchez2012, CidFernandes2013,Bundy2015,Cazzoli2020,Lin2020,Venturi2021,GarciaBernete2021,Riffel2021,Peralta2023,ChamorroCazorla2023,AH2024,GarciaBernete2024,GarciaBernete2024b,Speranza2024,Zhang2024,HM2024,HM2025}, such as circular motions, shocks, star formation (SF) processes, and/or the presence and effect of an active galactic nucleus (AGN). 
    All these processes can be studied through different tracers, such as molecular, ionised, or neutral gas, both in emission and/or in absorption depending on the wavelength of observation \citep[for AGN see e.g.][]{Cazzoli2016, Fiore2017, Fluetsch2019}. 

    From an observational perspective, a great effort has been made with optical surveys such as Mapping Nearby Galaxies at Apache Point Observatory \citep[MaNGA,][]{Bundy2015} or Calar Alto Legacy Integral Field Area \citep[CALIFA,][]{Sanchez2012}. These surveys use spectroscopic data for large statistical samples of galaxies to study their evolution, morphologies, internal and external physical processes, and kinematics, among others. Within these surveys, several works are dedicated to identifying the ionising sources of the gas through the well-known Baldwin-Philips-Terlevich \citep[BPT;][]{Baldwin1981} diagnostic diagrams, but applied in a spatially resolved way to locate the position of SF regions, shocked regions, and/or AGN, if present, etc. \citep{Belfiore2016,Gomes2016,Law2021}. 
    From a modelling perspective, there are available tools, such as pPXF \citep{Cappellari2004,Cappellari2017} or Pipe3D \citep{Sanchez2016}, that model the stellar continuum and the emission lines in the optical spectra. Also DeblendIRS \citep{HernanCaballero2015}, that separates between the AGN, interstellar medium (ISM), and the SF contributions in the mid-infrared (mid-IR) spectra of galaxies. Additionally, spectral decomposition tools as PAHFIT \citep{Smith2007}, or more recent tools such as the method presented in \cite{Donnan2024}, or CAFE \citep{DiazSantos2025}, identify polycyclic aromatic carbon (PAH) features in the infrared spectra and model them together with the dust and stellar continuum. These provide further insights into the nature and distribution of the gas and dust components. 
    
    Compared to the optical, the infrared spectral range provides a significantly broader variety of features that allow us to characterise the ISM. In particular, galaxy spectra include warm molecular lines (e.g. H$_{2}$), atomic fine-structure lines covering a large range of ionisation states (ionisation potentials, IPs, from $\sim7$ to $\sim190$\,eV), hydrogen recombination lines, PAH features, dust continuum, as well as absorption features from ices and several molecular species \citep{GarciaBernete2022,GarciaBernete2024c}. Thus thanks to the diversity of tracers, this wavelength range offers a more detailed view of the different ISM phases in galaxies.

    Large amounts of mid-IR data for nearby galaxies exist thanks to spectroscopic data from the Spitzer telescope Infrared Spectrograph \citep[IRS;][]{Houck2004}, with several works dedicated to understanding the relative contribution of AGN and SF \citep[see e.g.][]{Armus2007,Pope2008,Moustakas2010,AH2012}. Nevertheless, these data provide an integrated spectrum of the sources, with large apertures (slit sizes from $3.6\arcsec$ to $\sim12\arcsec$), a maximum resolution of $\sim 600$, and not always covering the complete mid-IR range. The large areas, as well as the low resolution, may dilute the detection of certain features within the circumnuclear regions beyond the AGN, such as SF regions, shocks, etc.

    With the launch of the James Webb Space Telescope \citep[JWST][]{Gardner2023}, we have significantly increased the sensitivity and resolution of the near-infrared (near-IR) and mid-IR data. Particularly in the mid-IR range, the JWST obtains IFS data with the medium-resolution spectrometer (MRS) of the Mid-Infrared Instrument (MIRI; \citealt{Rieke2015,Wright2015,Wright2023}). MIRI/MRS data are rich and complex, containing a wealth of information within a single data cube \citep[e.g.][]{PereiraSantaella2022,GarciaBernete2022,Armus2023,Davies2024,Donnan2023,Donnan2024,Dasyra2024,Zhang2024,Goold2024,EsparzaArredondo2025,HM2025,Riffel2025,RamosAlmeida2025,AH2025}. To fully exploit the information from these datasets and derive physical properties across large samples of galaxies, it is essential to develop automated methods to analyse the data. 
    
    The application of methods for astronomical classification started already in the 1990s with neural network structures to classify stellar spectra or galaxy types \citep[see e.g.][and \citealt{Smith2023} for a review]{vonHippel1994,Ball2004}. Some classical methods aimed at simplifying the data analysis are still used, such as Principal Component Analysis (PCA), that reduces the dimensionality problem of the data to obtain the main physical properties of the analysed objects \citep[see e.g.][and references therein]{Steiner2009}. However, the physical meaning of PCA components is often difficult to interpret, since they are a linear combination of various components. In recent years, several authors have started to develop alternative machine learning methods \citep[see e.g.][]{Baron2021,Chambon2024,deSouza2024,Lu2025}, using various techniques useful for handling complex data and identifying trends for different objects, some particularly focused on AGN \citep[see e.g.][]{Daoutis2025,Poitevineau2025,Nemer2025}. 
    For example, \cite{deSouza2024} analysed a sample of MaNGA galaxies using a clustering technique based on the spectral similarity within the cubes, named \textsc{Capivara}. This allowed them to easily separate distinct physical regions of galaxies, such as the nucleus, bulge, spiral arms, or bars. In a certain way, clustering is similar to spatial binning techniques, such as Voronoi tessellations \citep{Cappellari2003}, but based on the spectral physical properties rather than only in the signal-to-noise (S/N). While clustering itself groups spectra based on their similarity, the interpretation of these clusters to a specific physical mechanism (e.g. AGN, SF, etc.) requires additional labelling and/or the use of supervised methods. These techniques are independent of the galaxy type and could potentially be used to disentangle regions affected by different ionisation sources, providing a more efficient and automated way to separate SF processes, shocks, and AGN ionisation not only in the optical \citep[see e.g.][]{Daoutis2025}, but also in the mid-IR or other frequencies. Indeed, \cite{deSouza2024} classified their clusters into different categories using both the stellar continuum and emission line properties, based on the optical BPT diagrams \citep{Baldwin1981}. They reported an overall agreement between the cluster-based classification and the results from the traditional pixel-by-pixel analysis. This suggests that clustering tools can provide a simplified, but accurate method, to analyse complex data cubes.
	
	In this paper, we explore an unsupervised hierarchical clustering technique with a sample of galaxies, most containing an AGN, observed using the MIRI/MRS on board of the JWST. Part of this data set was observed within the Galaxy Activity, Torus, and Outflow Survey (\href{https://gatos.myportfolio.com/}{GATOS}) collaboration \citep{GarciaBurillo2021,AAH2021}. We would like to emphasize that this is an exploratory, empirical study that aims at evaluating this new analysis technique. We mainly focus on the search for empirical tracers that could potentially help to disentangle different ionising mechanisms and physical processes occurring in these galaxies using innovative machine learning techniques. To our knowledge, this is one of the early applications of a clustering method and automatic classification of the central regions of nearby galaxies using JWST spectroscopic data.
	
	The paper is organised as follows. Section~\ref{Sect2:Data} describes the observations, data reduction and the methodology, based on custom-made codes. In Sect.~\ref{Sect3:Results} we present the main results of the clustering technique, including the median spectra per cluster, and other empirical measurements, such as the line ratios. In Sect.~\ref{Sect4:Discussion} we compare and evaluate the performance of the method in different mid-IR wavelength ranges, and we discuss the main caveats of the methodology. Finally, we present the summary and main conclusions of this work in Sect.~\ref{Sect5:Conclusions}. 
	
	\section{Data and methodology}
	\label{Sect2:Data}

    We selected a total sample of 15 nearby (distances $<$100\,Mpc) galaxies (see Table~\ref{Table:1}) that primarily host different AGN types, observed with MIRI/MRS. The sources that will be used as the training dataset represent all the local AGN and starburst galaxies, from archival and proprietary time, that have been studied in detail with MIRI/MRS mid-IR spectroscopic data in recent works \citep[][]{AAH2019,PereiraSantaella2022,GarciaBernete2022,Zhang2023,Armus2023,GarciaBernete2024,GarciaBernete2024b,Dasyra2024,Davies2024,Goold2024,HM2024b,Zhang2024,Veenema2025}, providing prior knowledge of the physical processes (e.g. kinematics, ionisation, temperatures, etc.) at play in these systems. In this way, they can be used as a test bed to validate the technique, and then apply it to new MIRI/MRS unexplored data cubes. 

    \subsection{Data sample}
    \label{SubSect2:Data}
    
    We have made use of MIRI/MRS data coming mainly from the GATOS collaboration (see Table~\ref{Table:1}). The sample consists on four Seyfert (Sy) galaxies observed during JWST General Observer (GO) program Cycle 1 (NGC\,3081, NGC\,5506, NGC\,5728, and NGC\,7172; program ID 1670, PI T.~Shimizu, see details in \citealt{Zhang2024}), whose main mid-IR properties have already been analysed in several works from the collaboration \citep[\textbf{see e.g.}][]{PereiraSantaella2022,HM2024b,GarciaBernete2024,Davies2024,Zhang2024,Zhang2024b,EsparzaArredondo2025,Delaney2025}, and three Sy galaxies observed during JWST GO Cycle 2 (NGC\,3227, NGC\,4051, and NGC\,7582; ID 3535, PIs I.~Garc{\'i}a-Bernete \& D.~Rigopoulou), which will be used as a test-bed for the methodology (see Sect.~\ref{SubSect4:Disc_NGC7582}; \citealt{Veenema2025}). 
    
    We included the Sy galaxies Centaurus A (Cen~A from now on), IC\,5063, NGC\,7319, and NGC\,7469. These galaxies are publicly available in the archival and their MIRI/MRS data have been studied in detail in previous works. Cen~A was observed within the guaranteed time observation program MICONIC \citep[ID 1269, PI N.~Luetzgendorf, see][]{AH2025}, and IC\,5063 was observed in the cycle 1 program 2004 \citep[PI K.~M. Dasyra,][]{Dasyra2024}. The latter two objects were observed in the Early Release Observations program (ID 2732, PI K.M.~Pontoppidan, \citealt{Pontoppidan2022}) and the Early Release Science program (ID 1328, PI L.~Armus), respectively. Three of these objects, namely Cen~A, IC\,5063, and NGC\,7319, are known to have a radio jet that perturbes the ISM, but their AGN are not always the dominant ionising source \citep[][]{Williams2002,PereiraSantaella2022,Dasyra2024,AH2025}. NGC\,7469 hosts both a type-1.5 Sy and a nuclear starburst \citep{Cazzoli2020,GarciaBernete2022,Zhang2023,Armus2023}. We included two low luminosity AGN classified as low ionisation nuclear emission-line regions (LINERs), namely NGC\,1052 and NGC\,4594 \citep[ID 2016, PI A.~Seth,][]{Goold2024}, to compare with other AGN types (see Table~\ref{Table:1}). Finally, we included the pure starburst nuclei NGC\,3256~N \citep[ERS program ID 1328, PI A.~Lee, ][]{Bohn2024,Rigopoulou2024,GarciaBernete2025} and M\,83 \citep[ID 2219, PI S.S.~Hernandez, ][]{Hernandez2023,Hernandez2025}, to compare with the AGN systems. The MIRI/MRS data for all these galaxies have been already published in previous works. 
    
    For all the data, the reduction process was done following the standard MRS pipeline procedure (e.g. \citealt{Labiano2016}; \citealt{Bushouse2023} and references therein), with the pipeline release 1.11.4 and the calibration context 1130, except for Cen~A \citep[see details in][]{AH2024,AH2025}. The details of the procedure are fully explained in \cite{PereiraSantaella2022} and \cite{GarciaBernete2022,GarciaBernete2024c}. We subtracted the background from all the cubes by computing a median background at each wavelength.

    The MIRI/MRS covers a total wavelength range from 4.9 to 27.9\,$\mu$m, divided into four integral field units (referred to as channels) with different fields-of-view (FoVs), and spatial and spectral resolutions \citep[see more details in][]{Labiano2021,Argyriou2023}, namely channels 1 to 4 (ch1, ch2, ch3, and ch4). In this work we focus on ch3, that covers a range from 11.5 to 18$\mu$m, divided in three sub-channels (short, medium and long). In particular, we use the ch3-short cubes (11.55-13.47\,$\mu$m) and the combined ch3 spectral cubes (from now on, referred to as 'ch3-all'). The latter were produced using the tools from the MRS reduction pipeline (\textsc{cube\_build} module). We show the median maps of the combined ch3 cube for all the galaxies in Appendix~\ref{Appendix_MedianMaps} (see Fig.~\ref{FigAp:MedianMaps}). We select this channel mainly because it contains the three neon lines [Ne\,II] at 12.81$\mu$m, [Ne\,III] at 15.56$\mu$m, and [Ne\,V] at 14.32$\mu$m, that are typically used in the mid-IR to study the ionising source of the ionised gas \citep[see e.g.][]{Pereira2010}, as well as H$_{2}$ lines and PAH features. We discuss other channels in Sect.~\ref{Sect4:Discussion}.

    \begin{table*}
	    \caption{Basic information from the galaxies used in this work.}
	\label{Table:1}
	\centering          
	\begin{tabular}{llccccccc}
		\hline\hline
		Galaxy & Type & Distance & Redshift & Morph. type & Jet & Prop. ID & Reference & Sample \\ 
		&  & (Mpc) &  &  &  &  &  \\
		\hline           
		NGC\,1052       & LINER-1.9 & 19   & 0.0050 & E4           & Y & 2016 & [1]     & Train \\
		NGC\,3081$^{*}$ & Sy-2      & 34   & 0.0082 & (R)SAB0/a(r) & Y & 1670 & [2]     & Train \\ 
		NGC\,3227$^{*}$ & Sy-1.5    & 15   & 0.0038 & SAB(s)a pec  & Y & 3535 & --      & Test  \\ 
		NGC\,3256-N     & Starburst & 40   & 0.0094 & Merger pec   & N & 1328 & [3,4,5] & Train \\
		NGC\,4051$^{*}$ & NLS1      & 16.6 & 0.0023 & SAB(rs)bc    & N & 3535 & --      & Test  \\
		NGC\,4594       & LINER-2   & 10   & 0.0034 & SA(s)a       & Y & 2016 & [1]     & Train \\ 
		NGC\,5506$^{*}$ & Sy-2      & 26   & 0.0061 & Sa pec       & Y & 1670 & [2]     & Train \\ 
		NGC\,5728$^{*}$ & Sy-2      & 39   & 0.0092 & SAB(r)a?     & Y & 1670 & [6,7]   & Train \\ 
		NGC\,7172$^{*}$ & Sy-2      & 37   & 0.0087 & Sa pec       & N & 1670 & [7,8]   & Train \\ 
		NGC\,7319       & Sy-2      & 98   & 0.0225 & SB(s)bc pec  & Y & 2732 & [9]     & Train \\ 
		NGC\,7469       & Sy-1.5    & 71   & 0.0163 & (R')SAB(rs)a & N & 1328 & [10,11] & Train \\
		NGC\,7582$^{*}$ & Sy-2      & 22.7 & 0.0053 & (R')SB(s)ab  & - & 3535 & [12]    & Test  \\
		IC\,5063        & Sy-2      & 48.6 & 0.0114 & SA0\^\,+(s)? & Y & 2004 & [13]    & Train \\
		M\,83           & Starburst & 4.6  & 0.0017 & SAB(s)c      & N & 2219 & [14,15] & Train \\
		Centaurus A     & Sy-2      & 3.5  & 0.0018 & S0 pec       & Y & 1269 & [16]    & Train \\
		\hline        
	\end{tabular}\\
	    \tablefoot{$^{*}$ indicates the galaxies observed within the GATOS collaboration (see Sect.~\ref{Sect2:Data}). In Type: ``Sy" stands for Seyfert, and ``NLS1" for Narrow Line Seyfert-1. The redshift and morphological types have been obtained from NASA/IPAC Extragalactic Database (NED). In Jet: ``Y" stands for Yes, and ``N" stands for No; for NGC\,7582 it is unclear \citep{Veenema2025}. The cited works refer exclusively to analyses using JWST data, used to label the clusters (see Sect.~\ref{SubSect2:RFtecnhique}): [1] \cite{Goold2024}, [2] \cite{Delaney2025}; [3] \cite{Bohn2024}, [4] \cite{Rigopoulou2024}, [5] \cite{GarciaBernete2025}, [6] \cite{Davies2024}, [7] \cite{GarciaBernete2024b}, [8] \cite{HM2024b}, [9] \cite{PereiraSantaella2022}, [10] \cite{Armus2023}, [11] \cite{Feuillet2024}, [12] \cite{Veenema2025}, [13] \cite{Dasyra2024}, [14] \cite{Hernandez2023}, [15] \cite{Hernandez2025}, [16] \cite{AH2025}. The last column indicates if the target has been used for the training or testing samples within the analysis.}
\end{table*}

\subsection{Unsupervised hierarchical clustering technique}
\label{SubSect2:Clustering}

	\begin{figure*}
		\centering
		\includegraphics[width=.92\textwidth, trim={2cm 10.5cm 1.7cm 1cm}, clip]{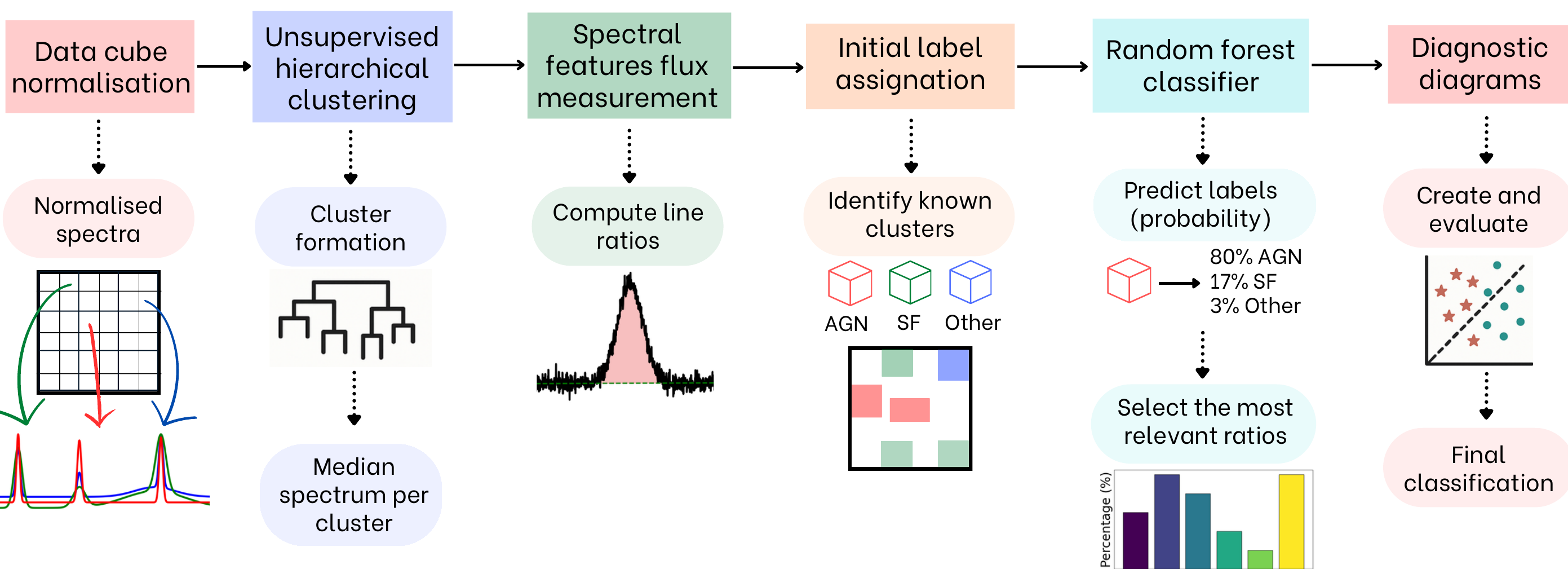}
		\caption{Flowchart of the methodology discussed in Sect.~\ref{Sect2:Data}.}
		\label{FigAp:flowchart}
	\end{figure*}

The complete methodology used in this paper is summarised in Fig.~\ref{FigAp:flowchart}. We first apply an unsupervised hierarchical clustering technique to analyse the data cubes. This step is similar to that used by \cite{deSouza2024}, so we refer the reader to this paper for more details (see also Sect.~\ref{Sect1:Introduction}). In short, this is a machine learning technique that is used to group spectra that are similar, without any prior assumption about their shape or composition. Our algorithm, implemented in \textsc{python}\footnote{All the packages used for the analysis are listed at the end of the acknowledgement section.}, takes as input the spectra of all spaxels within the cube, calculates the distances based on a metric (Euclidean distance, see below), and then clusters them together based on their similarity.

Most of the galaxies analysed here behave as a bright point source, where the nucleus is several times brighter than the circumnuclear regions. Thus, to apply our methodology, we first normalise each spaxel spectrum by dividing it by its total integrated flux. In this way, we remove absolute flux differences, and we can focus exclusively on the spectral shape and relative emission line and PAH strengths. 

The spectral similarity is evaluated by computing the Euclidean distance\footnote{For a full discussion on different distance metrics, we refer the reader to Appendix~A in \cite{Baron2021}.} between all spectra across the cube, defined as $d(x_{i},y_{i}) = \sum_{i} (|x_{i} -y_{i}|)^{2}$, where x and y are two different spectra, to quantify how different each pair is. This process is done iteratively, computing a global distance matrix from all possible pairs of spectra in the cube. Based on this matrix, the algorithm searches for the most similar spaxels and groups them together into clusters. The result of this process is a dendogram, a tree-like structure that visually represents the sequence of cluster mergers. Each spaxel is considered as an individual cluster at the lowest level, and they are progressively merged into larger clusters based on their spectral similarity (i.e. measured distances). 
The clustering process stops at a number of clusters that are selected by ``cutting" the tree at a chosen level, allowing the extraction of meaningful groupings for each galaxy. This number is chosen by visual inspection, stopping when adding more clusters either: 1) creates new concentric clusters from or around already formed clusters, or 2) creates new clusters from individual low S/N spaxels from the edges of the FoV. To account for the rotational velocity field of the galaxies, the algorithm accounts for a spectral shift of $\pm$6 spectral steps ($\sim 300$\,km\,s$^{-1}$) during the clustering, used to minimise the distance calculation. The resulting cluster maps for each galaxy are presented in Sect.~\ref{Sect3:Results}, Figs.~\ref{Fig:Cluster_NGC7172},~\ref{Fig:Cluster_NGC5728}, and in Appendix~\ref{Appendix_ClusterMaps}.

After the clustering process ends, we compute the median spectrum for each cluster to evaluate their features. By using the median, we mitigate possible PSF- and/or combination-driven systematics, providing a representative spectrum for each cluster that contains their dominant spectral trends. We select the median over the mean to avoid the appearance of double peaks in the emission lines due to possible velocity shifts within the cluster. We measure the slope of the continuum ($\alpha_{mIR}$) between 12 and 17\,$\mu$m for ch3-all, avoiding the lines and PAH features. We obtain the fluxes for the emission lines and the PAH features by integrating the profiles after subtracting a linear local continuum on either side of each feature. We consider the following lines: H$_{2}$ 0--0 S(2) at 12.28\,$\mu$m (from now on H$_{2}$ S(2)), HI (7-6) (Hu$_{\alpha}$) at 12.37\,$\mu$m, [Ne\,II] (IP of 21.6\,eV), and [Ar\,V] at 13.10\,$\mu$m (IP of $59.6$\,eV), and with the ch3-all cubes also [Mg\,V] at 13.52\,$\mu$m (IP of 109.2\,eV), [Ne\,V] (IP of 97.2\,eV), [Cl\,II] at 14.37\,$\mu$m (IP of 13.0\,eV), [Ne\,III] (IP of 41.0\,eV), and H$_{2}$ 0--0 S(1) at 17.03\,$\mu$m (from now on H$_{2}$ S(1)). We filter out all lines with low S/N ($< 3$ times the standard deviation of the continuum) before the integration. We also consider the following PAH features: at 12\,$\mu$m (PAH$_{12\mu m}$), the complex at 12.7\,$\mu$m (PAH$_{12.7\mu m}$), and at 16.43\,$\mu$m (PAH$_{17\mu m}$). There are additional PAH features in this wavelength range, but they are weaker and not present in all the sources \citep[see][]{Chown2024}. To integrate the PAHs, we defined the wavelength ranges using as reference their emission in the starburst galaxies, where they are stronger. We note that within this wavelength range, we also detect the red end of the PAH complex at 11.3\,$\mu$m. However, we do not consider it for the line ratio analysis, as it is only partially captured in ch3. For the PAH$_{12.7\mu m}$, which typically includes the [Ne\,II] line, in each spaxel we subtract the measured [Ne\,II] flux from the total flux of the feature to isolate the PAH emission. We cannot rule out the existence of residual line contribution to the total flux, or partial underestimation of the PAH flux due to possible extended wings.

To estimate the flux errors and account for possible flux variations within the individual spectra of a given cluster, we considered all the spectra contained within a single cluster and estimated the standard deviation per wavelength, computing an error spectrum. Then we use error propagation for the line integration and, later on, when computing the line ratios. 

We note that the computational time required for the clustering process is correlated with both a larger spatial extension and a larger wavelength range (average computing time in a laptop with 32\,GiB of RAM and 6 cores of $\sim 4$\,minutes for ch3-short, and $\sim 9$\,minutes for ch3-all). Because of the well-documented classification power of the mid-IR neon lines \citep[see e.g.][and references therein]{MartinezParedes2023,Feltre2023,Zhang2025}, we focus our method on the ch3 channel instead of using other channels or even the whole MIRI/MRS spectral range. Nevertheless, in Sect.~\ref{Sect4:Discussion} we discuss the possibility of applying this process to other channels.
The code is in GitHub available for download\footnote{\href{https://github.com/GonzalezFJR/agncluster}{https://github.com/GonzalezFJR/agncluster}}.

\subsection{Random forest classifier}
\label{SubSect2:RFtecnhique}

Unsupervised clustering techniques have been mainly applied to IFS cubes covering the complete galaxy, identifying the large-scale structures \citep[see][and Sect.~\ref{Sect1:Introduction}]{Chambon2024,deSouza2024}, but not applied specifically to the circumnuclear regions of local galaxies, where multiple processes are occurring simultaneously. 
In order to evaluate if this technique is able to separate the main ionising source for individual regions in complex systems, we developed a complementary method aimed at assigning a physical meaning to the resulting clusters. 

Most of the galaxies from our sample have already been analysed in previous works (see Sect.~\ref{SubSect2:Data} and Table~\ref{Table:1}). Thus, we could use that previous information to associate some of the clusters with particular regions from the galaxy for which we already know their physical nature (e.g. AGN, ionisation cone, SF region, disc, etc.). In particular, these come from evaluating their kinematics (line widths and velocities), emission features ratios (PAHs, warm molecular gas and/or mid-IR line ratios), and morphological (fluxes, SF circumnuclear rings, etc.) properties.
We can use this prior knowledge to create a subset of clusters that can be used as a training set for a supervised machine learning classifier. This way, we can identify those line ratios useful to separate between regions with different ionisation sources (see scheme in Fig.\ref{FigAp:flowchart}). 
Specifically, we manually assigned labels to the known clusters such that: we labelled as 0 the region where the AGN is located, the disc or SF regions as 1, and the interacting or outflowing regions (hereafter referred as 'Other') as 2. We left the clusters with an unclear physical origin, or those for which we did not have any prior information, unlabelled (see Table~\ref{Table:2}). 


We then used these labelled clusters to train a random forest (RF) classifier, a machine learning algorithm that combines the input data to create a large number of individual decision trees (RandomForestClassifier from the \textsc{scikit-learn} package in \textsc{python}, with the default parameters). It makes a final prediction based on the output that the majority of the trees agrees upon. Particularly for our case, the RF classifier was trained using the line ratios as the input features (see Sect.~\ref{SubSect3:Results_Ionisation} for details), allowing to automatically predict the most likely ionisation source for each cluster (i.e. the preferred label). This method provides a probabilistic classification of each cluster, assigning the final label to the category with the highest probability for each case. 
To take into account the flux errors of the measured line ratios, we run a Monte Carlo (MC) simulation on the RF classifier a total of 1000 times, to have an estimation of the uncertainties for all the probabilities derived from the trained models. We use as the final label for each cluster the average probability of the most likely category obtained from the trained models after the simulation.

We tested the accuracy of the model using cross-validation with the labelled points available in our sample (31 labelled clusters out of 65, see Sects.~\ref{SubSect2:RFtecnhique} and~\ref{SubSect3:Results_Ionisation}). For that we randomly splitted the initially-labelled clusters into training (80\%, i.e. $\sim$24 points) and validation (20\%, i.e. $\sim$7 points) subsamples, generating 100 different splits. For each division, we run 100 MC simulations to obtain the final RF model, as explained before. The average classification accuracy of the RF model on the validation dataset is 84\%.

From the resulting models, we can also evaluate the importance of each line ratio used for the classification process. This provides insights into which are the most relevant diagnostics that should be considered for distinguishing between ionising mechanisms in our sample. The RF classifier provides robust results and probabilities that allows us to evaluate the validity of the method (see Sect.~\ref{SubSubSect4:Disc_AutomaticClassification} for a discussion on the method). Finally, we tested the trained model in three GATOS Sy galaxies observed during cycle 2, namely, NGC\,3227, NGC\,4051, and NGC\,7582 (see Sect.~\ref{SubSect2:Data}).

The results of this methodology are presented in Sect.~\ref{SubSect3:Results_Ionisation} and discussed in Sect.~\ref{SubSect4:Disc_Association}. 

\section{Results}
\label{Sect3:Results}

In this section we present the results for the clustering process for all the galaxies in the training sample. In the main text we show two galaxies as examples, namely NGC\,7172 and NGC\,5728, and the rest of the sample is presented in the Appendix~\ref{Appendix_ClusterMaps}. We select these two galaxies as they are representative examples of the results (see Sects.~\ref{SubSect3:Results_clusters} and~\ref{SubSect4:Disc_Caveats}).

\begin{figure*}
	\includegraphics[width=.2\textwidth]{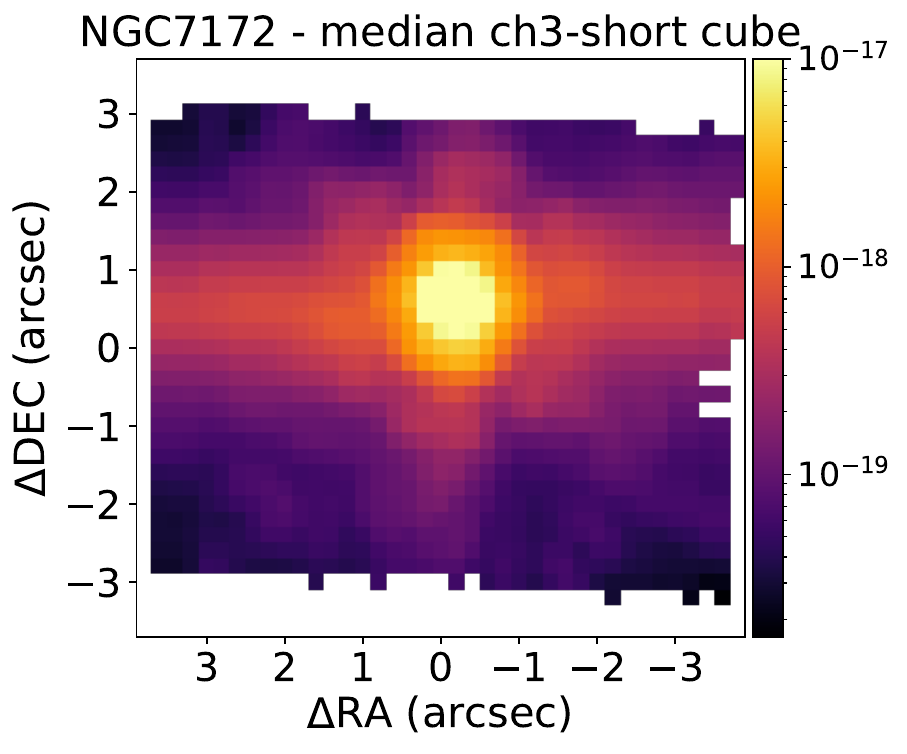}
	\includegraphics[width=.191\textwidth]{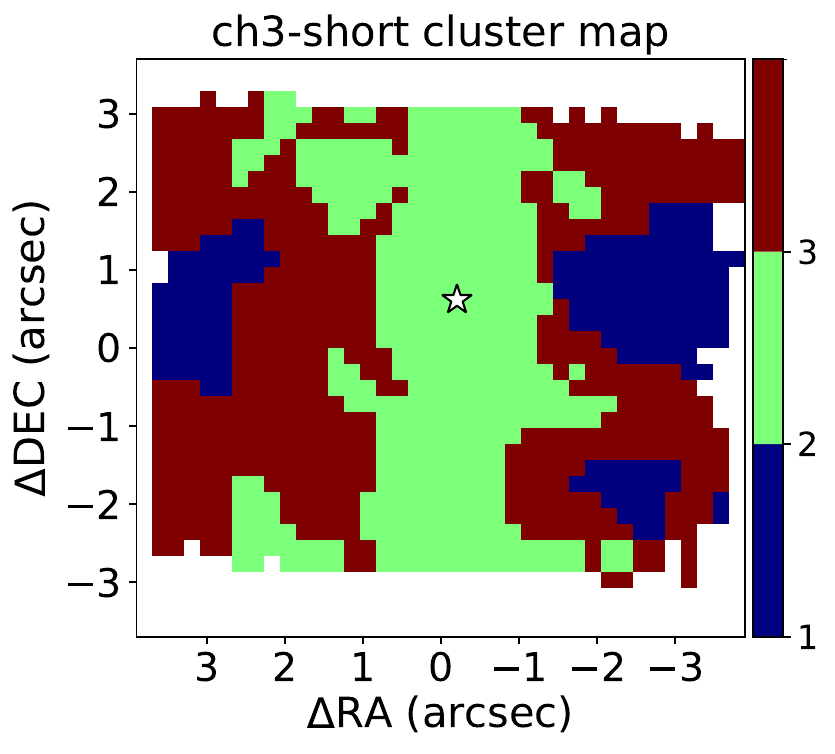}
	\includegraphics[width=.6\textwidth]{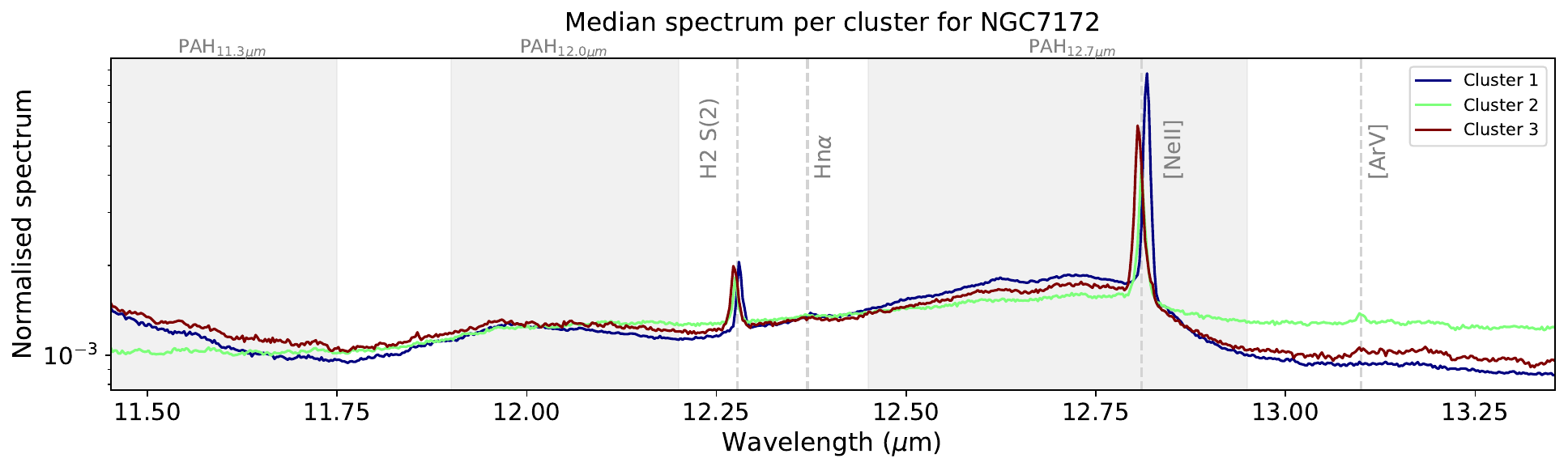}
	\includegraphics[width=.2\textwidth]{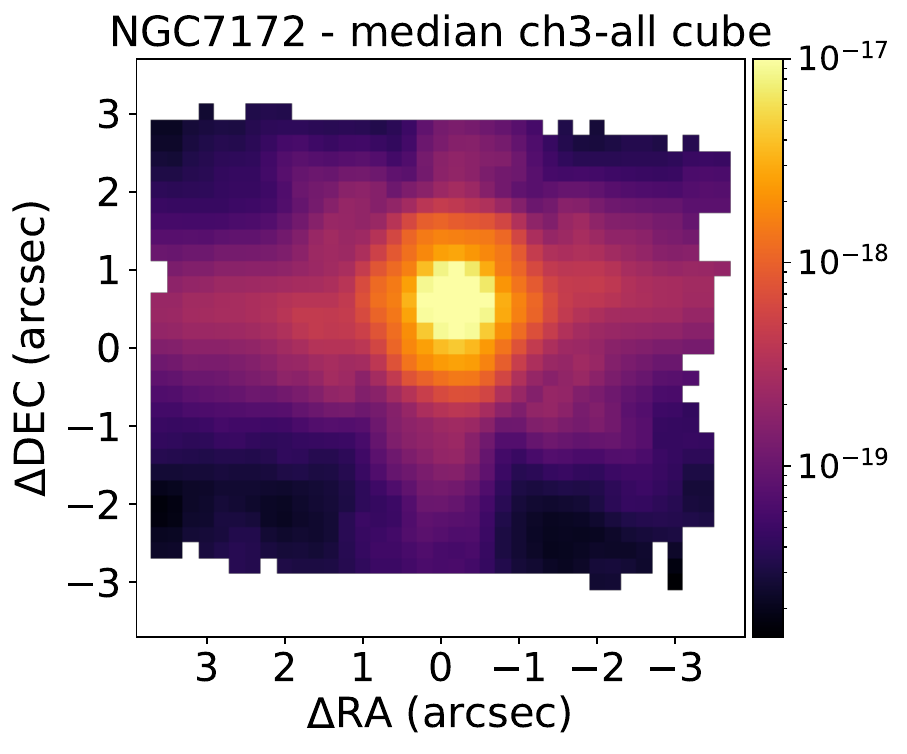}
    \includegraphics[width=.191\textwidth]{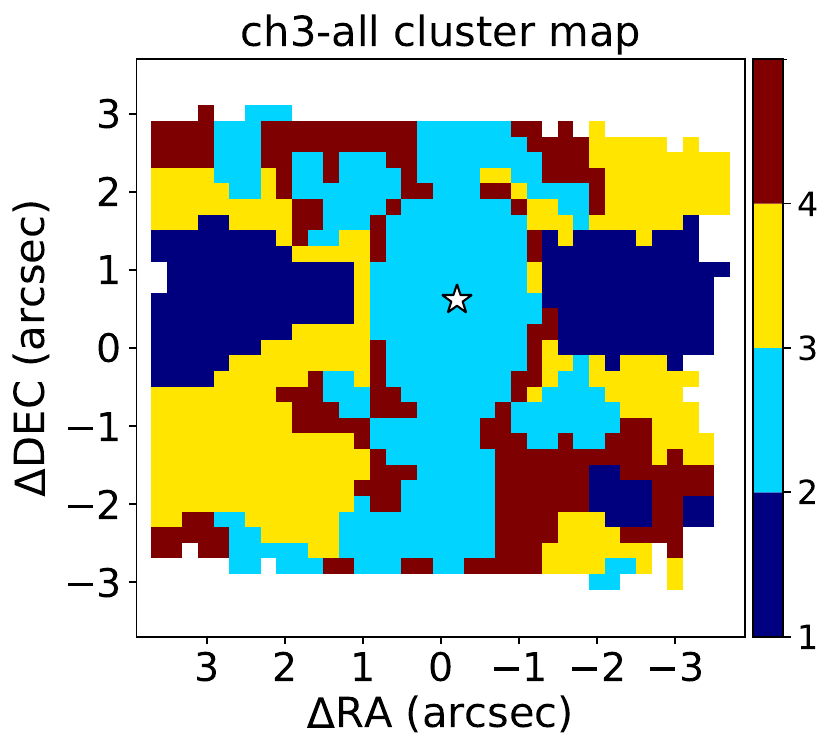}
	\includegraphics[width=.6\textwidth]{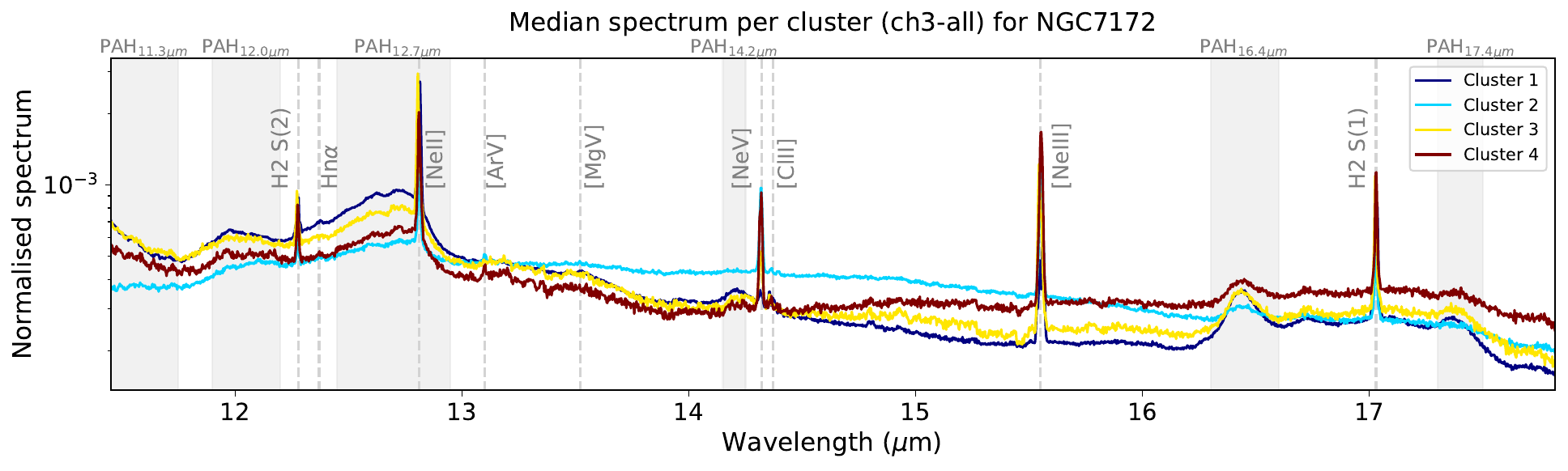}
	\caption{Clustering results of the ch3-short cube (top) and the complete ch3 channel cube (bottom) for NGC\,7172. We show in the left panels the median flux in logarithm scale of the ch3-short and ch3-all cube (top and bottom, respectively; see Appendix~\ref{Appendix_MedianMaps}). Middle panels show the cluster maps, while right panels show the median spectra per cluster in logarithm scale, normalised to the total integrated flux (see Sect.~\ref{SubSect2:Clustering}). The maps are centred in the original observed position (north is up and east to the left). The white star indicates the photometric centre. We assigned the same colours to the clusters and their respective spectrum. Colours are calculated automatically by dividing the `jet' palette in \textsc{matplotlib}. We note that both the cluster colours and numbering are arbitrary, have no physical meaning, and are assigned independently in the top and bottom panels. We mark with dashed, vertical, gray lines the main emission lines, and with grey bands the PAH features in the spectrum. The wavelength is in rest frame.}
	\label{Fig:Cluster_NGC7172}
\end{figure*}

\begin{figure*}
	\includegraphics[width=.205\textwidth]{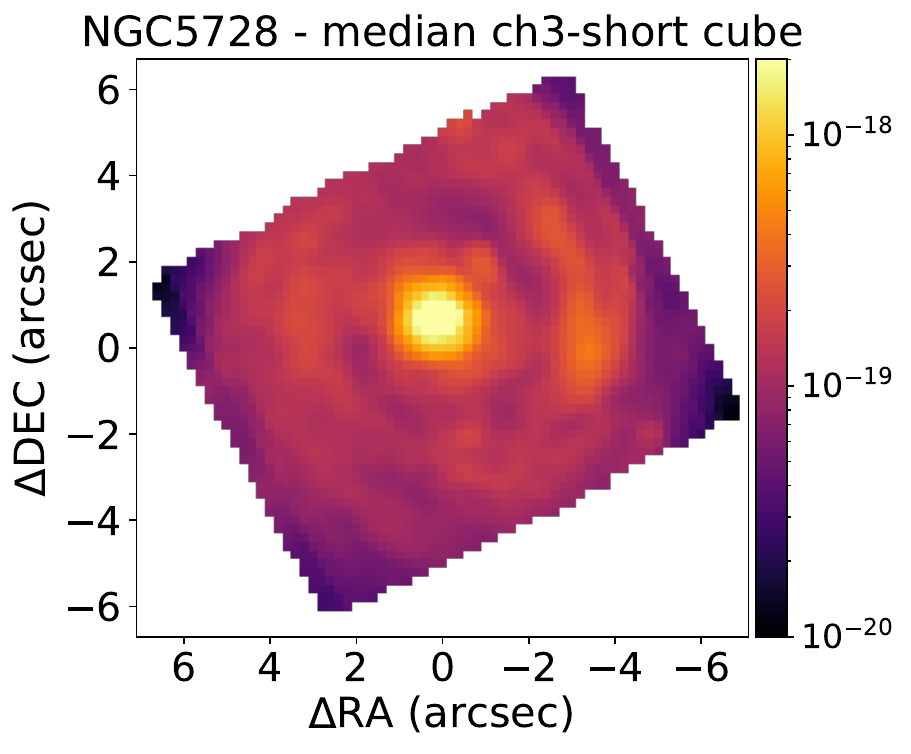}
	\includegraphics[width=.193\textwidth]{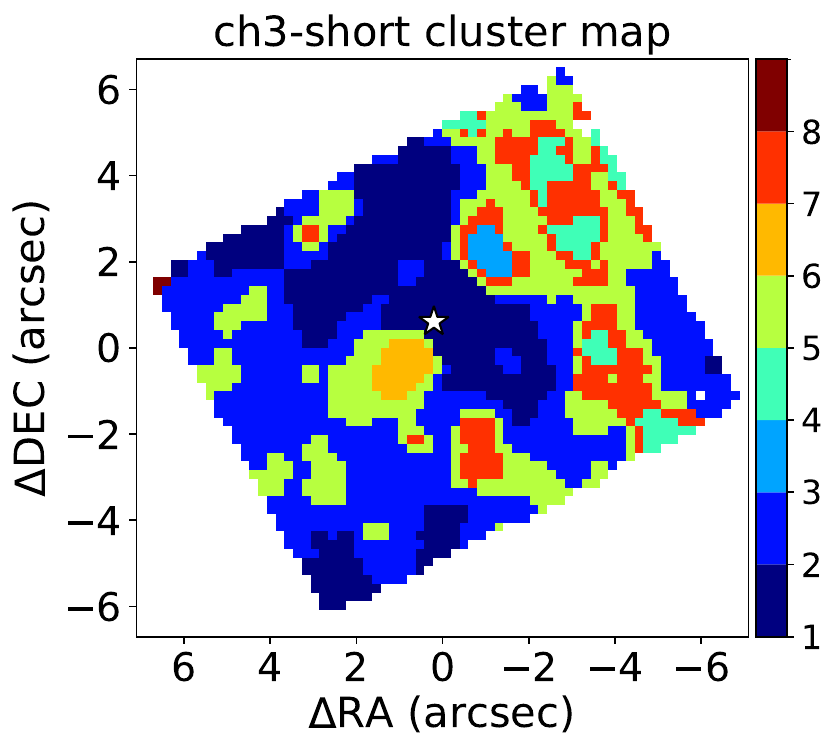}
	\includegraphics[width=.595\textwidth]{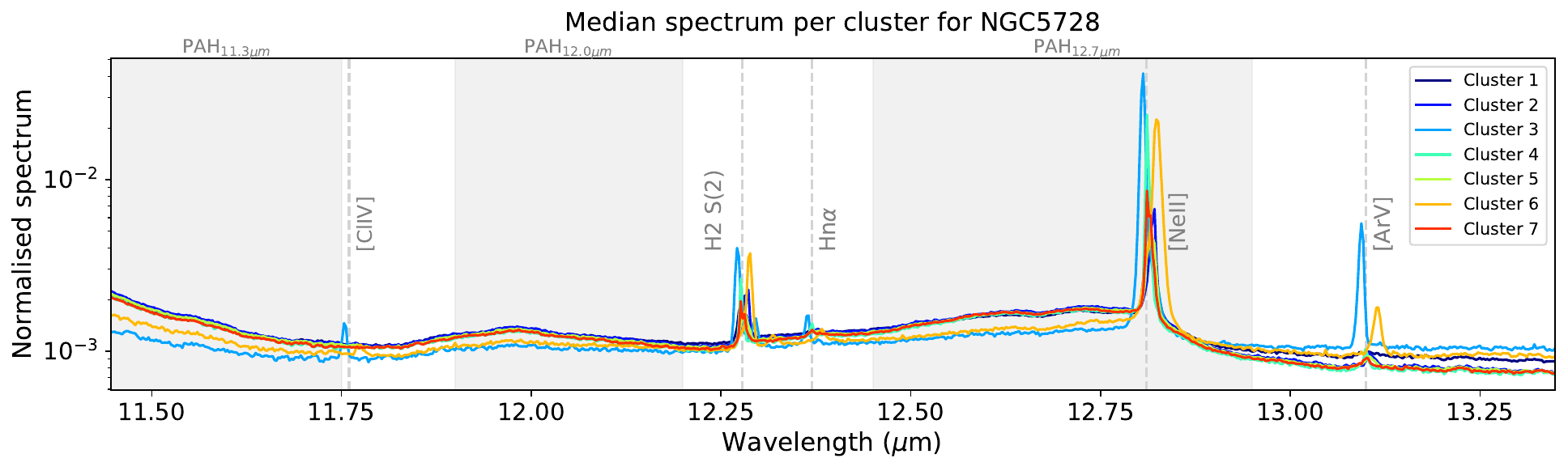}
	\includegraphics[width=.205\textwidth]{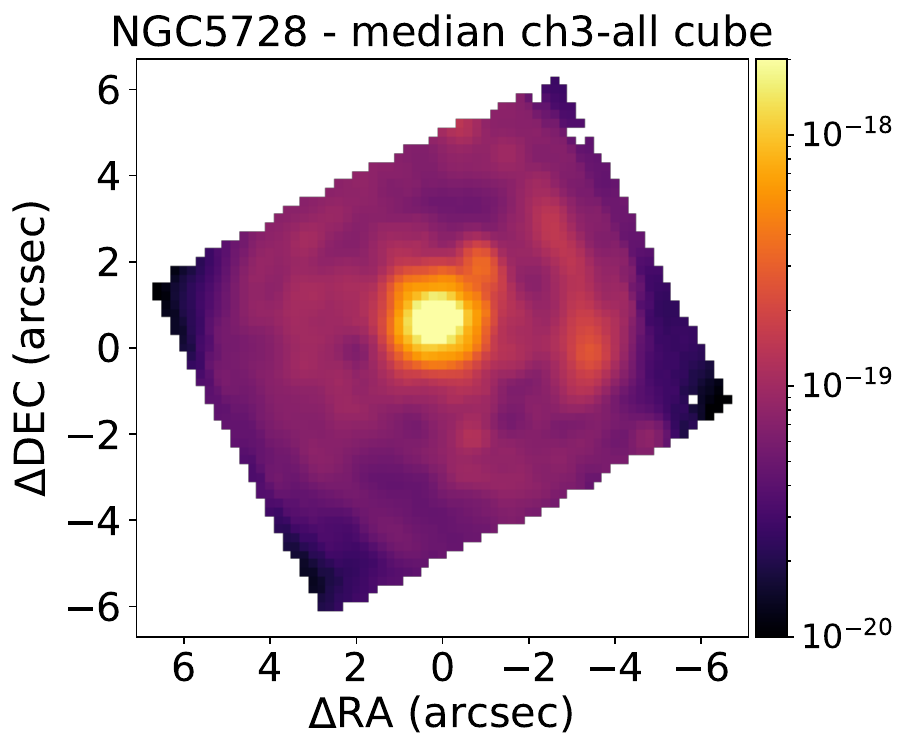}
    \includegraphics[width=.193\textwidth]{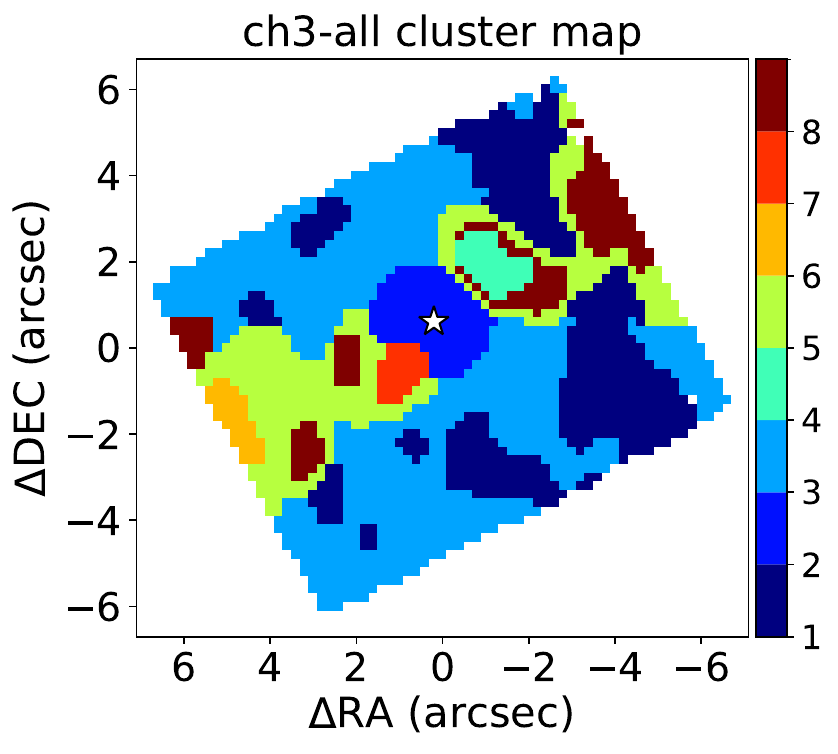}
	\includegraphics[width=.595\textwidth]{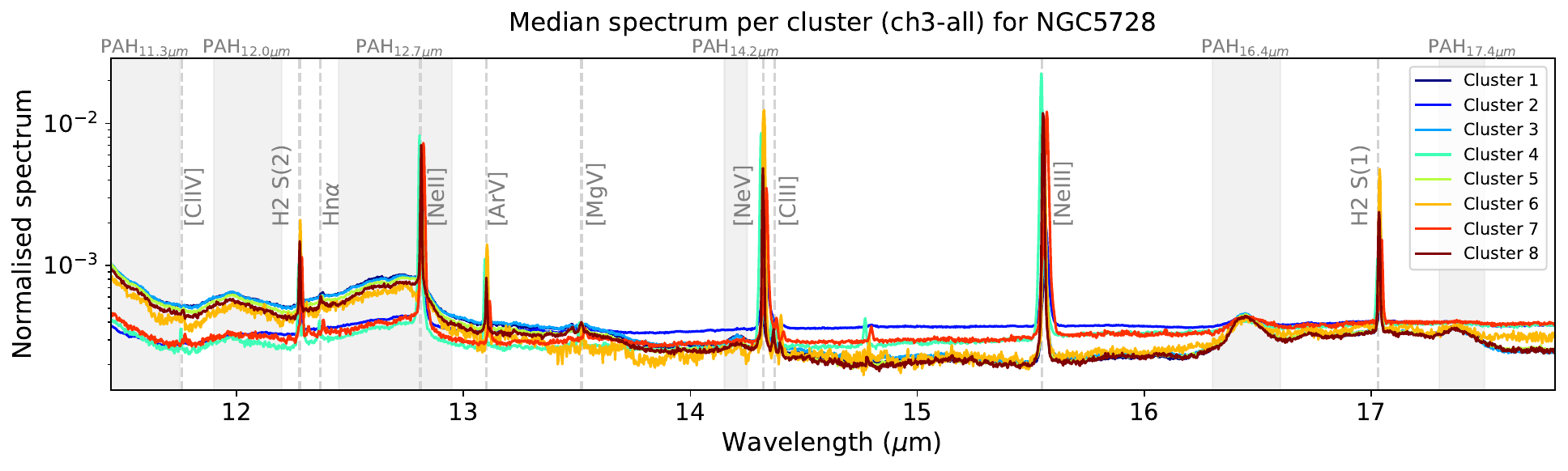}
	\caption{Same as Fig.~\ref{Fig:Cluster_NGC7172} but for NGC\,5728. We note that, for the top, right panel, we do not show the spectrum for cluster 8, as it is a low S/N cluster.}
	\label{Fig:Cluster_NGC5728}
\end{figure*}

\subsection{Main properties of the clustering maps}
\label{SubSect3:Results_clusters}

We made use of the cubes corresponding to ch3-all and ch3-short for applying the clustering technique. When using only ch3-short, the most important features for the clustering are [Ne\,II], H$_{2}$ S(2), [Ar\,V], and the PAH features at 11.3 (partly), 12, and 12.7\,$\mu$m. When considering the ch3-all cube, a similar trend is seen with all the lines and features present in this wavelength range (see Sect.~\ref{SubSect2:Clustering}), but the increased amount of information could lead to identify other structures dominated by different ionisation processes. 
This can be seen in Figs.~\ref{Fig:Cluster_NGC7172} and~\ref{Fig:Cluster_NGC5728}, where we show the results for the clustering of the ch3-short (top panels) and ch3-all cubes (bottom panels) for the galaxies NGC\,7172 and NGC\,5728, respectively. 

NGC\,7172 is a nearly edge-on galaxy with a circumnuclear ring \citep{AH2023} and a prominent ionisation cone extending almost perpendicular to the disc, previously detected with the MIRI/MRS data \citep{HM2024b,Zhang2024,GarciaBernete2024b}. These morphological features are reflected in the clustering maps (see Fig.~\ref{Fig:Cluster_NGC7172}, left). Using both ch3-short and ch3-all cubes, the system is well divided with a total of 3 and 4 clusters, respectively. In both cases, the AGN and the ionisation cone are included together within a single cluster (cluster 2 for ch3-short and ch3-all). The disc and the SF clumps detected in \cite{HM2024b} to the south-west from the nucleus, associated with positive feedback produced by the interaction of the outflow with the ISM, are clustered together in both cases (cluster 1). Finally, there is an intermediate region between the ionisation cone and the disc in both cubes (cluster 3 for ch3-short and 3 and 4 for ch3-all).

NGC\,5728, on the other hand, shows noticeably different results when comparing both cubes (see Fig.~\ref{Fig:Cluster_NGC5728}). This galaxy hosts a circumnuclear SF ring, a radio jet, and a known outflow \citep{Shimizu2019,Davies2024,GarciaBernete2024b}. In the ch3-short cube, clusters clearly trace the ring and several SF regions (clusters 4, 5 and 7), the AGN-dominated region (cluster 1), and an intermediate region (cluster 2). However, in the ch3-all cube only some of these SF regions are traced (cluster 1), whereas the majority of the clusters are aligned with the direction of the outflow and the radio jet. This suggests that different physical processes dominate the spectra, and thus the clustering, in each spectral range. In both cases, we identify the hotspot detected by \cite{Davies2024} as an independent cluster (3 for ch3-short and 4 for ch3-all), indicating that it is a differentiated, particular region of this galaxy.

From the results for the other galaxies (see figures in Appendix~\ref{Appendix_ClusterMaps}), it is clear that the clustering is affected by the strong point-spread function (PSF) of the JWST. This is particularly noticeable when using the ch3-all cube, as the PSF size increases with wavelength, but also in the ch3-short cube for some sources (e.g. NGC\,1052, see Fig.~\ref{Fig:Cluster_NGC1052}). We decided not to subtract the PSF in this work, as the main interest is to explore the results and caveats of this new technique, but its importance will be evaluated in a future work \citep[see also][]{GonzalezMartin2025}. Despite this, we are able to differentiate regions of interest for each galaxy, specially when there is extended emission. Following what was observed for NGC\,7172 and NGC\,5728, in general the AGN-PSF, ionisation cone-outflow, intermediate, and SF-disc regions are isolated for almost all the galaxies (see figures in Appendix~\ref{Appendix_ClusterMaps}). 

\subsection{Median spectrum per cluster}
\label{SubSect3:Results_MedianSpectra}

We produced the median spectrum per cluster in all galaxies (right panels in Figs.~\ref{Fig:Cluster_NGC7172},~\ref{Fig:Cluster_NGC5728}, and in Appendix~\ref{Appendix_ClusterMaps}), and only plotted those clusters with enough S/N in the continuum ($> 3$ times the median standard deviation of all the spectra for all the clusters in a given source). These spectra allow us to evaluate the most relevant features driving the clustering. 

Focusing on the spectra for NGC\,7172 (right panels in Fig.~\ref{Fig:Cluster_NGC7172}), it is clear that the AGN+ionisation cone region (cluster 2 in ch3-short and ch3-all) has faint PAH features, while they are stronger in the disc region (cluster 1 in ch3-short and ch3-all). The intermediate region (cluster 3 in ch3-short and clusters 3 and 4 in ch3-all) has moderate PAHs, and more complex emission line profiles compared to the disc region. Although we have applied a velocity correction to the spectra (see Sect.~\ref{SubSect2:Clustering}), we also detect shifts in the peak of the emission lines, that suggest velocity differences between the clusters.
For the ch3-all cube, these differences on the PAH features and the emission lines are also seen, and additionally some variations in the shape of the continuum between clusters. 

When focusing on the median spectra for NGC\,5728 (see right panels of Fig.~\ref{Fig:Cluster_NGC5728}), the differences in the ch3-all spectra are less evident. In this case, we increased the number of clusters to capture the emission from the SF regions in the galaxy (see Fig.~7 in \citealt{Shimizu2019}), resulting in a subdivision along the jet/outflow axis, with clusters that have similar spectral properties. For a more in depth analysis of a particular source, if these apparently similar clusters share the same physical properties, they should be merged together to simplify the maps. In contrast, the ch3-short spectra are more different, mainly due to the emission lines, as found for NGC\,7172. Cluster 3 is the most distinct cluster, with very prominent high excitation lines (i.e. [Ar\,V] and [Cl\,IV] at 11.76\,$\mu$m), coinciding with the hotspot \citep{Davies2024}. Velocity differences for the clusters are seen as in NGC\,7172. These could be related to physical differences such as the presence of multiple kinematic components, as already noted in previous works \citep[e.g.][]{Davies2024}.

In general, these trends are repeated for all the AGN galaxies in the sample. We see that the nuclei tend to have faint PAH emission and strongest high excitation lines. These characteristics are consistent with what is typically observed when comparing to spectra of discs or SF regions (see e.g. NGC\,7469 in Fig.~\ref{Fig:Cluster_NGC7469}). We also detect low S/N spectra in certain clusters in some galaxies (see e.g. Figs.~\ref{Fig:Cluster_NGC1052} and~\ref{Fig:Cluster_NGC3081}), that mostly correspond to a few spaxels located at the edges of the FoV. 
In the LINERs NGC\,1052 and NGC\,4594 (see Figs.~\ref{Fig:Cluster_NGC1052} and~\ref{Fig:Cluster_NGC4594}, respectively) the spectra are mostly flat, with the main feature separating the regions being the emission lines, which are quite broad in all cases, as already discussed in previous works \citep{Goold2024}. This is likely because these type of objects host mainly old stellar populations. A similar trend is seen for IC\,5063 (see Fig.~\ref{Fig:Cluster_IC5063}) and for NGC\,7319 (it has little gas due to past interactions, \citealt{PereiraSantaella2022} and references therein; see Fig.~\ref{Fig:Cluster_NGC7319}), although in these cases, the spectra of some clusters do show a mild contribution from the PAH features. Both galaxies host a low-intermediate power radio jet that is interacting with the ISM, with several radio hotspots differently affected by the interaction \citep{PereiraSantaella2022, Dasyra2024}. Finally, for the starburst galaxies NGC\,3256\,N and M\,83 (see Figs.~\ref{Fig:Cluster_NGC3256} and~\ref{Fig:Cluster_M83}) the spectra for all the clusters are very similar, mainly separated by the shape of the PAH features in ch3-all. In general, they do not show high excitation lines, except for clusters 1 and 3 in M\,83. While the median spectrum shows only a faint [Ne\,V] line, the error spectrum, computed as the standard deviation of all the spectra within the cluster, reveals the line more clearly (see the insets in Fig.~\ref{FigAp:M83_NeV}), in agreement with the regions highlighted in \cite{Hernandez2025}.

\subsection{Line ratios per cluster}
\label{SubSect3:Results_Ionisation}

We measured the fluxes of all the emission lines and PAH features present in the spectra for all the clusters, as well as the slope of the continuum (see Sect.~\ref{SubSect2:Clustering}). We created line ratios accounting for all the available possibilities both for the ch3-short and ch3-all cubes to compare how the clusters behave for the different galaxies. By selecting lines close in wavelength, the differential extinction effects are minimised \citep{HernanCaballero2020,Donnan2024}. The histograms showing the distribution of some of these ratios measured in the clusters obtained with the ch3-all cubes are presented in Fig.~\ref{FigAp:Histograms}. From previous works, there are promising line ratios in the 11.5 to 17.5$\mu$m mid-IR range that could help to disentangle between AGNs, starburst, and/or sources affected by other ionisation mechanisms, such as [Ne\,III]/[Ne\,II] or [Ne\,V]/[Ne\,II], among others, for nearby galaxies \citep[see e.g.][]{Pereira2010,Inami2013,MartinezParedes2023,Feltre2023,GarciaBernete2024b,Feuillet2024,RamosAlmeida2025,AH2025}.  

Considering only the ch3-all cubes, we have a total of 65 selected clusters for all galaxies, after excluding those with low S/N in the continuum. We estimated their line ratios, taking into account that, as expected, some clusters lack some features in their spectra, such as high excitation lines or PAHs. For those clusters that are associated with regions whose physical origin was already known from previous analysis of the MIRI/MRS data (see Table~\ref{Table:1} for the references), we assigned them labels as explained in Sect.~\ref{SubSect2:RFtecnhique}. In total, we set the initial labels for 49\% of the clusters, as shown in Table~\ref{Table:2}. 

\begin{figure}
	\centering
	\includegraphics[width=.48\textwidth]{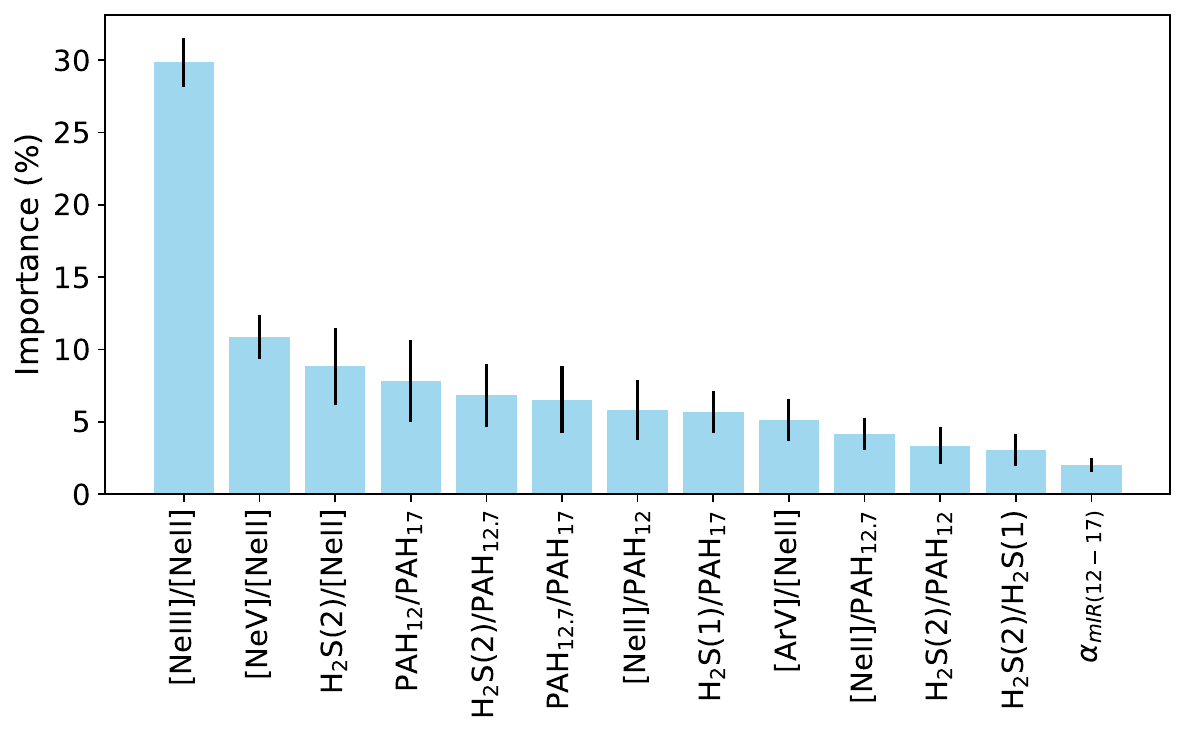}
	\caption{Histogram of the average, relative importance of the features measured in the ch3-all cubes obtained from the automatic classification of the clusters (see Sect.~\ref{SubSect3:Results_Ionisation}). The errorbars are estimated as the standard deviation of all the importances for each feature calculated using Monte Carlo simulations (n$=$1000, see Sect.~\ref{SubSect2:RFtecnhique}).}
	\label{Fig:HistogramRatiosImportance}
\end{figure}

From the resulting RF model, we obtained the relevance of each line ratio to classify the clusters. This output quantifies the relative weight of each feature compared to the rest for the trained model (see Sect.~\ref{SubSect2:RFtecnhique}). Figure~\ref{Fig:HistogramRatiosImportance} presents all the features evaluated by the model (ratios and $\alpha_{mIR}$) ordered by their importance. The most relevant ratios for ch3-all are: [Ne\,III]/[Ne\,II] ($30\pm 2\%$), [Ne\,V]/[Ne\,II] ($11\pm 2\%$), H$_{2}$ S(2)/[Ne\,II] ($9\pm 3\%$), PAH$_{12\mu m}$/PAH$_{17\mu m}$ ($8\pm3\%$), and H$_{2}$ S(2)/PAH$_{12.7\mu m}$ ($7\pm2\%$). The slope of the continuum, $\alpha_{mIR}$, shows the lowest importance for the classification ($\sim$2\%). In fact, considering the uncertainties of the relevance for each feature (see Fig.~\ref{Fig:HistogramRatiosImportance}), ratios with importance below 10\% are equally (un)important for the model, meaning that they are exchangeable in terms of classifying the clusters. With this in mind, we prioritise the H$_{2}$-based ratios to create the diagrams, as they allow us to construct diagnostic diagrams for the ch3-short cubes as well, as explained below. The diagrams with the PAH$_{12\mu{\rm m}}$/PAH$_{17\mu{\rm m}}$ ratio are included in Appendix~\ref{Appendix} (see Fig.~\ref{FigAp:DiagnosticsPAHs}) and discussed in Sect.~\ref{SubSect4:Disc_Association}.

\begin{figure*}
	\centering
	\includegraphics[width=.267\textwidth, trim={0 0 10cm 0},clip]{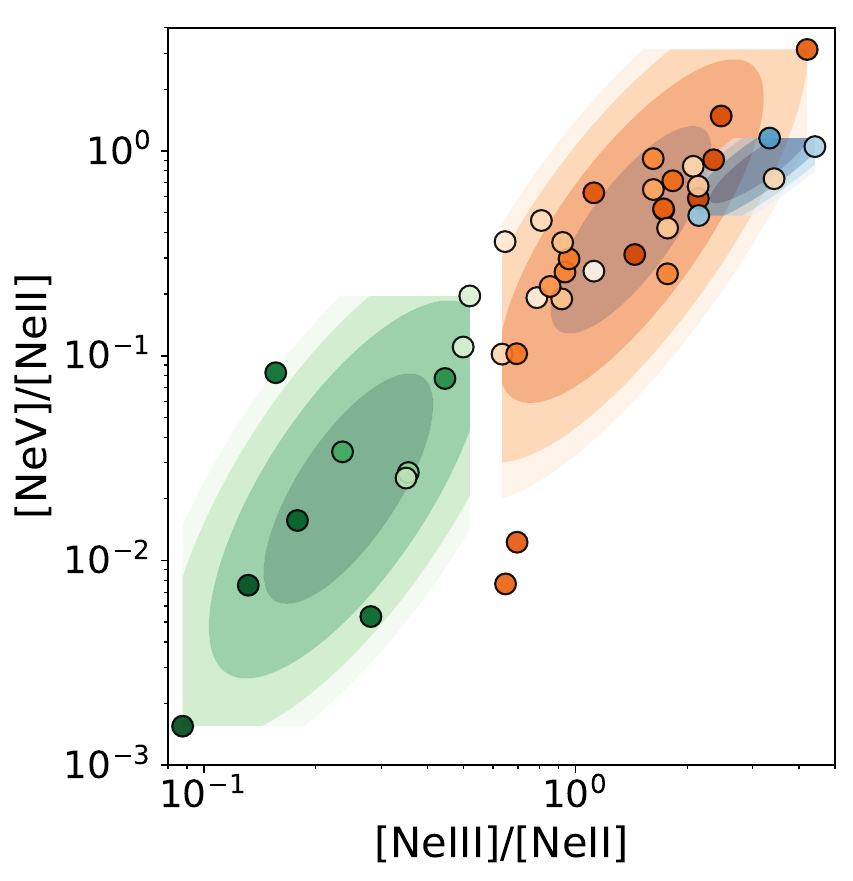}
	\includegraphics[width=.268\textwidth, trim={0 0 10cm 0},clip]{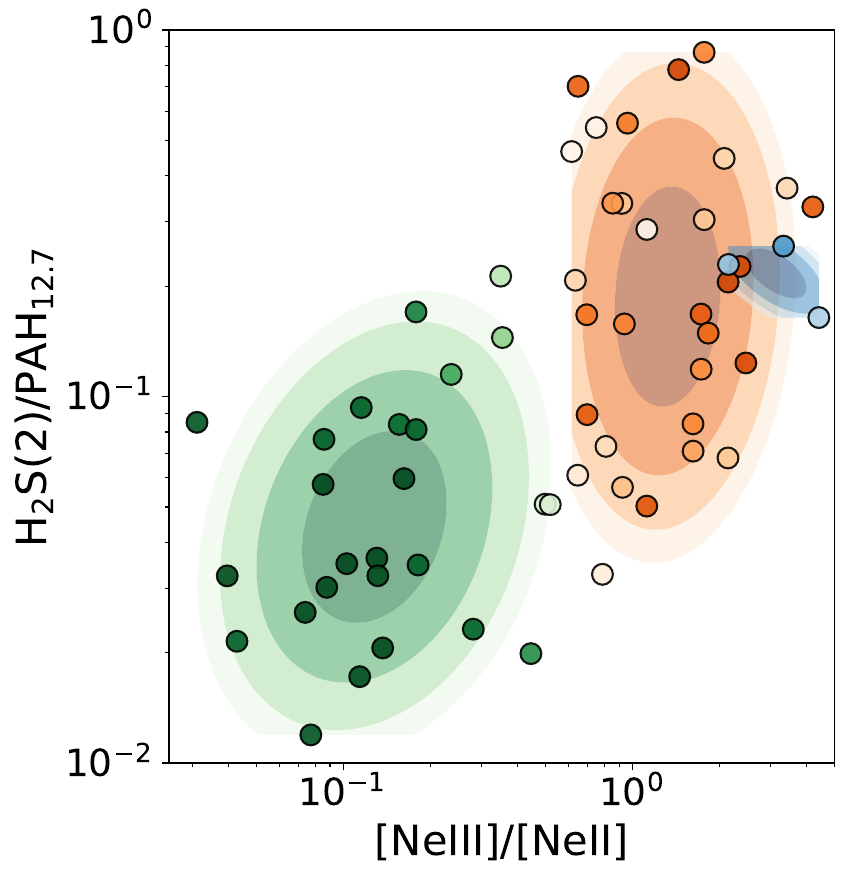}
    \includegraphics[width=.45\textwidth]{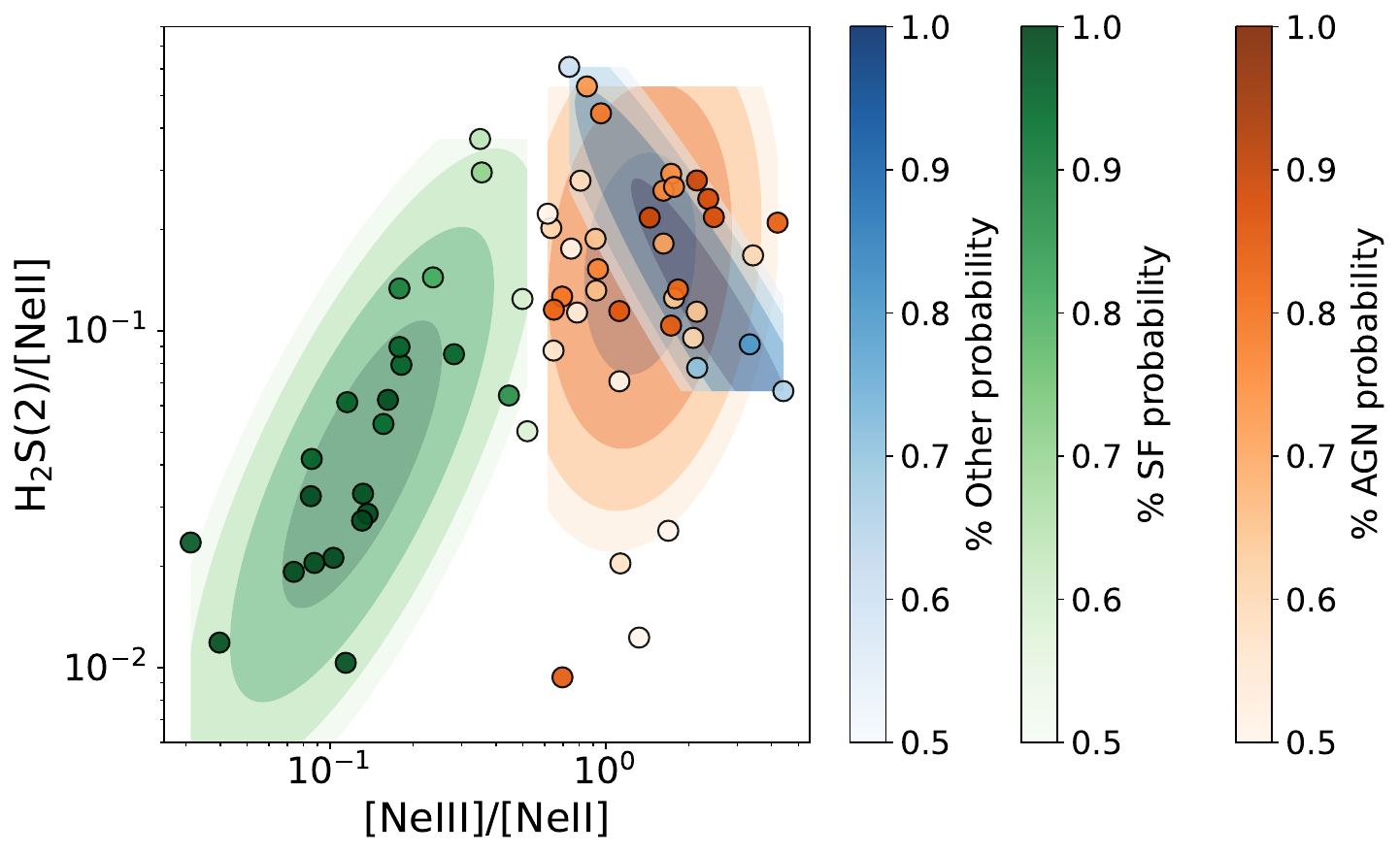}
	\caption{Diagnostic diagrams based on the best preferred line ratios using the ch3-all cubes, in a logarithmic scale (see Fig.~\ref{Fig:HistogramRatiosImportance} and details in Sect.~\ref{SubSect3:Results_Ionisation}). Each point is a cluster from the galaxies used as the training sample, colour-coded by their assigned class probability (AGN in orange, SF in green, and Other in blue; see details in Sect.~\ref{SubSect3:Results_Ionisation}), with darker colours indicating a higher probability. The initial training labels of the clusters (see Table~\ref{Table:2}) were obtained from previous detailed JWST MIRI/MRS analysis of the sources (see references in Table~\ref{Table:1}, and details in Sect.~\ref{SubSect2:RFtecnhique}). Contours show the kernel density estimate (KDE) of the distribution for each class at four probability levels: 0.5, 0.6, 0.75, and 0.9.}
	\label{Fig:LineRatios1}
\end{figure*}

In Fig.~\ref{Fig:LineRatios1}, we present the diagnostic diagrams created combining the most relevant ratios, with their probabilistic classification, for all the galaxies from the training sample. We detect a separation in the [Ne\,III]/[Ne\,II] ratio between regions dominated by SF and the rest of the clusters (see Fig.~\ref{Fig:LineRatios1} and histogram in Fig.~\ref{FigAp:Histograms}). Clusters corresponding to NGC\,3256-N, M\,83, NGC\,7469, and the disc in NGC\,7172, are classified with the largest probabilities of being SF regions (greenish points), and have log([Ne\,III]/[Ne\,II]) $< -0.5$. These clusters have little to no emission of high IP gas, such as [Ne\,V], as they are SF dominated, so most do not appear in the [Ne\,V]/[Ne\,II] diagram (see Figs.~\ref{Fig:LineRatios1} and~\ref{FigAp:Histograms}). 
The H$_{2}$ S(2)/PAH$_{12.7\mu m}$ ratio also shows a bimodality, particularly when log([Ne\,III]/[Ne\,II]) $> -0.5$ (see Fig.~\ref{Fig:LineRatios1} and~\ref{FigAp:Histograms}). A composite region with AGN-like and Other-like clusters is found at larger values of log(H$_{2}$ S(2)/PAH$_{12.7\mu m}$) and log([Ne\,III]/[Ne\,II]). This would be in agreement with previous works showing that the ratio is approximately constant for starbursts, while it is increased in the presence of an AGN or shocks \citep{Roussel2007,Lambrides2019,Riffel2020,Zhang2022,Riffel2023,GarciaBernete2024}. A similar trend is found for the H$_{2}$ S(2)/[Ne\,II] ratio, although the clusters are more concentrated and mixed than in the previous case at log([Ne\,III]/[Ne\,II]) $>-0.5$. 

In contrast, and as predicted by the random forest classifier, there are other line ratios such as H$_{2}$ S(2)/H$_{2}$ S(1) (see Fig.~\ref{FigAp:Histograms}), related to the excitation temperature of the warm molecular gas, that show a similar distribution for all galaxies, and thus are not useful for separating regions. We note however that these two warm molecular gas lines have relatively close upper level energies. We cannot discard that combining other H$_2$ lines at shorter mid-IR wavelengths could help to disentangle different ionisation regions, as proposed in previous works both with Spitzer and JWST data \citep[see e.g. Fig.~18 in][]{Lambrides2019, Togi2016, CostaSouza2024, RamosAlmeida2025}.

\begin{figure}
	\centering
	\includegraphics[width=.95\columnwidth]{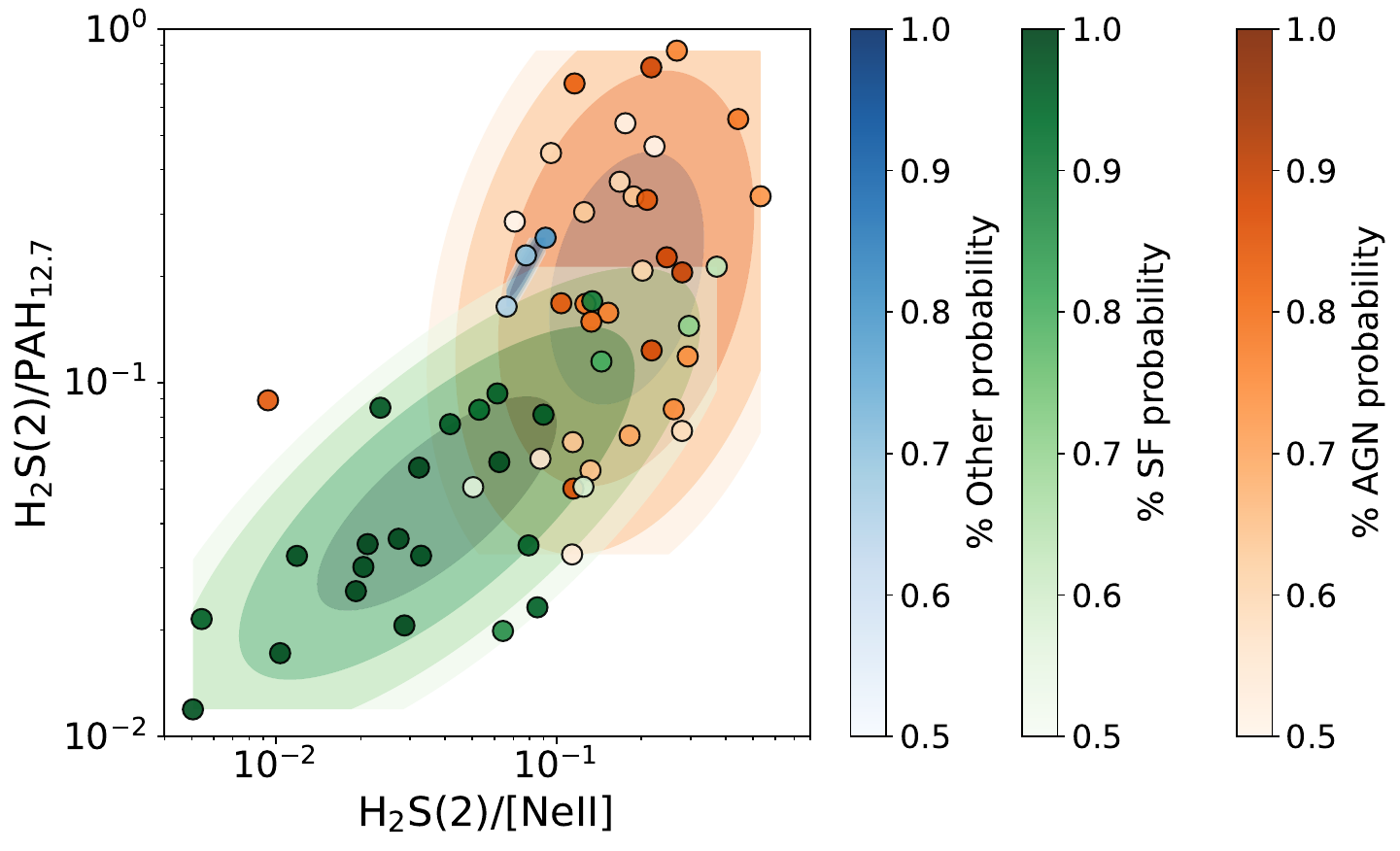}
	\caption{Diagnostic diagram in a logarithmic scale similar to Fig.~\ref{Fig:LineRatios1}, but using the most relevant line ratios available for both ch3-short and ch3-all cubes for all the galaxies from the training sample (see Sect.~\ref{SubSect3:Results_Ionisation}).}
	\label{FigAp:LineRatios_Ch3-short}
\end{figure}

Additionally, we created equivalent diagnostic diagrams for the lines detected in ch3-short (see Fig.~\ref{FigAp:LineRatios_Ch3-short}). In particular, based on the results from Fig.~\ref{Fig:HistogramRatiosImportance}, the most relevant lines that can be used in both data cubes are H$_{2}$ S(2)/[Ne\,II] and H$_{2}$ S(2)/PAH$_{12.7\mu m}$. The clusters associated with SF regions, such as those of NGC\,7469 or M\,83, are found in the lower left part of the diagram, while those associated with AGNs, such as the nuclei, are in the upper right part. Similarly to what was found for ch3-all (see Fig.~\ref{Fig:LineRatios1}), we see a bi-modality with this diagram, that seems to be able to separate between both SF and AGN ionised regions, although with a large composite region.

\section{Discussion}
\label{Sect4:Discussion}

With the clustering process we aimed at providing a method to identify physically distinct regions of a galaxy. In this section we discuss about the main aspects to be considered when applying this methodology. In Sect.~\ref{SubSect4:Disc_ShortVSAll}, we explore the possible differences on the results of the clustering when using the ch3-short or the ch3-all cubes. In Sect.~\ref{SubSect4:Disc_Caveats} we present the caveats for the applied methodology, particularly focusing on exploring other MIRI/MRS channels, and in the automatic classification process of the clusters. Finally, we discuss the diagnostic diagrams in Sect.~\ref{SubSect4:Disc_Association}, and use the trained model for evaluating three galaxies, NGC\,3227, NGC\,4051 and NGC\,7582, in Sect.~\ref{SubSect4:Disc_NGC7582}. 

\subsection{The importance of the wavelength range: ch3-short vs ch3-all}
\label{SubSect4:Disc_ShortVSAll}

As seen in Sect.~\ref{Sect3:Results}, the clustering results derived from using only the ch3-short cube versus the complete ch3 spectra can differ. This implies that the selected wavelength range can impact the clustering process results (see a discussion for other MRS channels in Sect.~\ref{SubSubSect4:Disc_Clustering}). The ch3-all cube includes more features that can be used to evaluate the performance of the method, such as several neon transitions ([Ne\,II], [Ne\,III], and [Ne\,V]). These are the brightest lines in this wavelength range, and trace gas with different ionisation levels (see Sect.~\ref{SubSect2:Clustering}). This is in contrast with the wavelength range covered by ch3-short, where [Ar\,V] traces the high excitation regions, although it is much weaker than [Ne\,V], and there are no intermediate excitation lines (IPs $\sim$30\,eV to 90\,eV), except for [Cl\,IV] at 11.76 $\mu$m (IP $53.5$\,eV), which is only detected in the hotspot of NGC\,5728 \citep[see][]{Davies2024}. Additionally, ch3-all cubes have a broader continuum range than ch3-short cubes. This is relevant, as more features are available to create the clusters, as well as possible changes in the continuum, thus helping to separate potentially physically distinct regions. 

Despite this, there are some objects, such as NGC\,7172 and NGC\,7469 (see Figs.~\ref{Fig:Cluster_NGC7172} and~\ref{Fig:Cluster_NGC7469}), with similar clustering results using both the ch3-short and ch3-all cubes. This suggests that in these objects the different physical regions (i.e. nuclei, disc, etc.) are more clearly separated, and thus there is a clear dominant ionising mechanism. This is in contrast to the results for more complex systems, such as NGC\,5728, where the differences between both cubes are evident, see Fig.~\ref{Fig:Cluster_NGC5728} and Sect.~\ref{SubSect3:Results_clusters}). For this galaxy, \cite{GarciaBernete2024b} reported a strong coupling between the outflow and the disc, which significantly disturbs the ISM, suggesting that the gas is highly mixed and inhomogeneous. In these cases, the broadest wavelength range is to be preferred, as the larger variety of spectral features could be used to obtain a more comprehensive view of the physical processes at play.

Based on the results obtained with the random forest classifier, among the most important line ratios for classifying the clusters (see Fig.~\ref{Fig:HistogramRatiosImportance} and Sect.~\ref{SubSect3:Results_Ionisation}), there are spectral features than can be computed with both wavelength ranges. Specifically, those including H$_{2}$\,S(2), PAH$_{12.7\mu m}$, and [Ne\,II]. In fact, as shown in Fig.~\ref{FigAp:LineRatios_Ch3-short} (see also Sect.~\ref{SubSect3:Results_Ionisation}), the ch3-short range alone is useful for separating between SF and AGN/Other dominated regions. However, the [Ne\,III]/[Ne\,II] ratio, only available when using the complete ch3 cube, is the preferred ratio to separate SF and AGN-ionised regions not only in this work, but also as shown previously in several works in the literature using Spitzer/IRS spectroscopy \citep[see e.g.][]{Pereira2010,Inami2013,MartinezParedes2023}.

\subsection{Caveats of the methodology}
\label{SubSect4:Disc_Caveats}

\subsubsection{Clustering technique}
\label{SubSubSect4:Disc_Clustering}

The unsupervised hierarchical clustering can be applied to a variety of data cubes observed at different wavelengths beyond the mid-IR traced by ch3 of MIRI/MRS. Nevertheless, to correctly interpret the result of this clustering technique when applying it to a new, untested dataset, a number of possible caveats need to be considered. 

The choice of the spectral range significantly affects the clustering results, as the dominant spectral features change (see e.g. NGC\,5728 in Sect.~\ref{SubSect3:Results_clusters} and Fig.~\ref{Fig:Cluster_NGC5728}). This is particularly relevant when all the spectral features are equally dominant. Our tests with MIRI/MRS channel 1 (from 4.9 to 7.6\,$\mu$m) revealed that the presence of many different features (low, intermediate, and high excitation lines, as well as H$_2$, PAHs, and ices), without any particularly dominant line, made it difficult for the algorithm to separate physically distinct regions. More specifically, the regions that are identified with ch3, such as SF regions or the ionisation cones, are not clearly detected by using ch1 for some galaxies. This happens despite the presence of [Fe\,II] lines, typically used to identify shocks, and high excitation lines such as [Mg\,V], most likely produced by AGN ionisation. In channel 2, there are no low excitation lines, which means that the SF regions are not distinctively differentiated as clusters, thus they remain undetected with this method. Channel-2 should be evaluated carefully, as prominent silicate absorption features may introduce extinction effects on the observed spectral features, such as the H$_{2}$ S(5) line at 9.66\,$\mu$m, and it may also show additional physical effects that are not recovered in the other channels. 
In channel 4 we have the lowest spatial and spectral resolution, and the largest PSF contribution, which may affect the detection of some interesting regions, as we have already encountered in ch3 for some galaxies (see Sect.~\ref{SubSect3:Results_clusters}). In a future work, the PSF subtraction tool created by \cite{GonzalezMartin2025} could be used to subtract the PSF and test the methodology for the most affected data cubes, and evaluate in detail all of the other channels (including the complete MIRI/MRS cube with the whole mid-IR wavelength range), or even other wavelength ranges such as the optical, or near-IR with NIRSpec. 

The combination of MIRI/MRS subchannels to create the ch3-all cubes, using the pipeline (see Sect.~\ref{SubSect2:Data}), can introduce systematic errors. These are translated into small flux discontinuities at the wavelengths where the cubes are combined, as well as possibly introducing errors in the spaxels located at the edges of the FoV due to shifts of the centroids. We detected such flux offsets for only two objects, NGC\,5506 and IC\,5063 (see bottom panels in Figs.~\ref{Fig:Cluster_NGC5506} and~\ref{Fig:Cluster_IC5063}). When present, such flux differences are expected to affect the majority of the spaxels in the corresponding ch3-all cube. Given that each datacube is clustered independently, based exclusively on its internal structure, any offset will affect all spaxels uniformly. As a result, the offsets will not bias the clustering results. As for the potential shifting errors, most of the clusters excluded due to low S/N lie close to the FoV limits, and therefore have been likely excluded from the analysis. We also note that combining all MIRI/MRS channels will imply a significant reduction of the covered FoV (3.2$\arcsec\times$3.7$\arcsec$ for ch1 vs 5.2$\arcsec\times$6.2$\arcsec$ for ch3; \citealt{Labiano2021}), reducing the information currently recovered for the outermost regions.

The initial normalisation of the spectra is equally important. If we were to normalise in a particular wavelength region instead of using the integrated flux per spaxel (see Sect.~\ref{SubSect2:Clustering}), the slope of the continuum would change significantly, especially at the ends of the wavelength range, thus altering the clustering. Also, variations on the results would appear depending on where this normalisation region is selected. Implicitly, this method assumes that there is a continuum to normalise, which may not always be the case (for example, extended gas emission in the outer parts of a galaxy). 

Finally, the selection of the number of clusters is currently done through visual inspection (see Sect.\ref{SubSect2:Clustering}). This could be refined in the future by using, for example, PCA \citep[see e.g.][]{Steiner2009}, that will allow for a more robust and quantitative estimation of the most suitable number per each galaxy. 

The method in this work should serve as a tool to identify regions of interest within a cube, which can help to guide a future, in-depth analysis of a specific galaxy. 

\subsubsection{Automatic classification of the ionising source of the clusters}
\label{SubSubSect4:Disc_AutomaticClassification}

As mentioned in Sect.~\ref{SubSect3:Results_Ionisation}, we have prior information about the physical origin of some clusters associated with particular regions of the galaxies, but we could not assign labels to all of them. Thus from the dataset, only a relatively small subset of clusters (49\%, see Sect.\ref{SubSect3:Results_Ionisation}) could be used to train the classifier, potentially reducing the robustness of the classification results. 

In addition, the relative distribution of the labels across classes plays an important role in the performance of the classifier. Our initial sample is unbalanced, with an underrepresentation of the ``Other" class ($\sim13$\%, i.e. 4 of the initial labels, see Fig.~\ref{FigAp:BalanceClasses}). This limited representation may affect the generalisation capability of the RF classifier for this specific class, particularly given the relatively small size of the labelled dataset. 

Despite this, as shown in Sect.~\ref{SubSect3:Results_Ionisation}, clusters associated with known discs, starbursts, and SF regions are all classified as SF with high probabilities ($\geq85\%$), occupying a well-defined part of the diagnostic diagrams (see Fig.~\ref{Fig:LineRatios1} and Sect.~\ref{SubSect4:Disc_Association}). This suggests that, in our case, misclassified clusters are probably associated with composite regions, maybe affected by a combination of AGN ionisation, shocks, and/or other processes, making automatic classification challenging. If we were to use other MRS wavelength ranges, thus accounting for other emission lines, we could create further diagnostic diagrams potentially useful, as those discussed in \cite{Feltre2023} \citep[see also][]{Zhang2025,Ceci2025}.

The results of this machine learning approach should be considered as a preliminary test, but a larger dataset is needed to further probe and better constrain the results suggested by the diagnostic diagrams proposed in Fig.~\ref{Fig:LineRatios1}.  

\subsection{Identifying the ionising source of the clusters}
\label{SubSect4:Disc_Association}

Figure~\ref{Fig:LineRatios1} shows the best preferred diagnostic diagrams to classify the clusters. The [Ne\,III]/[Ne\,II] ratio has been proposed to separate SF and AGN ionisation in several previous works in the mid-IR, includying spatially-resolved studies \citep[see e.g.][]{Groves2006,Pereira2010,Inami2013,GarciaBernete2024b,HM2024b,HM2025,Feuillet2024,Zhang2025}. This ratio depends on the intensity and hardness of the radiation field, meaning that a larger value is associated with a more energetic ionising source, such as an AGN. We find a clear separation at $\sim -0.5$ in all diagrams for the most probable SF regions ($> 90\%$), although there are other SF regions at larger values, together with other AGN-classified clusters, in what we can define as composite regions (generally between log([Ne\,III]/[Ne\,II]) $\sim -0.5$ and $3$, see Fig.~\ref{Fig:LineRatios1}). This ratio compared to the [Ne\,V]/[Ne\,II] is a well-known estimator of AGN versus SF regions. Indeed, the nuclear regions of Sy galaxies and quasars tend to fall in the upper right part of this diagram \citep{Zhang2024,HM2025,RamosAlmeida2025,AH2025}. The location of shocked regions in this diagram is uncertain, although photoionisation model predictions put it between the AGN and SF regions \citep[see e.g.][]{Feltre2023,Zhang2024b,Zhang2025,Ceci2025}. It is thus likely coincident with the composite region seen in our diagram at log([Ne\,V]/[Ne\,II]) below $\sim -0.3$ and log([Ne\,III]/[Ne\,II]) above $\sim -0.5$. In fact, the AGN distribution resembles that shown in Fig.~4 in \cite{Zhang2024b}, although their models predict higher values of [Ne\,III]/[Ne\,II] than what we find. This could be a combination of them using larger apertures (3\arcsec$\times$3\arcsec) than the average sizes of our clusters (average area of 6.9\arcsec$^{2}$), and also probably because we are still lacking a representative AGN distribution (see also Fig.~B.7 in \citealt{Zhang2025}). 

\cite{Riffel2025} compared the distributions of H$_2$S(3)/PAH$_{11.3\mu m}$ for AGN and non-AGN galaxies observed with Spitzer \citep{Lambrides2019} with their MIRI/MRS galaxies. They systematically detected higher values of this ratio for the JWST-observed AGN. In general, SF and AGN dominated systems can also be distinguished using other warm H$_2$ transitions, such as H$_2$S(1)/PAH$_{11.3\mu m}$ (\citealt{GarciaBernete2024b}, see also \citealt{Pereira2010b}), with the largest values associated with AGN ionisation, similar to what we detect (see middle panel in Fig.~\ref{Fig:LineRatios1}). These ratios are particularly high for the outflow region of NGC\,5728, as it is strongly coupled with the jet and the host galaxy \citep{GarciaBernete2024b,Davies2024}. 

As for the H$_2$S(2)/[Ne\,II], SF emission tends to increase [Ne\,II], whereas in LINERs, where shocks are present, this ratio appears increased \citep{Roussel2007}. This is consistent with our results, although the regions are more mixed up than for the previous ratio (see bottom panel of Fig.~\ref{Fig:LineRatios1}). 

PAH ratios and diagnostic diagrams have been widely discussed in previous works from the literature to disentangle regions with different ionisation conditions \citep{Draine2007,Draine2021,Rigopoulou2021, AH2014, GarciaBernete2022a, GarciaBernete2022c, GarciaBernete2024b, Zhang2024}, although normally different species are considered, such as PAHs at 7.7\,$\mu$m or 11.3\,$\mu$m (associated to neutral and ionised molecules, respectively). PAHs are well-known tracers of SF activity and they can be destroyed due to the intense radiation field of the AGN. However, recent works with JWST observations have seen that neutral PAHs are more resilient near Seyfert-like AGN \citep{GarciaBernete2022a,GarciaBernete2024b}. In our diagram (see Fig.~\ref{FigAp:DiagnosticsPAHs}), the main separation between SF and AGN is given by the neon ratio, although large values of PAH$_{12\mu m}$/PAH$_{17\mu m}$ ($> 2$ in log) are indicative of AGN ionisation. Within our sample, the points falling in this regime are mainly the nuclear- and ionisation cone-assigned clusters (e.g. cluster 2 in NGC\,7172, cluster 5 in IC\,5063, or cluster 2 in NGC\,3081).


We note that it is possible that no purely shocked regions are detected in our data cubes. This would imply that even the most shocked regions would be contaminated by either SF or AGN ionisation, thus preventing a robust cluster classification for this type of regions in this particular wavelength range. This could explain the composite regions that are detected in all diagrams in Fig.~\ref{Fig:LineRatios1}, formed by SF and AGN classified clusters, mainly with lower probabilities. 

The number of local known galaxies selected in the analysis is still small. A larger sample could increase the confidence on the classification of the clusters in a particular category (see Sect.~\ref{SubSubSect4:Disc_AutomaticClassification}), and/or provide hints of additional categories that can be added to the model. The label assignation was done such that we consider as ``Other" clusters previously identified as interaction, shocked, and composite regions (see Sect.~\ref{SubSect2:RFtecnhique}). While using a single category simplifies the classification, it contains physically distinct regions that may have very different properties and ratios (e.g. a region illuminated by an AGN versus those where a jet and an outflow are interacting with the ISM). This would make the algorithm to classify such composite regions as either AGN or SF, likely with a lower probability, instead of ``Other", where the dispersion in the measured features is larger. Nevertheless, the clusters assigned to ``Other" all tend to fall at larger values of [Ne\,III]/[Ne\,II] ratios (see Fig.~\ref{Fig:LineRatios1}). These points correspond mostly to the jet-outflow-ISM interacting regions of IC\,5063 and NGC\,5728 (see Figs.~\ref{Fig:Cluster_IC5063} and~\ref{Fig:Cluster_NGC5728}, respectively), and one cluster in NGC\,1052 (see Fig.~\ref{Fig:Cluster_NGC1052}). This classification, especially for the interacting regions, indicates that these behave differently from regular AGN-ionised regions. Given that within the sample there are other radio galaxies, also with known outflows, this suggests that additional processes are occurring, maybe related to the geometrical coupling \citep{RamosAlmeida2022,GarciaBernete2024,Harrison2024,Audibert2025} or with the power of the jet. However, there are few points in this category to draw any firm conclusion. With a larger sample of objects with well-identified regions, we could introduce other categories (such an specific ``jet" category) that could capture the true physical nature of these clusters in a more reliable way. The addition of other lines in the MIRI mid-IR range, or even in the near-IR data with NIRSpec, such as [Fe\,II], that is believed to be a good tracer of shocks, could also help to this purpose \citep[e.g.][]{AH2025}.

\subsection{Testing the methodology: NGC 3227, NGC 4051, and NGC 7582}
\label{SubSect4:Disc_NGC7582}

\begin{figure*}
\centering
	\includegraphics[width=.24\textwidth]{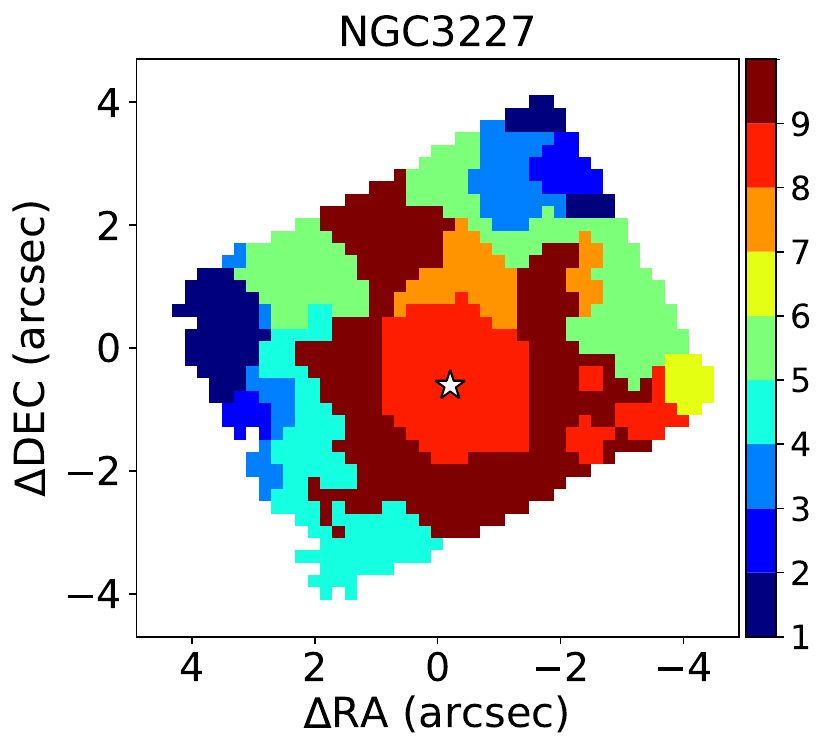}
	\includegraphics[width=.736\textwidth]{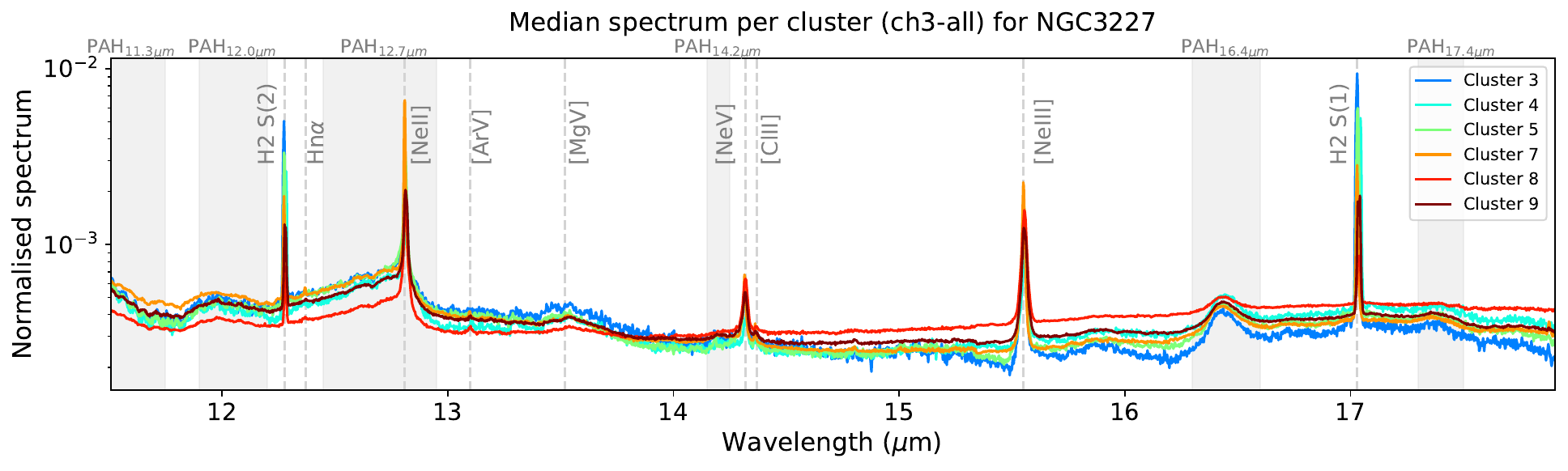}
    \includegraphics[width=.24\textwidth]{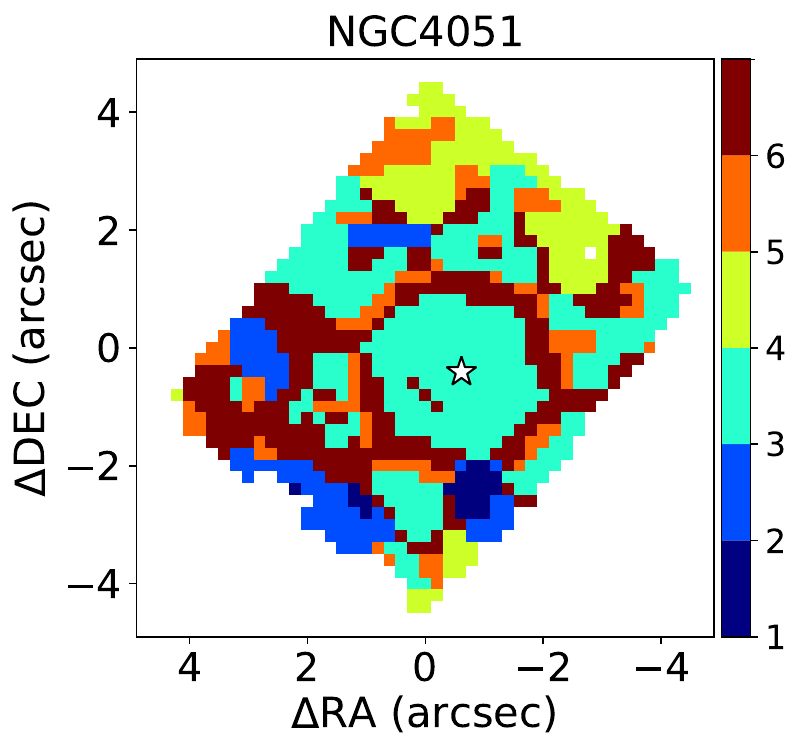}
	\includegraphics[width=.735\textwidth]{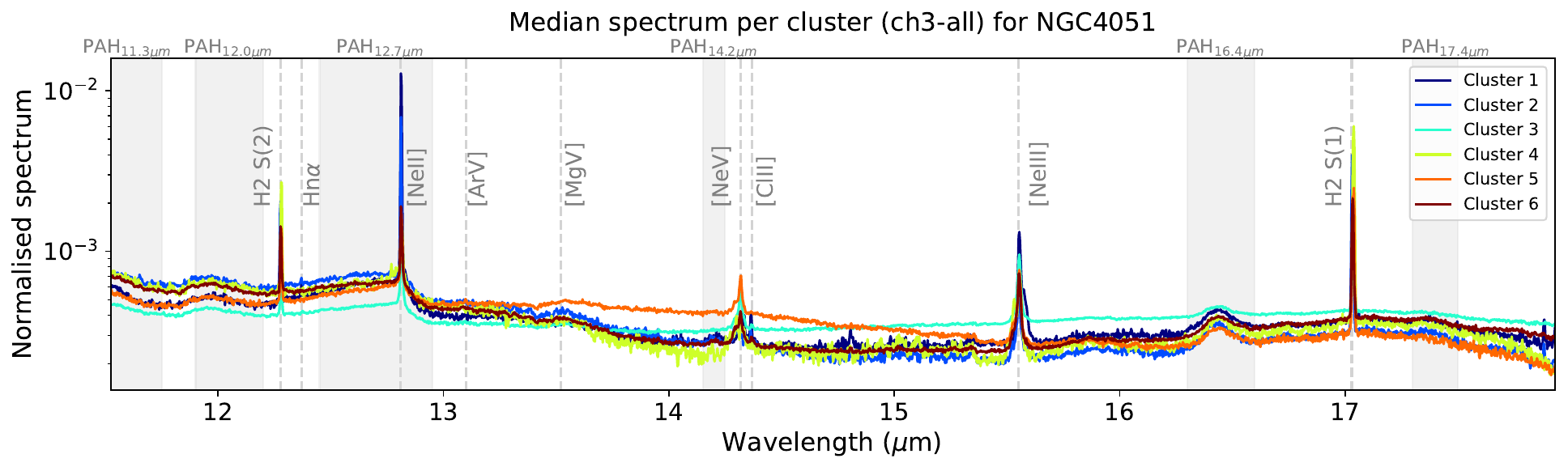}
    \includegraphics[width=.24\textwidth]{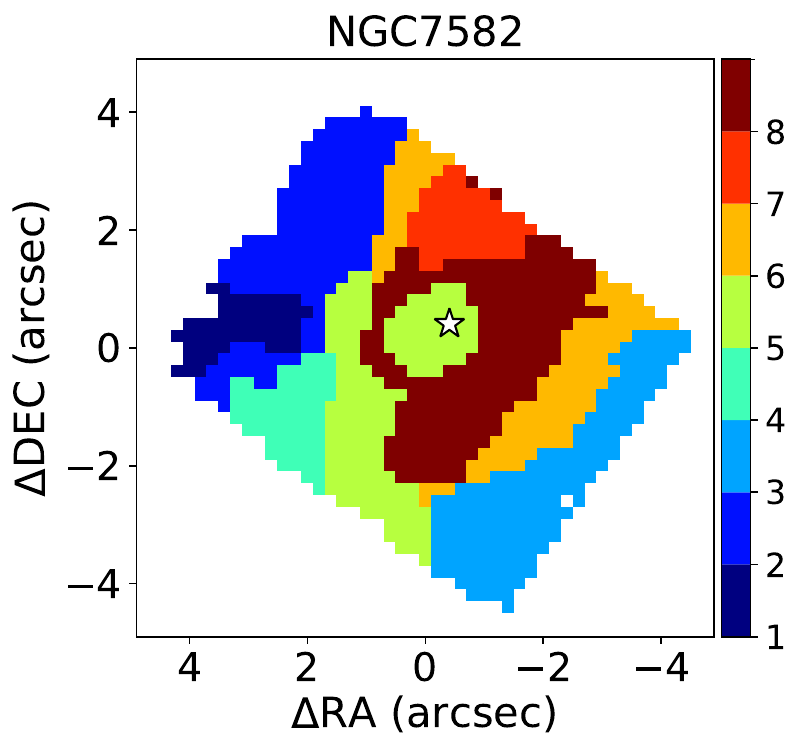}
	\includegraphics[width=.74\textwidth]{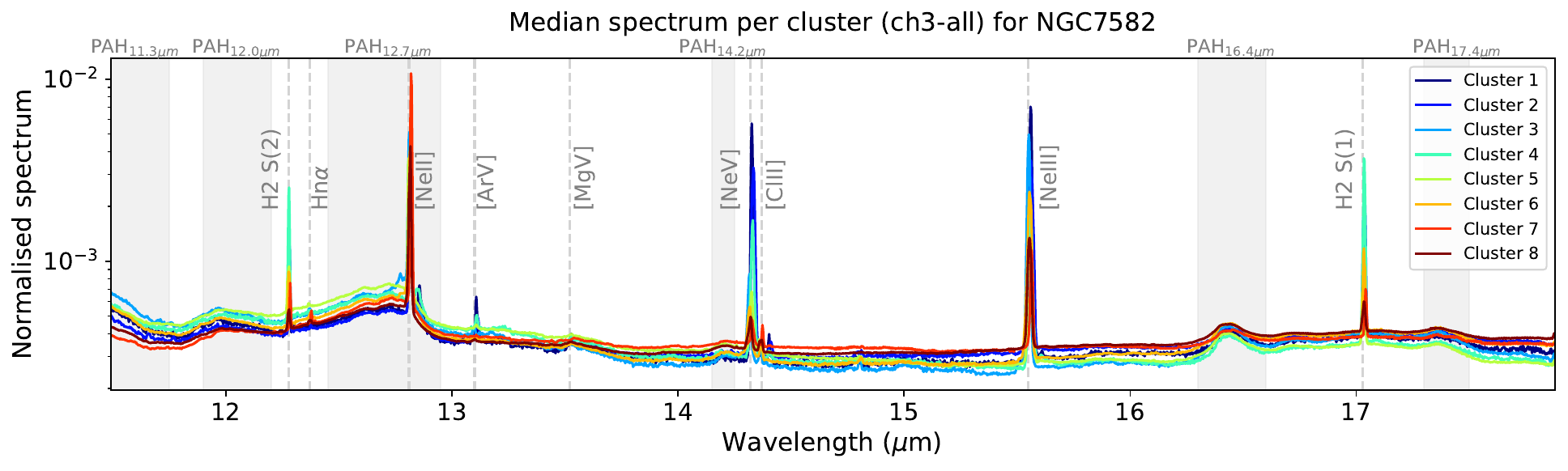}
	\caption{Same as Fig.~\ref{Fig:Cluster_NGC7172} but for the ch3-all cubes of NGC\,3227, NGC\,4051, and NGC\,7582. Their corresponding median maps are in Fig.~\ref{FigAp:MedianMaps}.}
	\label{Fig:Cluster_NGC7582}
\end{figure*}

In order to test the validity of the method, in this section we apply the clustering technique and the RF model to the new MIRI/MRS data of the galaxies NGC\,3227, NGC\,4051, and NGC\,7582, observed within the GATOS collaboration during cycle 2 (see Sect.~\ref{SubSect2:Data}). An in-depth analysis of the last source will be presented in \cite{Veenema2025}.

NGC\,3227 also shows dominance of the PSF in the clustering, but several other regions in the north-east towards west of the nucleus are identified (see top panel in Fig.~\ref{Fig:Cluster_NGC7582}). These could be related to the extended component identified by \cite{AAH2019} with ALMA data, attributed to radial streaming motions produced by gas being funnelled inwards, or to the [O\,III] ionised gas outflow extending up to 7\arcsec\,north-east from the nucleus \citep[see][and also \citealt{Mundell1995}]{Falcone2024}. This galaxy has recent SF both in the nuclear and circumnuclear regions, inferred from the near-IR properties and the detection of PAH at 11.3\,$\mu$m \citep{Davies2006,Hoenig2010,AH2016}. The regions $\sim 3-4$\arcsec\,south-west of the nucleus, corresponding to clusters 6 and 8 (partially), could be related to a SF region previously detected through ionised gas \citep[see][]{AAH2019}. With the RF classifier, all the clusters are classified as AGN, although with median probability ($\sim$48 to 52\%) except for clusters 6, 8, and 9 ($\sim$68\%, 60\%, and $\sim$64\%, respectively). Clusters 5 and 9, classified as AGN ($\sim$48\% and $\sim$67\%, respectively), are extended in the direction of the identified AGN ionisation cone where the non-circular motions were detected in previous works, which supports their AGN-origin \citep[see also][]{AAH2019,Riffel2021,Falcone2024}. In some cases such as cluster 3, the probability for being classified as an AGN is almost equal to being classified as SF, which could be a consequence of the recent SF and the AGN acting simultaneously in the (circum)nuclear region. A further in-depth analysis of these individual regions with the MIRI/MRS data are needed. 

\begin{figure*}
	\centering
	\includegraphics[width=.28\textwidth, trim={0 0 6cm 0},clip]{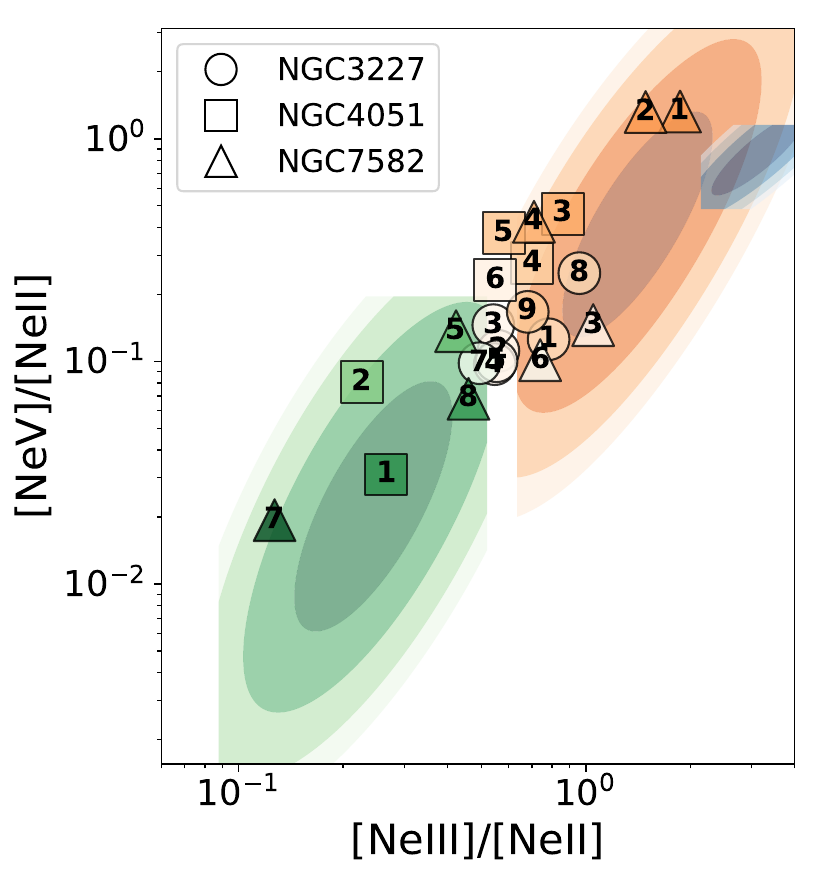}
	\includegraphics[width=.28\textwidth, trim={0 0 6cm 0},clip]{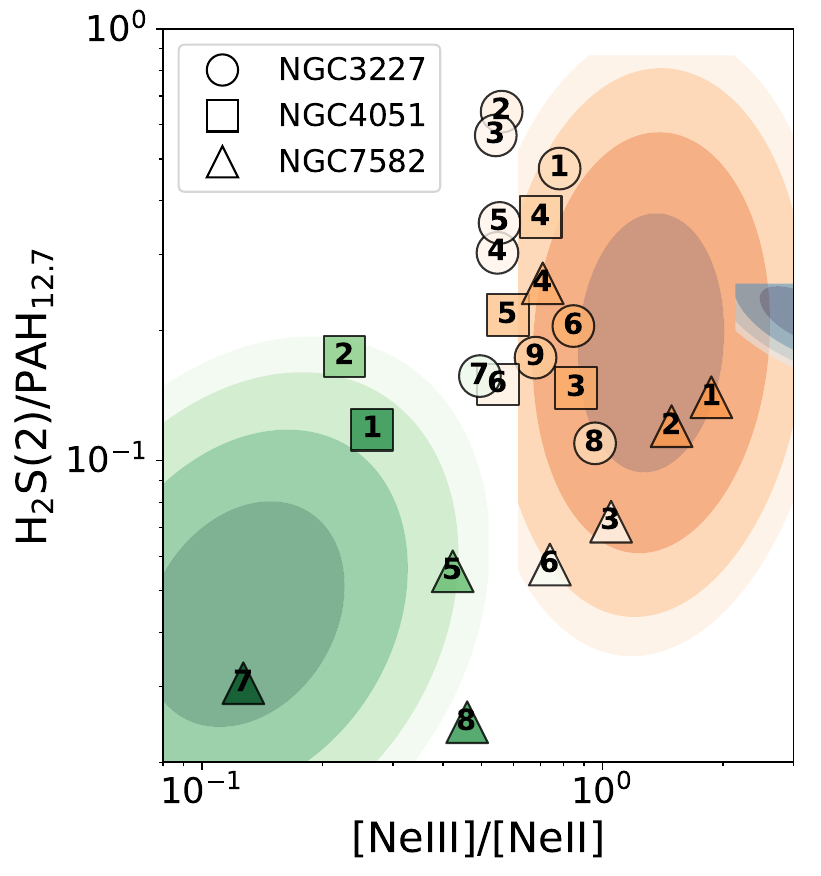}
	\includegraphics[width=.40\textwidth]{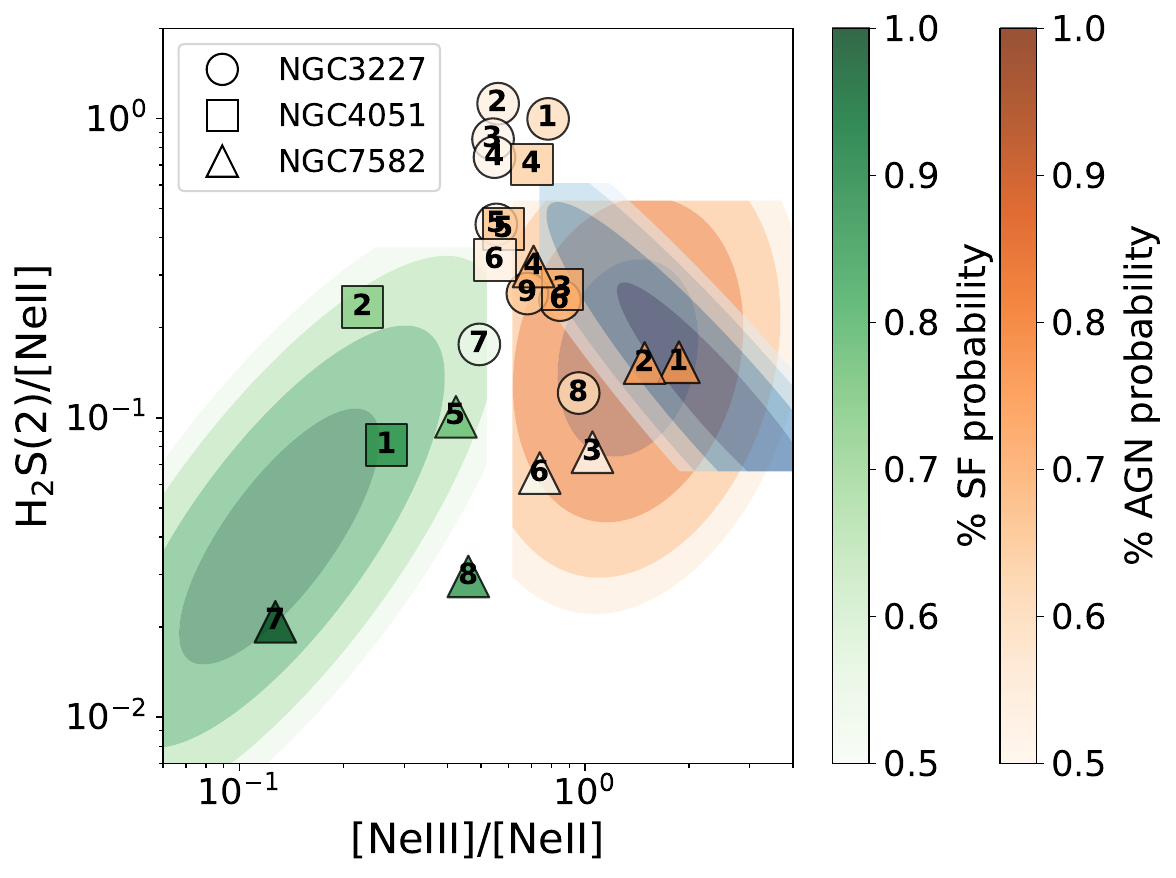}
	\caption{Same diagrams as Fig.~\ref{Fig:LineRatios1}, with the KDE contours of the AGN (orange), SF (green), and Other (blue) distributions, with the predictions by the random forest classifier (see Sect.~\ref{SubSect4:Disc_NGC7582}) for the clusters derived from the ch3-all cubes of the galaxies from the testing sample: NGC\,3227 (circles), NGC\,4051 (squares), and NGC\,7582 (triangles). We indicate the cluster number for each case (see the clustering maps in Fig.~\ref{Fig:Cluster_NGC7582}).}
	\label{Fig:LineRatios_NGC7582}
\end{figure*}

NGC\,4051 is a narrow-line Sy-1 galaxy with an almost face-on ionised gas outflow ($12^{\circ}$ with respect to the line-of-sight) detected by \cite{Fischer2013} and \cite{Meena2021} with optical data \citep[see also][]{Christopoulou1997}. \cite{Riffel2008} detected evidence for non-circular motions associated with a molecular gas inflow, using near-IR data from Gemini. These previously detected non-circular motions are not evident in our clustering maps. The strong PSF dominates the clustering results of the MIRI/MRS ch3-all data cube (see middle panel in Fig.~\ref{Fig:Cluster_NGC7582}). However, we recovered two regions south from the nucleus, clustered together, (clusters 1 and 2 in ch3-all maps), that are classified as SF regions in Fig.~\ref{Fig:LineRatios_NGC7582}. All the remaining clusters are classified as AGN, and are located in all the diagrams within the AGN contours in Fig.~\ref{Fig:LineRatios_NGC7582}. A prior PSF-subtraction of the cube \citep[see the tool by][]{GonzalezMartin2025} could help to disentangle the previously detected, underlying physical processes, such as the ionised and molecular outflows.

Finally, NGC\,7582 has been studied in great detail in the optical with MUSE data by \cite{Juneau2022}, where they observed mainly the approaching part of a biconical ionised gas outflow (opening angles for the north-western edge of $115^{\circ}$ and for the southern edge of $15^{\circ}$) traced with [O\,III]\,5007\AA\,\citep{Juneau2022}. The receding part was partially covered by dust from the galaxy disc \citep[see also][]{Riffel2009,Veenema2025}. With the clustering technique applied to the MIRI/MRS cube, we detect the outflow cone (receding side: clusters 1 and 2 in Fig.~\ref{Fig:Cluster_NGC7582}; and, probably, the approaching side: clusters 5 and 6 in Fig.~\ref{Fig:Cluster_NGC7582}). 
We also captured part of what appears to be the SF ring previously detected with ALMA \citep{AH2020,GarciaBurillo2021}. This region (clusters 7 and 8 in ch3-all map, bottom panel of Fig.~\ref{Fig:Cluster_NGC7582}) coincides with the SF-composite region detected with the BPT diagrams in \cite{Juneau2022}, and is in fact classified as a SF region with the RF model in Fig.~\ref{Fig:LineRatios_NGC7582}. For the ch3-all cube, however, the results from the RF classifier put the nucleus and its surrounding regions (clusters 5 and 6, respectively) as SF ionisation, and the receding part of the galaxy (north and east from the nucleus, clusters 1, 2, and 4) and the approaching part (cluster 3, although with low probability, $\sim$48\%) as AGN (see Fig.~\ref{Fig:LineRatios_NGC7582}). This receding part also correspond to AGN ionisation based on the optical BPT diagram \citep[see Fig.~13 in][]{Juneau2022}. We note that the clusters 5 and 6 (the nucleus and the approaching part of the galaxy, respectively; see Fig.~\ref{Fig:Cluster_NGC7582}), although classified as SF, fall in the composite regions for all the diagrams. This indicates (as mentioned in Sect.~\ref{SubSubSect4:Disc_AutomaticClassification}), that these regions are probably affected by several physical processes simultaneously, maybe produced by the superposition of the outflow and the disc along the line of sight, and thus the SF classification is not correct. Indeed, their derived probabilities are not as high as those corresponding to the disc clusters. 

These examples demonstrate the potential of the clustering method to identify regions of interest, therefore facilitating the analysis of new data cubes. It is important to note that using exclusively the line ratios and considering three categories to train the RF classifier is a simplistic way of classifying the clusters. This method does not consider more complex scenarios that may be present in the galaxies, such as the different coupling situations, obscuration (could still be significant in some sources), the power of the jets, or the overlap of multiple physical processes. This highlights the need for further investigation of these methods and diagnostic diagrams.

\section{Summary and conclusions}
\label{Sect5:Conclusions}

In this work, we have presented a method based on an unsupervised hierarchical clustering technique to automatically identify regions of interest in data cubes of nearby galaxies based on spectral similarity. We have used data for 15 galaxies, mostly nearby AGN, observed with MIRI/MRS, on board of the JWST, from the GATOS collaboration and the JWST archive (see Sect.~\ref{SubSect2:Data}). We used the channel-3 data cubes, covering a wavelength range from $\sim$11.5 to 18\,$\mu$m. We applied the clustering technique to all the cubes, obtaining a median spectrum per cluster for each of the galaxies. We measured the fluxes of several lines of interest (e.g. [Ne\,II], [Ne\,V], and several H$_{2}$ transitions) as well as the PAH features in this range. We then estimated line ratios with these features, and with them we trained a random forest classifier to try to automatically identify the main ionising source for each cluster (AGN, SF, or Other). Here we present the main results of the analysis:

\begin{itemize}
    \item \textit{Clustering technique:} The proposed methodology is useful to identify potentially interesting regions of galaxies, such as SF or disc regions. The nuclei for all the active galaxies are always identified as an independent cluster, although sometimes they are identified together with the ionisation cones. We have checked the validity of the method for the circumnuclear regions of galaxies with MIRI/MRS data cubes, but it can be applicable to any cube observed with any instrument (considering the wavelength range and the normalisation). We note that this methodology is limited for objects with a bright point-like source, as the PSF dominates the clustering. In these cases, a prior PSF subtraction should be performed. 

    \item \textit{Dependence on the wavelength range:} Using both ch3-short and ch3-all cubes we detected mainly consistent results in the clustering results, except for a few galaxies, such as NGC\,5728. Despite this, to better evaluate the performance of the method, the whole wavelength range is preferred here. This is motivated by the larger amount of features available (i.e. low, intermediate, and high excitation, warm molecular, and neutral gas lines, and PAH features), as well as the continuum, that allow for further characterisation of the clusters.

    \item \textit{Mid-IR diagnostic diagrams:} We have found that the most relevant line ratios to be used to classify the clusters using exclusively the ch3 cubes are [Ne\,V]/[Ne\,II], H$_{2}$ S(2)/[Ne\,II], [Ne\,III]/[Ne\,II], PAH$_{12\mu m}$/PAH$_{17\mu m}$, and H$_{2}$ S(2)/PAH$_{12.7\mu m}$. Using the complete MRS ch3 wavelength range, the diagrams formed with these ratios can distinguish between SF and AGN ionisation in all cases, especially that involving the H$_{2}$ S(2)/PAH$_{12.7\mu m}$ and [Ne\,III]/[Ne\,II]. We find composite regions in all the diagrams, which probably trace clusters with mixed ionisation.

    \item \textit{'Other' regions:} With the RF classifier we identified a group of clusters with larger [Ne\,III]/[Ne\,II] than regions with regular AGN ionisation (e.g. the nuclei). Although the sample is still small to draw any strong conclusion, we detected that most of these clusters correspond to interacting regions along the jet and outflow of IC\,5063 and NGC\,5728 \citep[see][for detailed analyses of these galaxies]{Dasyra2024,Davies2024}. Potentially, this means that the processes occurring in the ISM for these galaxies differ from the interactions happening in other galaxies that also have a radio jet within our sample. This suggests that additional physical mechanisms are at play for these two galaxies (e.g. ISM-outflow-jet coupling, power of the jet, or inclination effects).
    
\end{itemize}

Machine learning techniques are a powerful tool that should be explored to simplify the data analysis of IFS data cubes. The method presented here can be used as a test-bed for further and larger analyses that incorporate additional MRS channels containing other emission lines and features (such as [Fe\,II] lines or other PAH ratios) that could be used to enhance the diagnostic power of the method. With a larger galaxy sample observed with the resolution of instruments such as MIRI/JWST, in the future we will expand and put more constraints on the method, in order to classify the different physically distinct regions with more precision.  

\begin{acknowledgements}
	We thank the referee for his/her comments that have helped to improve the manuscript. LHM thanks M. Cervi{\~n}o for useful discussions. LHM and AAH acknowledge financial support by the grant PID2021-124665NB-I00 funded by the Spanish Ministry of Science and Innovation and the State Agency of Research MCIN/AEI/10.13039/501100011033 PID2021-124665NB-I00 and ERDF A way of making Europe. IGB is supported by the Programa de Atracci{\'o}n de Talento Investigador ``C{\'e}sar Nombela" via grant 2023-T1/TEC-29030 funded by the Community of Madrid. OG-M acknowledge financial support from Ciencia de Frontera project number CF2023-G100 (SECIHTI) and PAPIIT project IN109123 (UNAM). MPS acknowledges support under grants RYC2021-033094-I, CNS2023-145506, and PID2023-146667NB-I00 funded by MCIN/AEI/10.13039/501100011033 and the European Union NextGenerationEU/PRTR. CRA acknowledges support from the Agencia Estatal de Investigaci{\'o}n of the Ministerio de Ciencia, Innovaci{\'o}n y Universidades (MCIU/AEI) under the grant ``Tracking active galactic nuclei feedback from parsec to kiloparsec scales", with reference PID2022-141105NB-I00 and the European Regional Development Fund (ERDF). LZ, EKSH, CP, and JS acknowlege grant support from the Space Telescope Science Institute (ID: JWST-GO-01670). AA acknowledges funding from the European Union (WIDERA ExGal-Twin, GA 101158446). EB acknowledges support from the Spanish grants PID2022-138621NB-I00 and PID2021-123417OB-I00, funded by MCIN/AEI/10.13039/501100011033/FEDER, EU. DEA is supported by the ``Becas Estancia Postdoctorales por M{\'e}xico" EPM(1) 2024 (CVU:592884) program of SECIHTI, and acknowledges financial support from PAPIIT UNAM IN109123 and ``Ciencia de Frontera” CONAHCyT CF2023-G100. RAR acknowledges the support from Conselho Nacional de Desenvolvimento Cient\'ifico e Tecnol\'ogico (CNPq; Proj.~303450/2022-3, 403398/2023-1 \& 441722/2023-7) and Coordena\c c\~ao de Aperfei\c coamento de Pessoal de N\'ivel Superior (CAPES; Proj.~88887.894973/2023-00).
	This work is based on observations made with the NASA/ESA/CSA James Webb Space Telescope. The data were obtained from the Mikulski Archive for Space Telescopes at the Space Telescope Science Institute, which is operated by the Association of Universities for Research in Astronomy, Inc., under NASA contract NAS 5-03127 for JWST; and from the European JWST archive (eJWST) operated by the ESDC. These observations are associated with programs 1269, 1328, 1670, 2004, 2016, 2219, 2721, 2732, and 3535.
	This research has made use of the NASA/IPAC Extragalactic Database (NED), which is operated by the Jet Propulsion Laboratory, California Institute of Technology, under contract with the National Aeronautics and Space Administration.
	This work has made extensive use of Python (v3.9.12), particularly with \textsc{astropy} \citep[v5.3.3;][]{astropy:2013, astropy:2018}, \textsc{lmfit} \citep[v1.2.2;][]{NewvilleLMFIT2014}, \textsc{matplotlib} \citep[v3.8.0;][]{Hunter:2007}, \textsc{seaborn} \citep[v0.13.2;][]{Waskom2021}, \textsc{scipy} \citep[v1.11.2;][]{2020SciPy-NMeth}, \textsc{numpy} \citep[v1.26.0;][]{Harris2020}, \textsc{scikit-learn} \citep[v1.4.1;][]{scikit-learn}, and \textsc{pandas} (v2.2.3).
\end{acknowledgements}

%
\bibliographystyle{aa} 
\bibliography{bibliography.bib} 

%

\begin{appendix}

	\section{Median maps of the ch3-all cubes for all the sample}
	\label{Appendix_MedianMaps}

	We include here the median flux intensity maps of the ch3-all cubes for all the galaxies from the sample (see Table~\ref{Table:1}). They have been calculated by getting the median value of the cube along the spectral axis. We note that the continuum for most galaxies (9 out of 15) is dominated by a strong PSF pattern, clearly visible in the images.

	\begin{figure*}
                \centering
                \includegraphics[width=.245\textwidth]{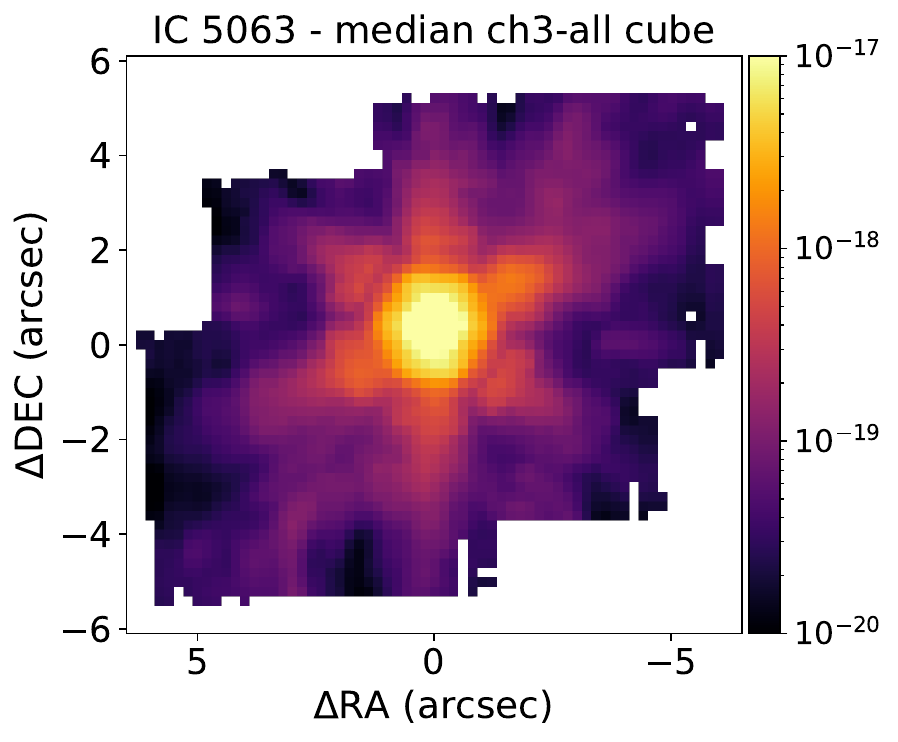}
                \includegraphics[width=.248\textwidth]{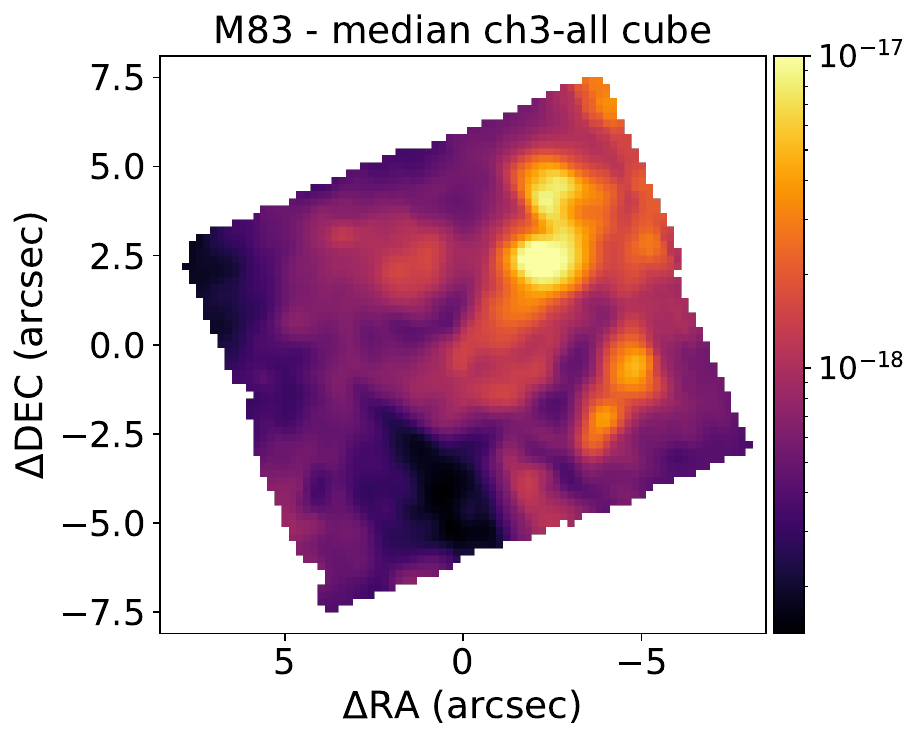}
                \includegraphics[width=.245\textwidth]{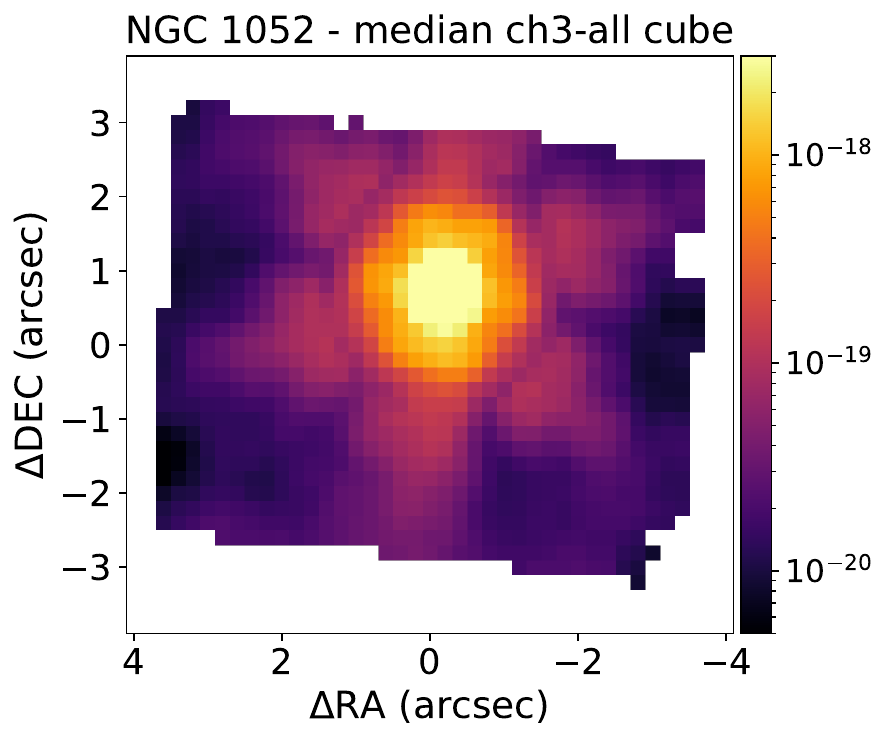}
                \includegraphics[width=.24\textwidth]{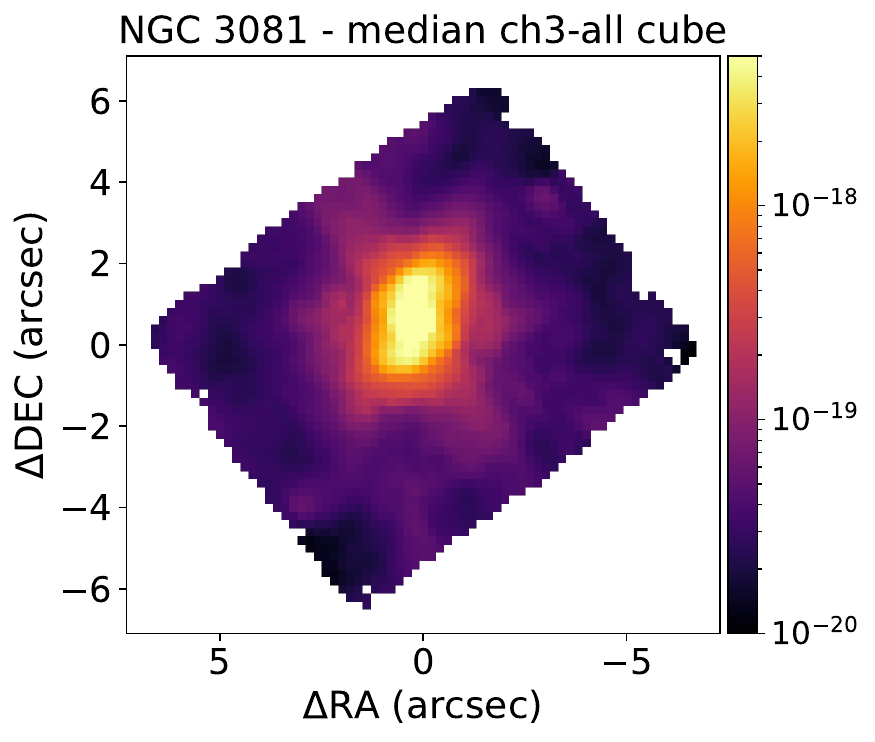}
                \includegraphics[width=.245\textwidth]{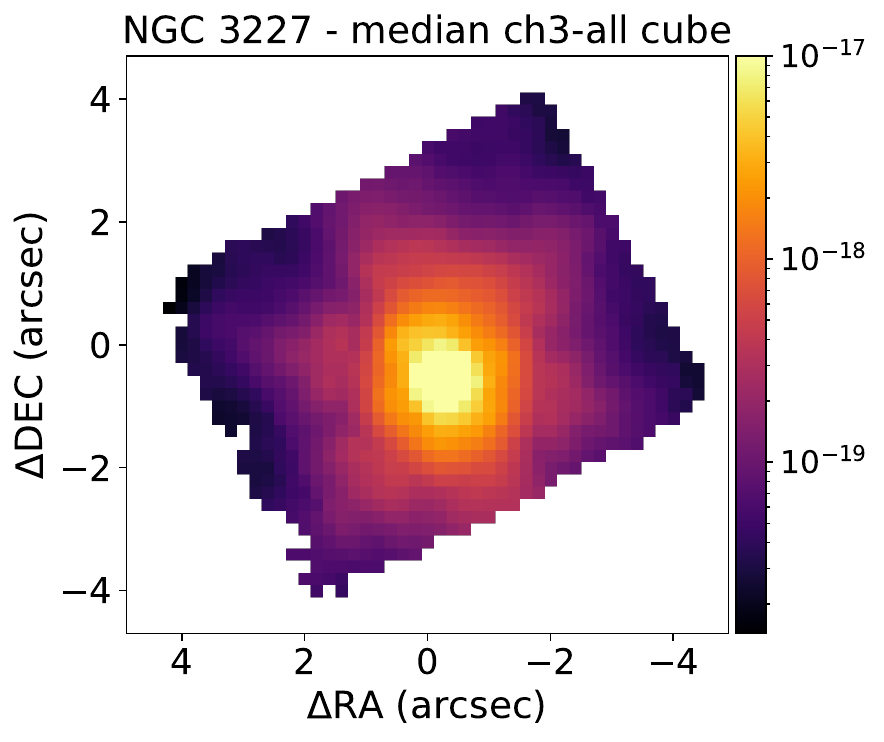}
                \includegraphics[width=.245\textwidth]{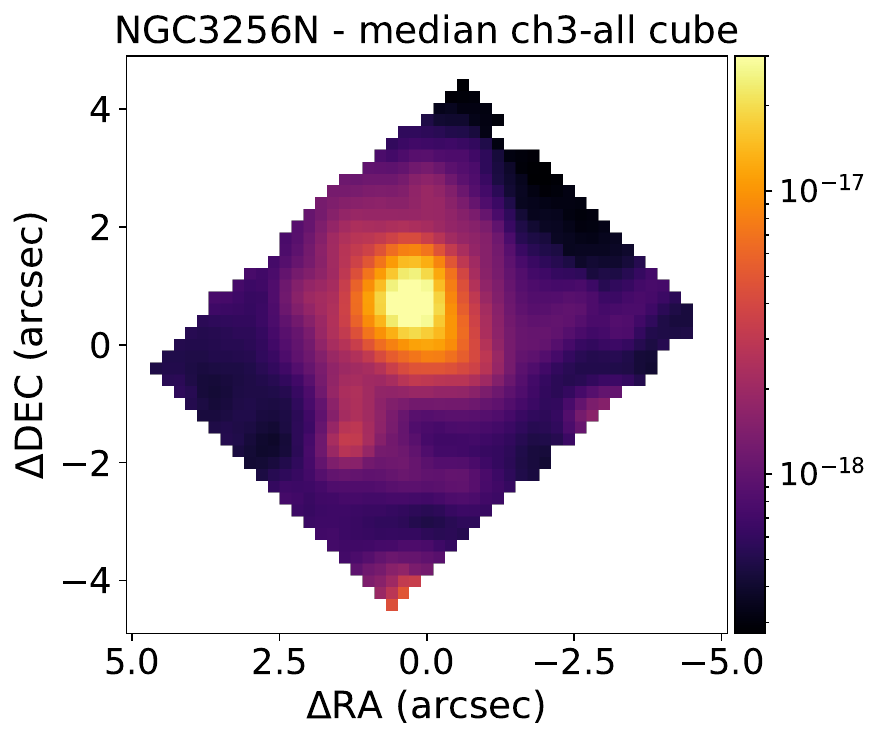}
                \includegraphics[width=.245\textwidth]{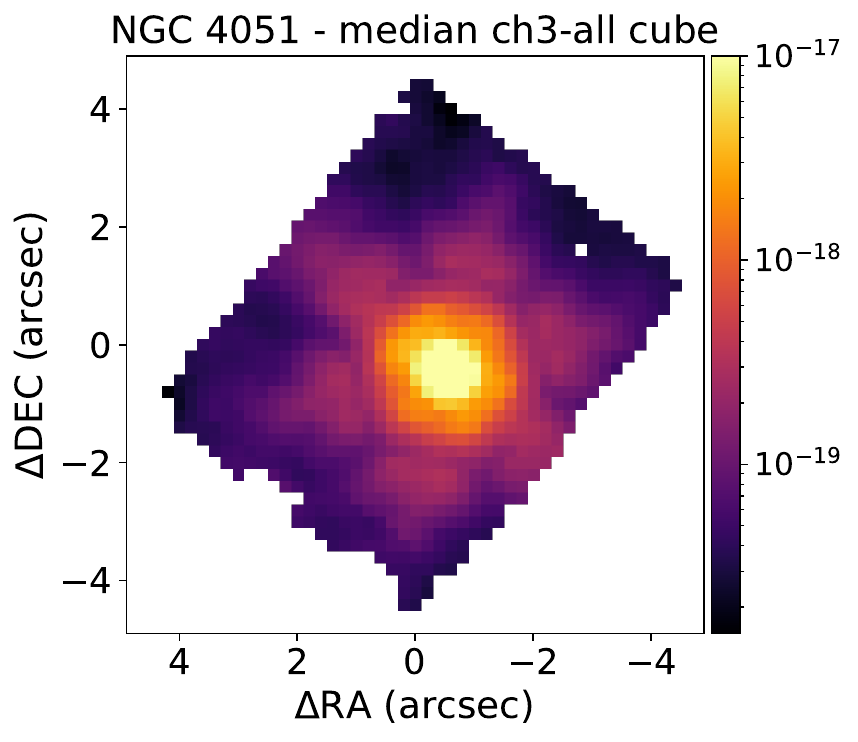}
                \includegraphics[width=.245\textwidth]{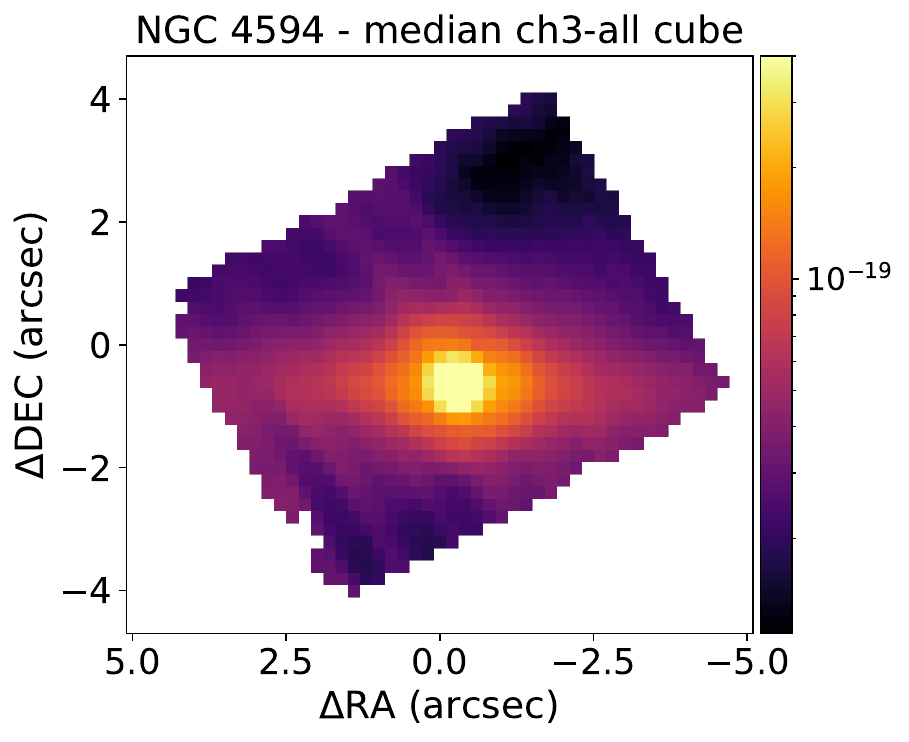}
                \includegraphics[width=.25\textwidth]{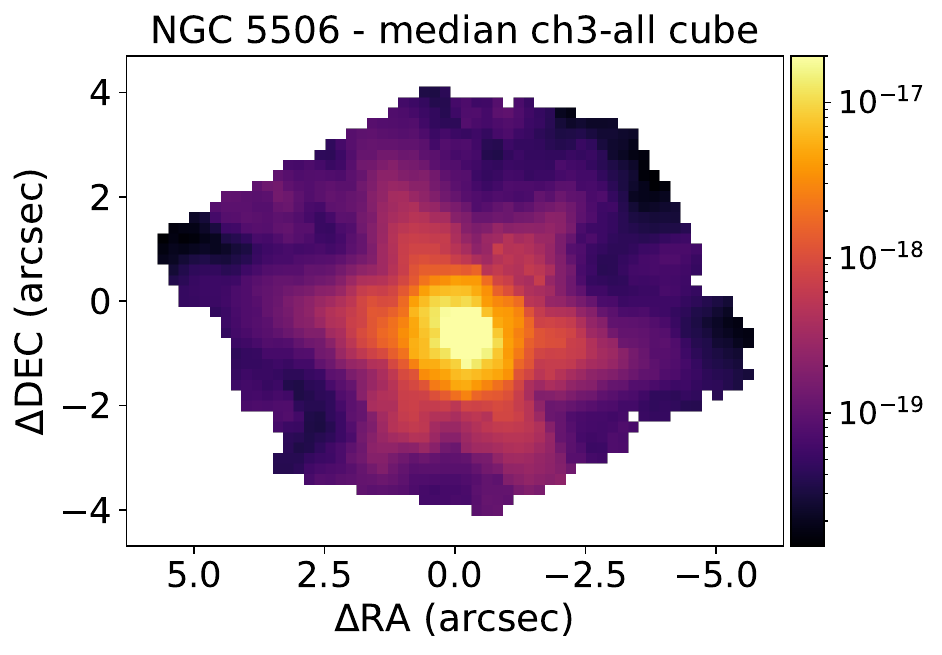}
                \includegraphics[width=.235\textwidth]{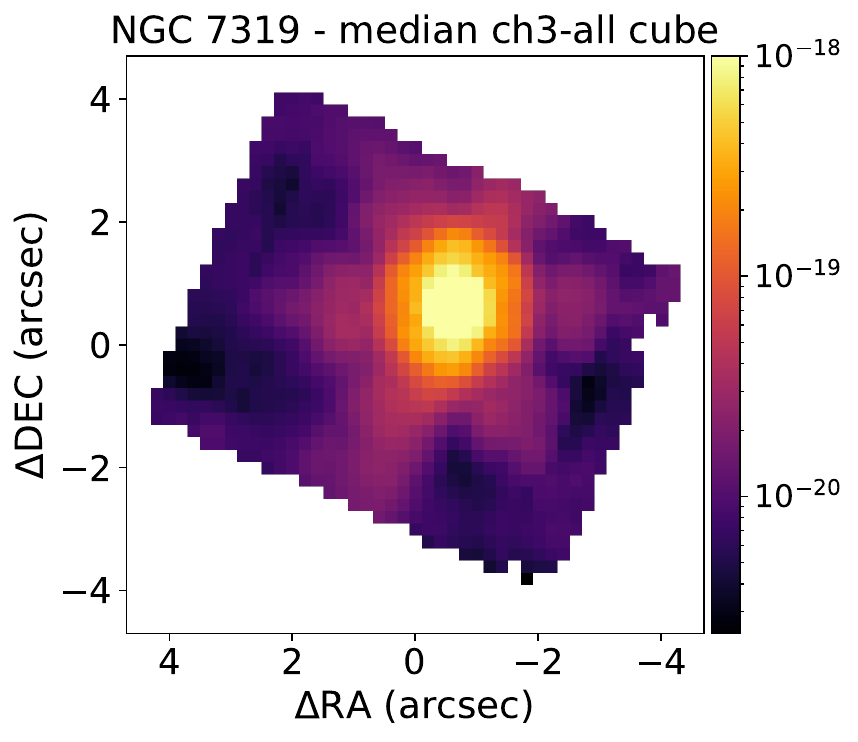}
                \includegraphics[width=.242\textwidth]{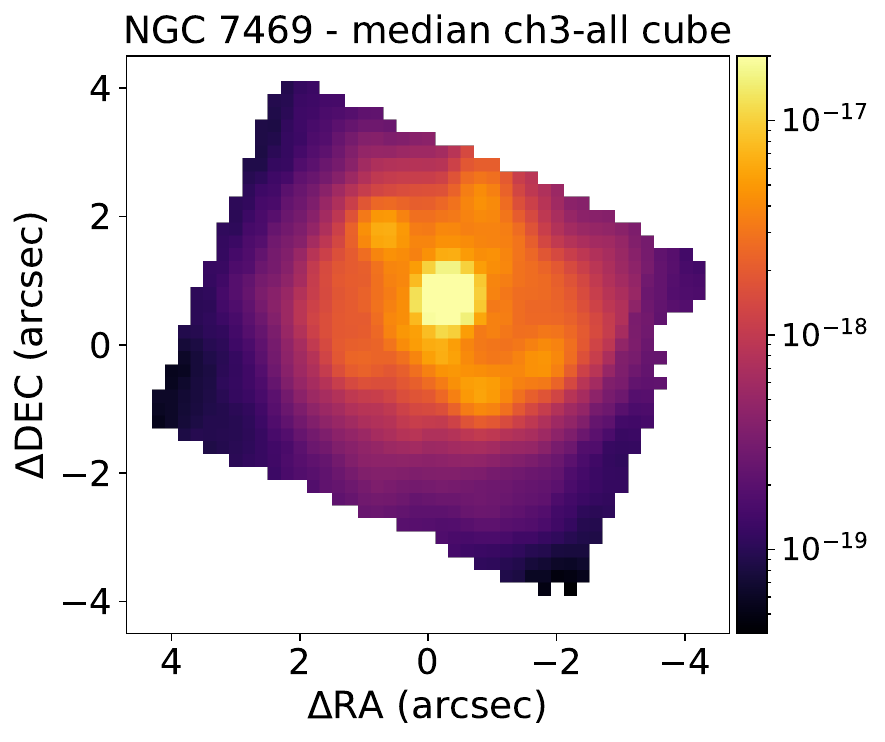}
                \includegraphics[width=.24\textwidth]{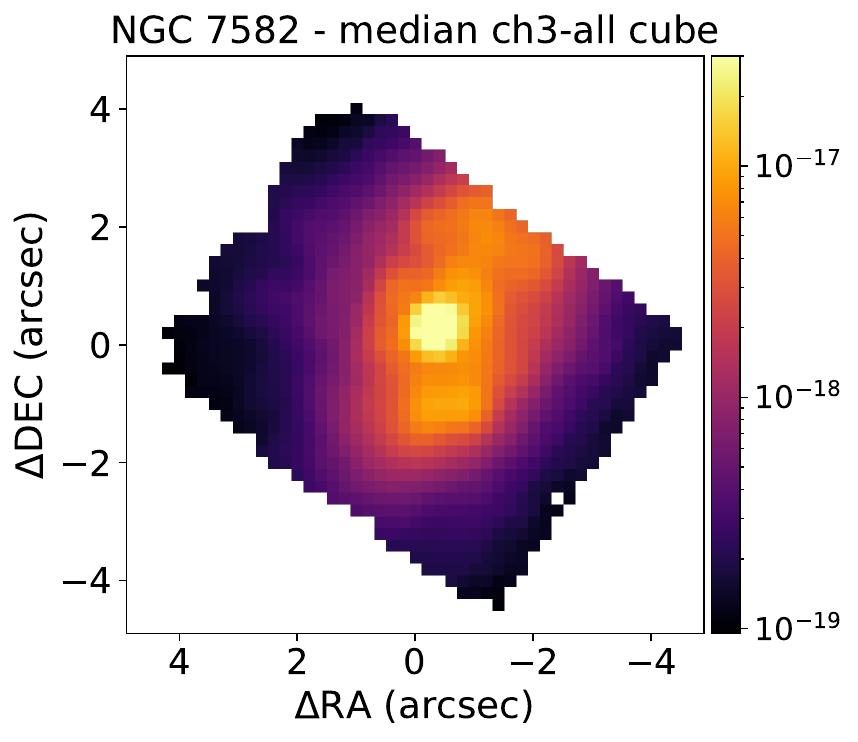}
                \includegraphics[width=.225\textwidth]{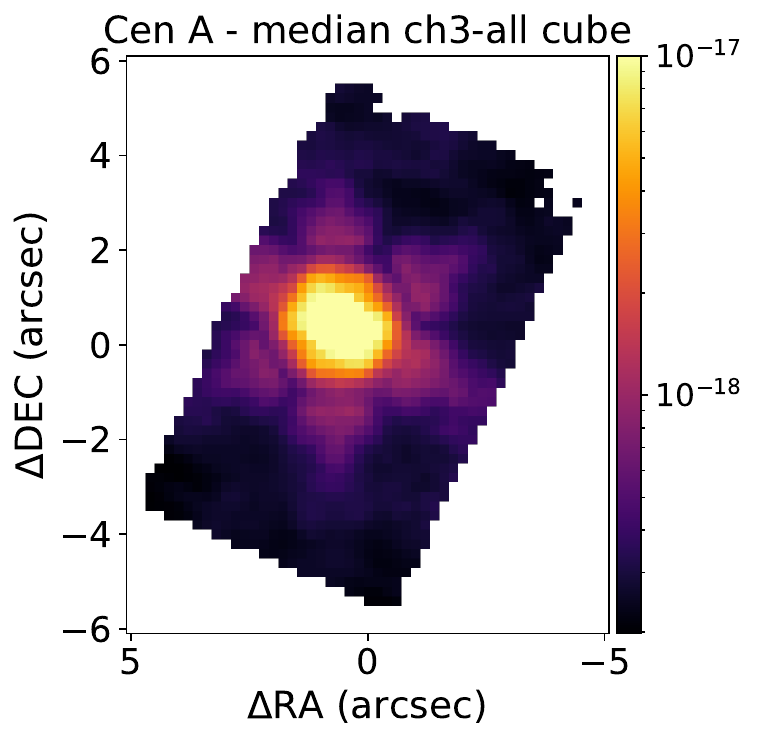}
		\caption{Median flux maps of the ch3-all cubes for all the galaxies from the sample, in logaritmic scale (see also left panels in Figs.~\ref{Fig:Cluster_NGC7172} and~\ref{Fig:Cluster_NGC5728}).}
                \label{FigAp:MedianMaps}
        \end{figure*}

	
	\section{Clustering results for the remaining galaxies}
	\label{Appendix_ClusterMaps}

    In this appendix we show the maps and spectra for all the galaxies not discussed in the main text, following Fig.~\ref{Fig:Cluster_NGC7172}. We include Table~\ref{Table:2}, that contains the initial and final classification assigned with our methodology to all the individual clusters for all the galaxies used to train the RF model.

    \begin{figure*}
	\includegraphics[width=.255\textwidth]{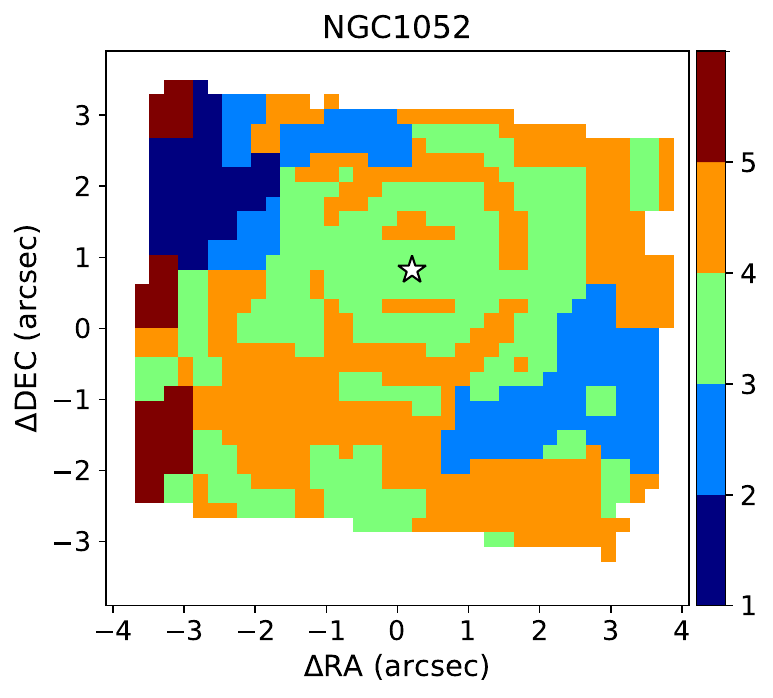}
	\includegraphics[width=.745\textwidth]{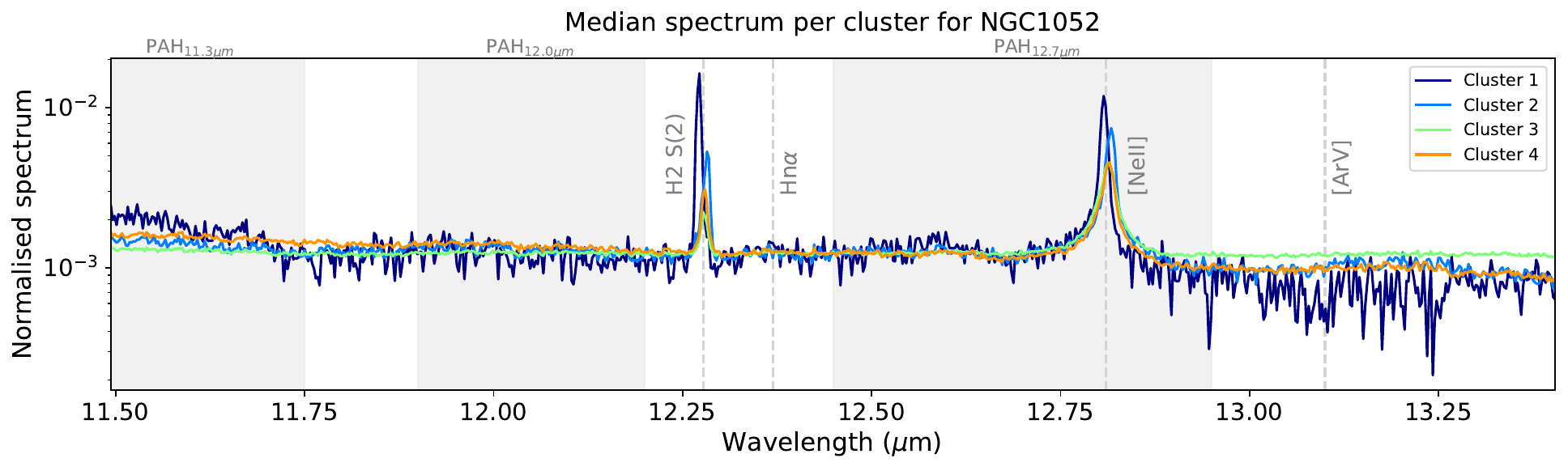}
    \includegraphics[width=.255\textwidth]{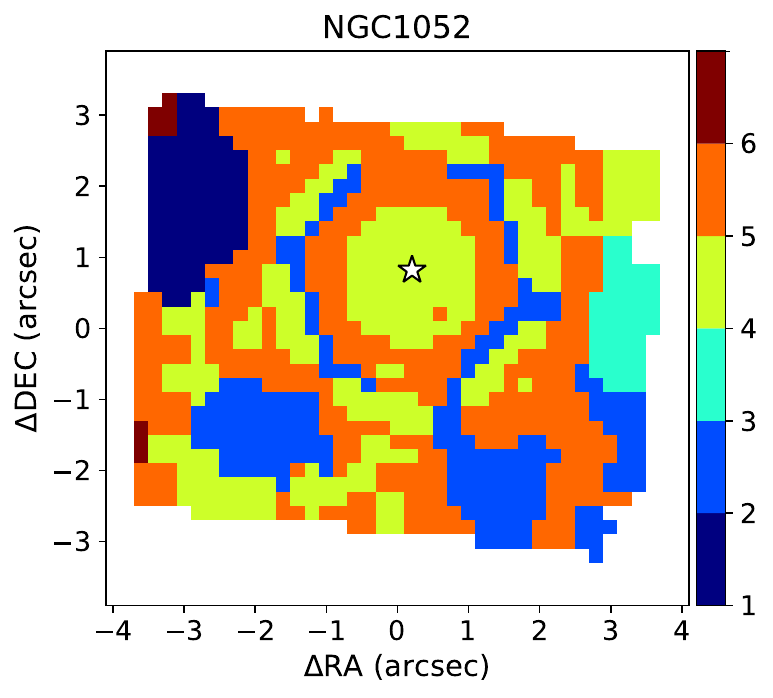}
	\includegraphics[width=.745\textwidth]{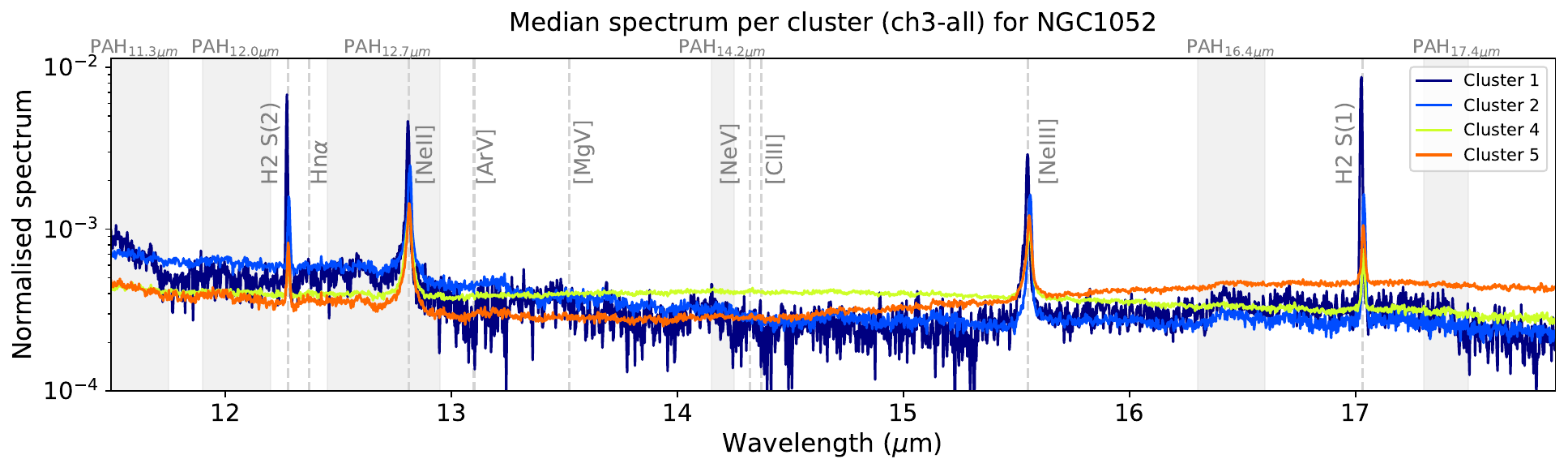}
	    \caption{Same as Fig.~\ref{Fig:Cluster_NGC7172} but for NGC\,1052. We note that, for the top (bottom) panel, we do not show the spectrum for cluster 5 (3 and 6), as it has low S/N.}
	\label{Fig:Cluster_NGC1052}
\end{figure*}

\begin{figure*}
	\includegraphics[width=.255\textwidth]{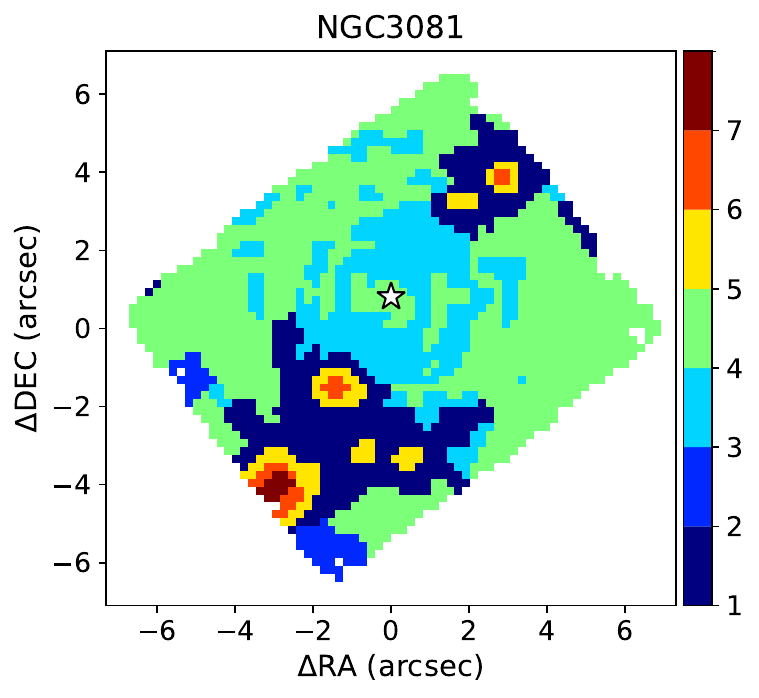}
	\includegraphics[width=.745\textwidth]{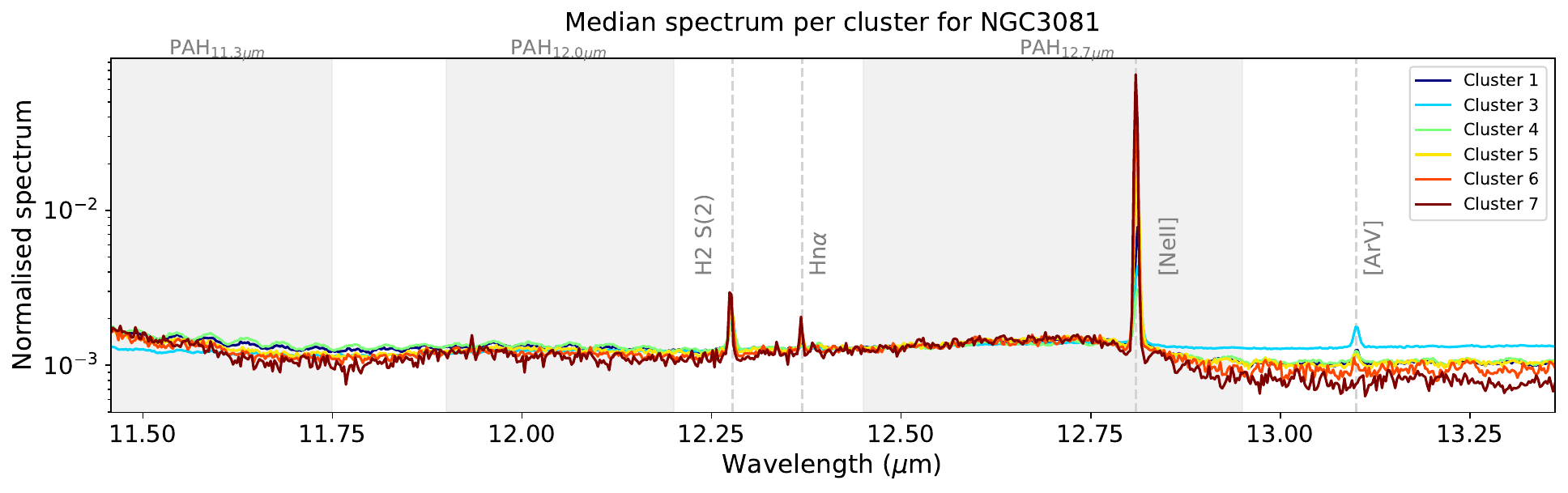}
    \includegraphics[width=.255\textwidth]{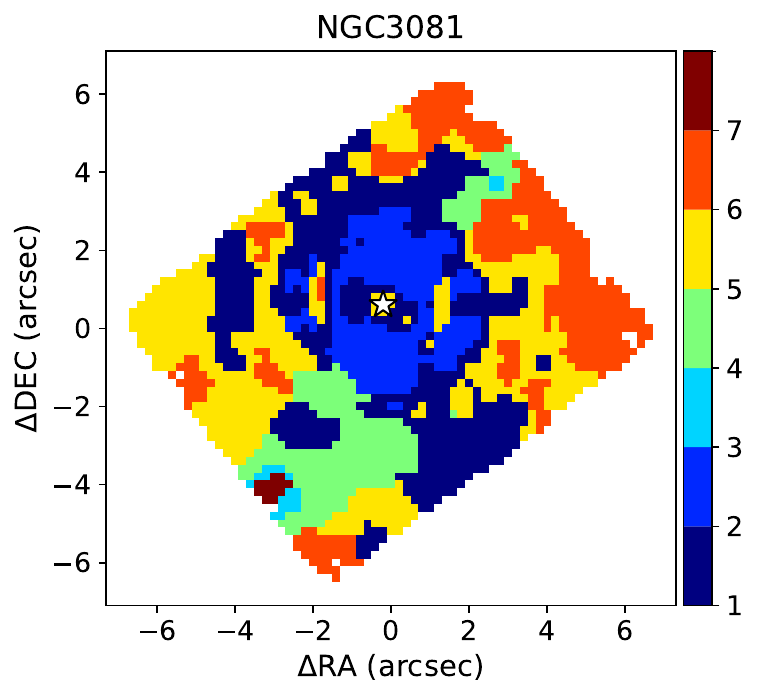}
	\includegraphics[width=.745\textwidth]{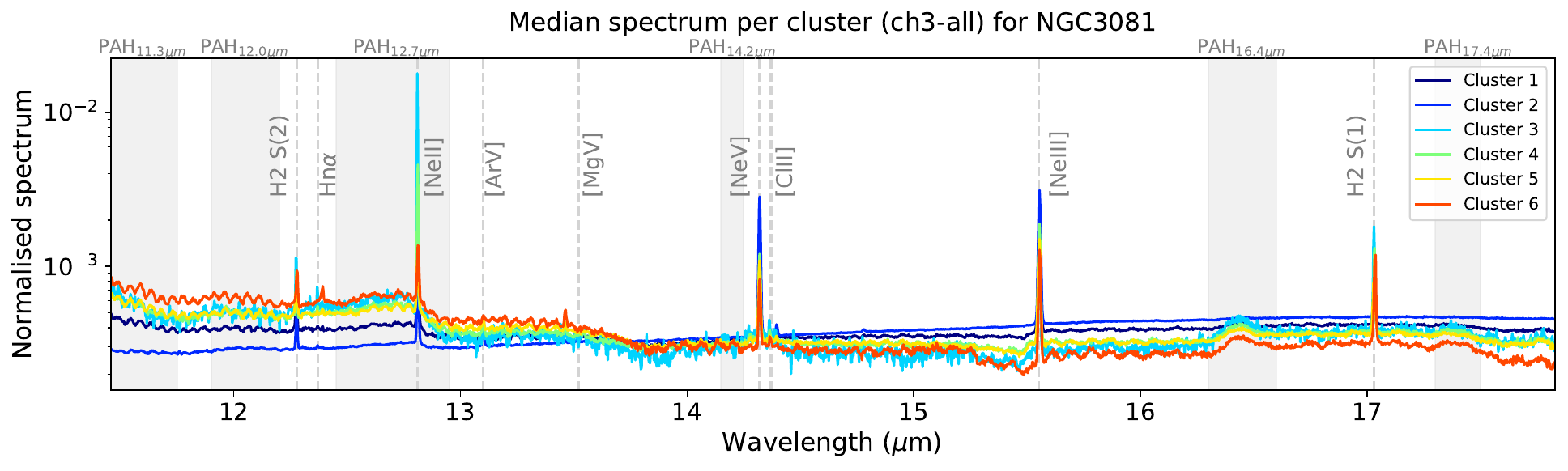}
	\caption{Same as Fig.~\ref{Fig:Cluster_NGC7172} but for NGC\,3081. We note that, for the top (bottom) panel, we do not show the spectrum for cluster 2 (7), as it has low S/N.}
	\label{Fig:Cluster_NGC3081}
\end{figure*}

\begin{figure*}
	\includegraphics[width=.253\textwidth]{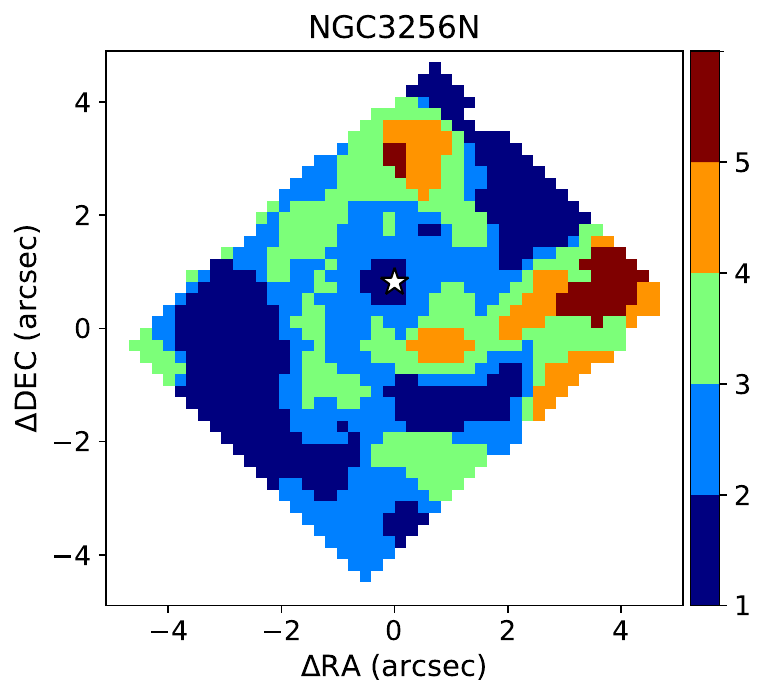}
	\includegraphics[width=.747\textwidth]{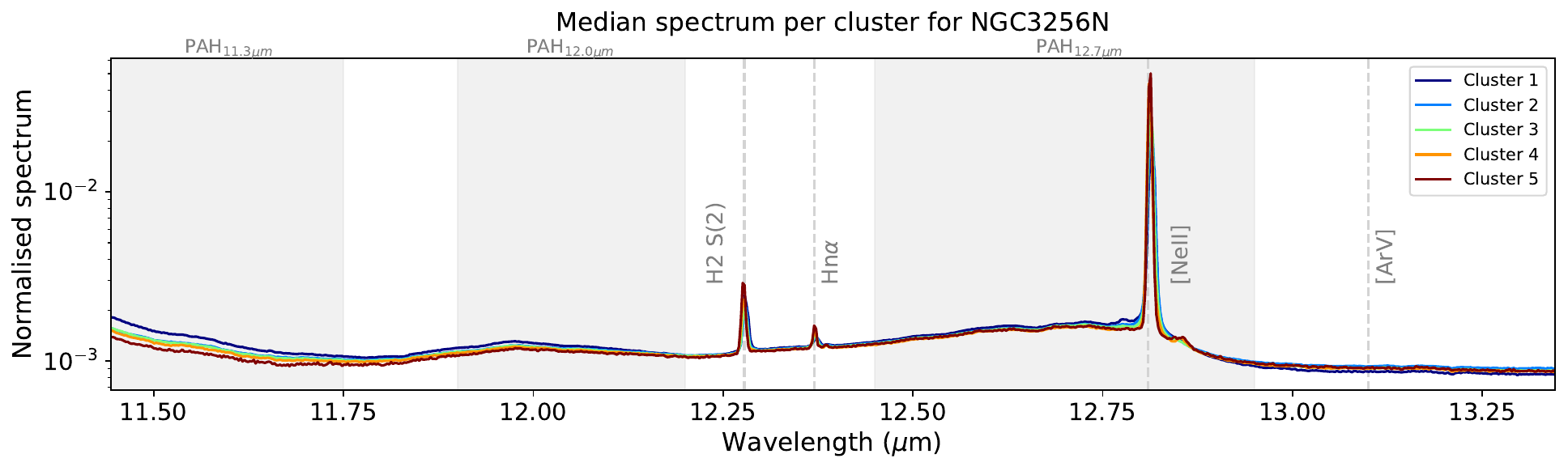}
    \includegraphics[width=.253\textwidth]{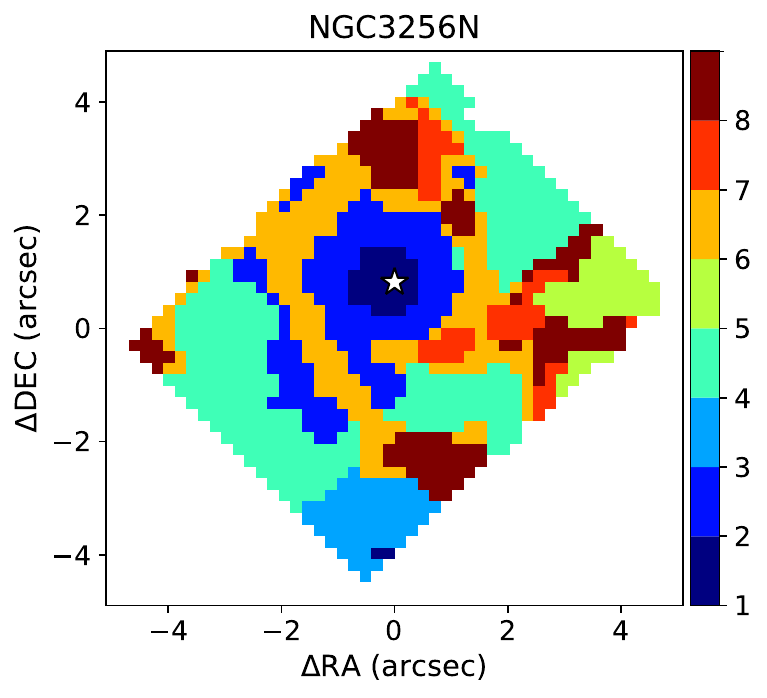}
	\includegraphics[width=.747\textwidth]{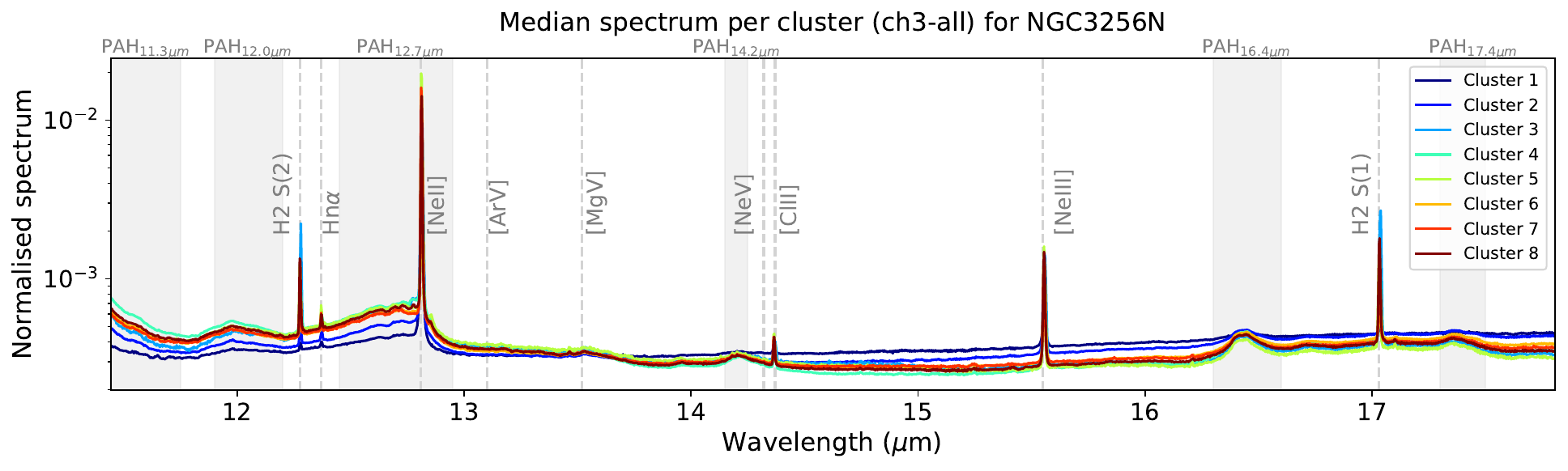}
	\caption{Same as Fig.~\ref{Fig:Cluster_NGC7172} but for NGC\,3256-N.}
	\label{Fig:Cluster_NGC3256}
\end{figure*}

    \begin{figure*}
	\includegraphics[width=.258\textwidth]{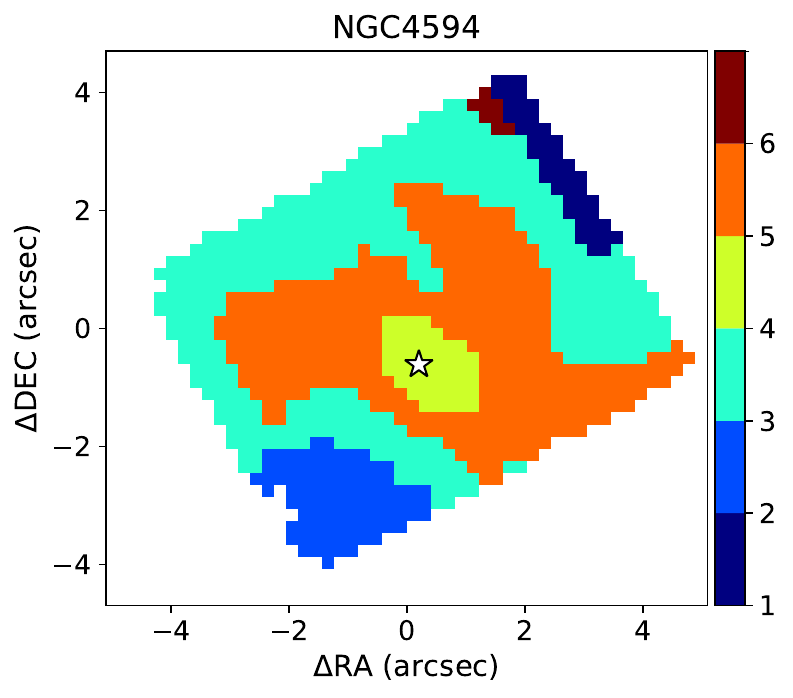}
	\includegraphics[width=.742\textwidth]{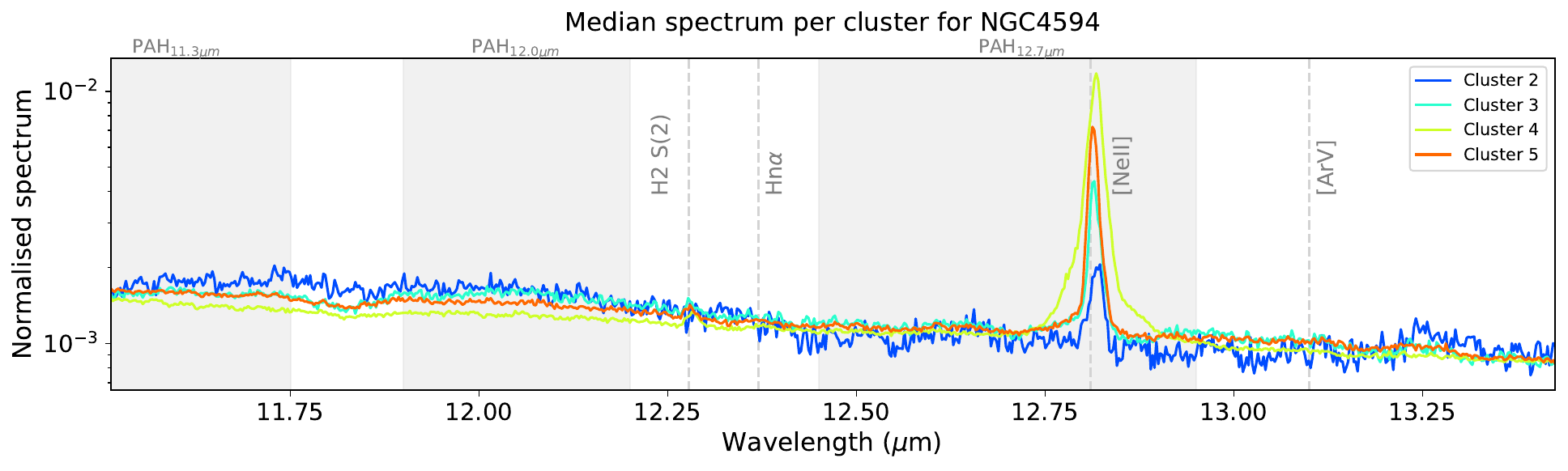}
    \includegraphics[width=.258\textwidth]{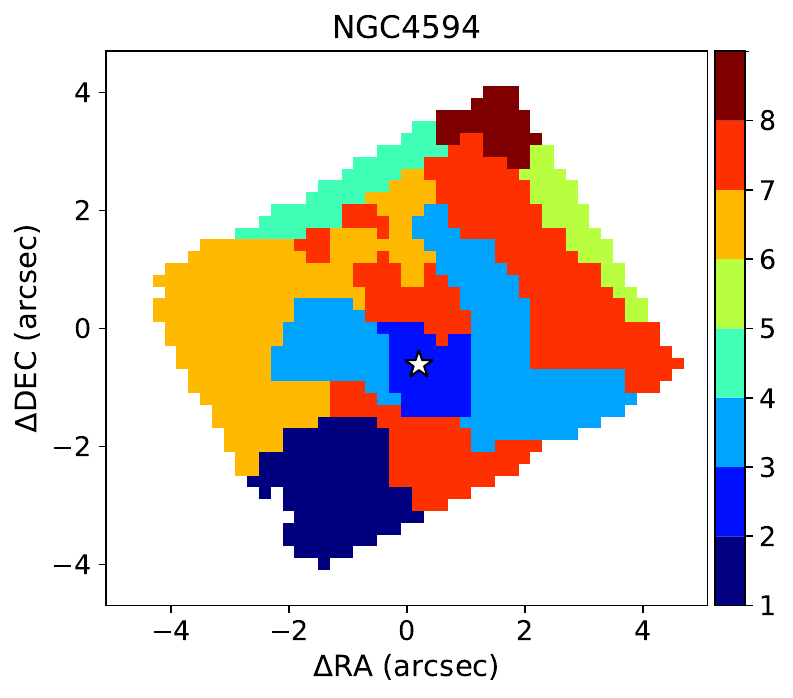}
	\includegraphics[width=.742\textwidth]{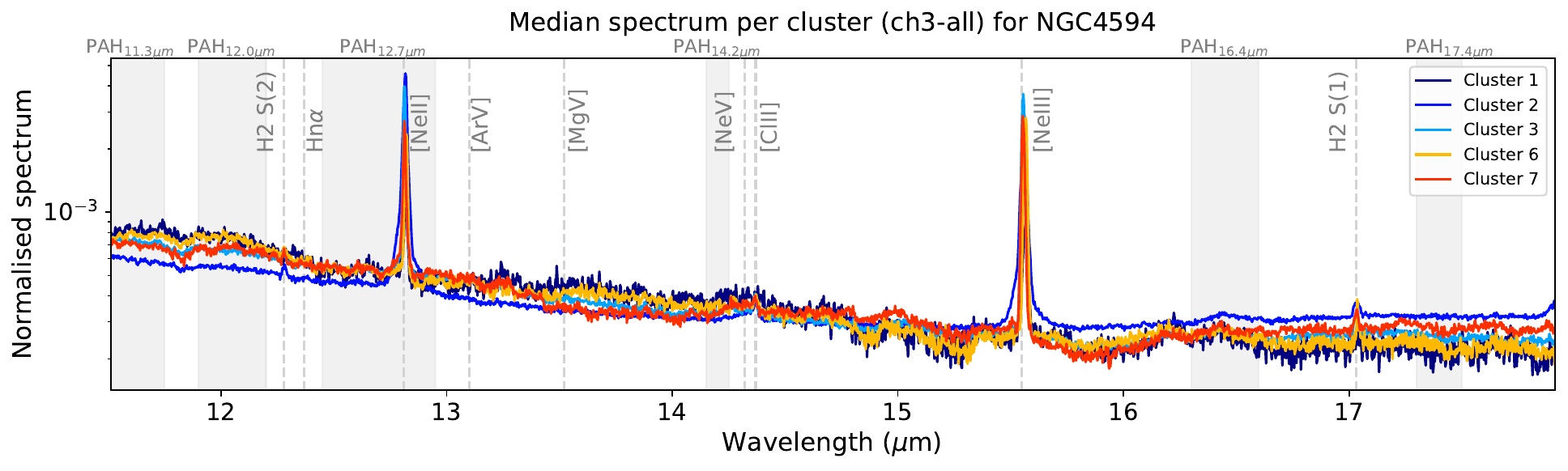}
	\caption{Same as Fig.~\ref{Fig:Cluster_NGC7172} but for NGC\,4594. We note that, for the top (bottom) panel, we do not show the spectrum for clusters 1 and 6 (4, 5, and 8), as they are low S/N clusters.}
	\label{Fig:Cluster_NGC4594}
\end{figure*}

    \begin{figure*}
	\includegraphics[width=.29\textwidth]{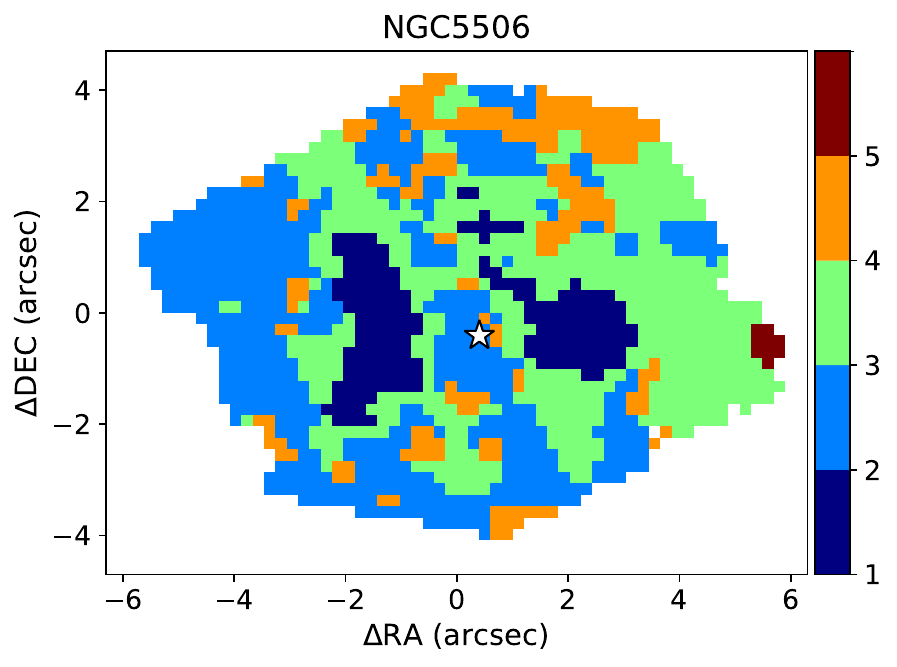}
	\includegraphics[width=.71\textwidth]{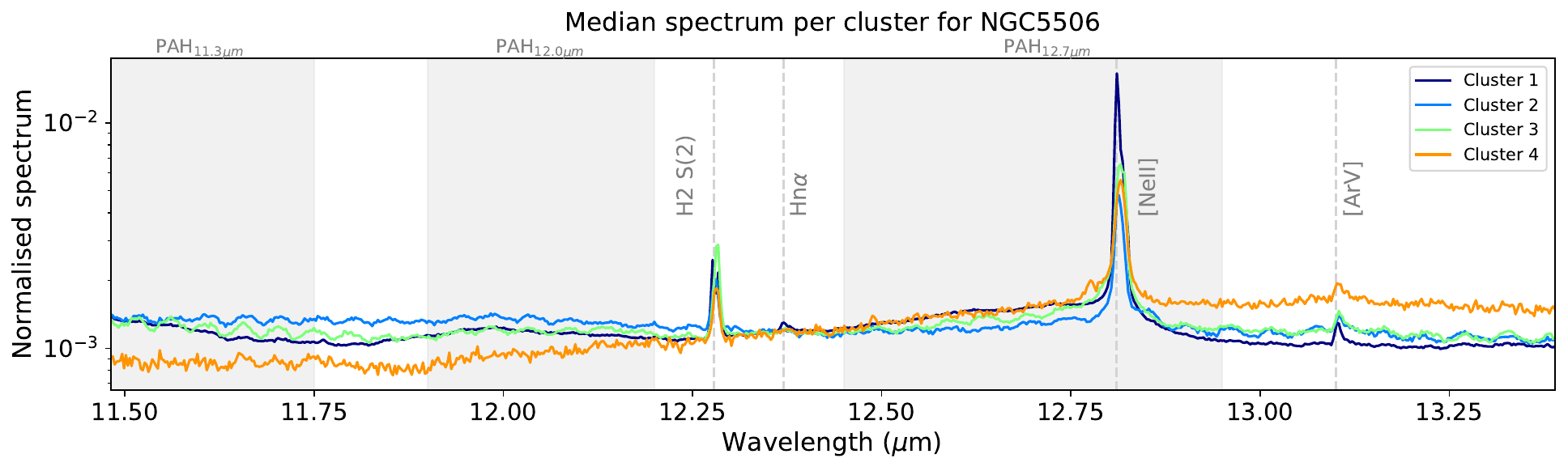}
    \includegraphics[width=.29\textwidth]{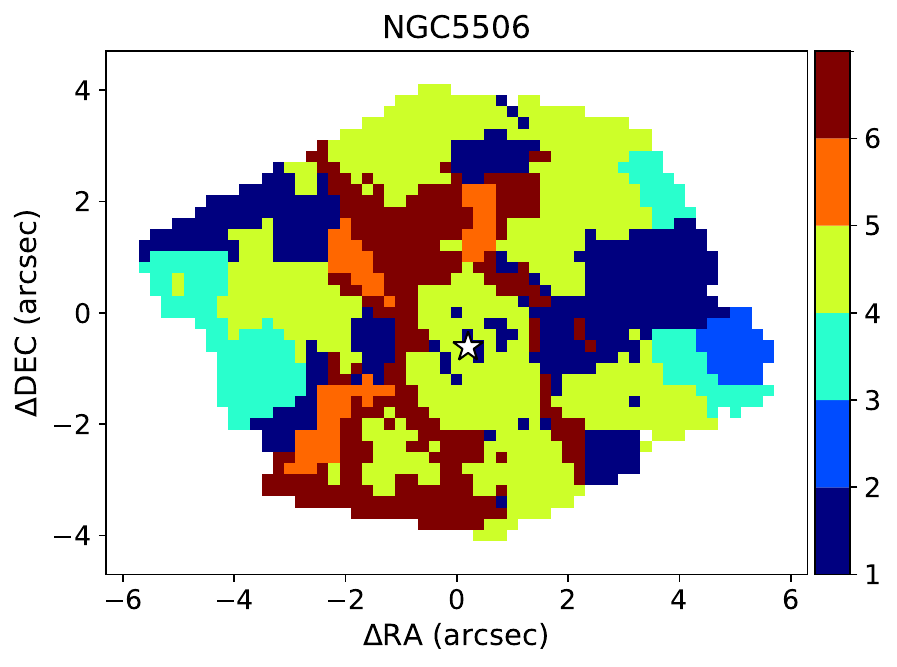}
	\includegraphics[width=.71\textwidth]{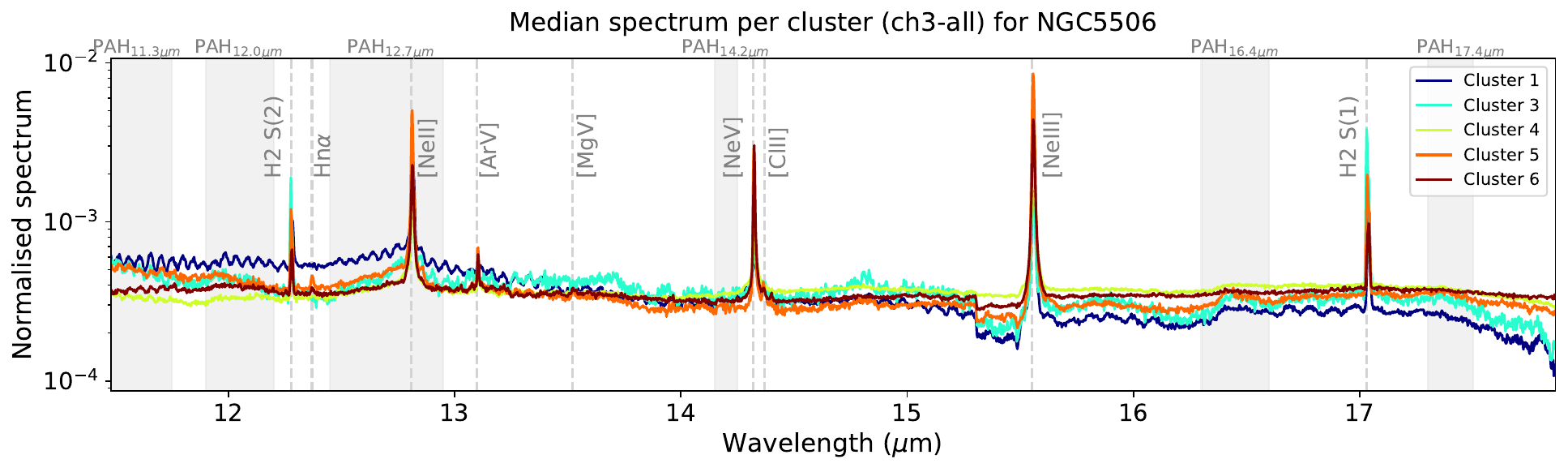}
	\caption{Same as Fig.~\ref{Fig:Cluster_NGC7172} but for NGC\,5506. We note that, for the top (bottom) panel, we do not show the spectrum for cluster 5 (2), as it is a low S/N cluster.}
	\label{Fig:Cluster_NGC5506}
\end{figure*}

    \begin{figure*}
	\includegraphics[width=.247\textwidth]{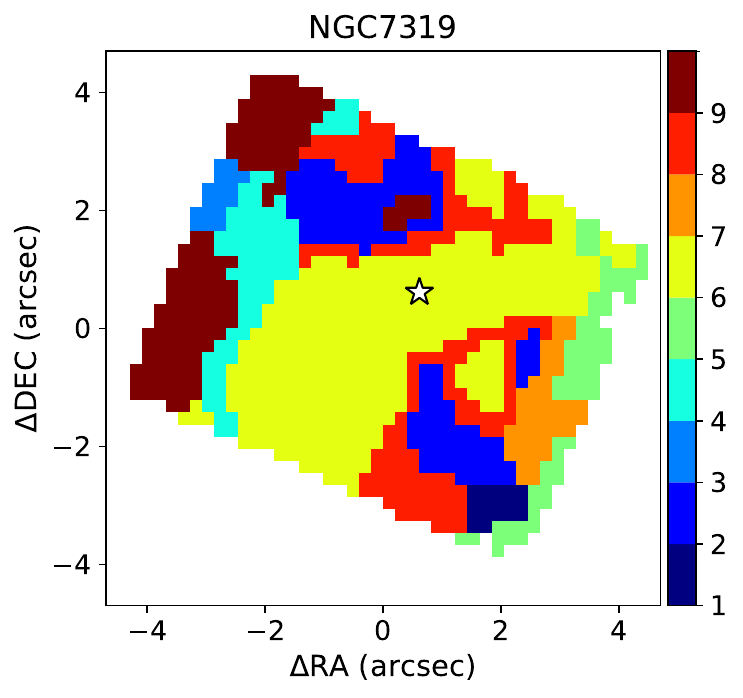}
	\includegraphics[width=.755\textwidth]{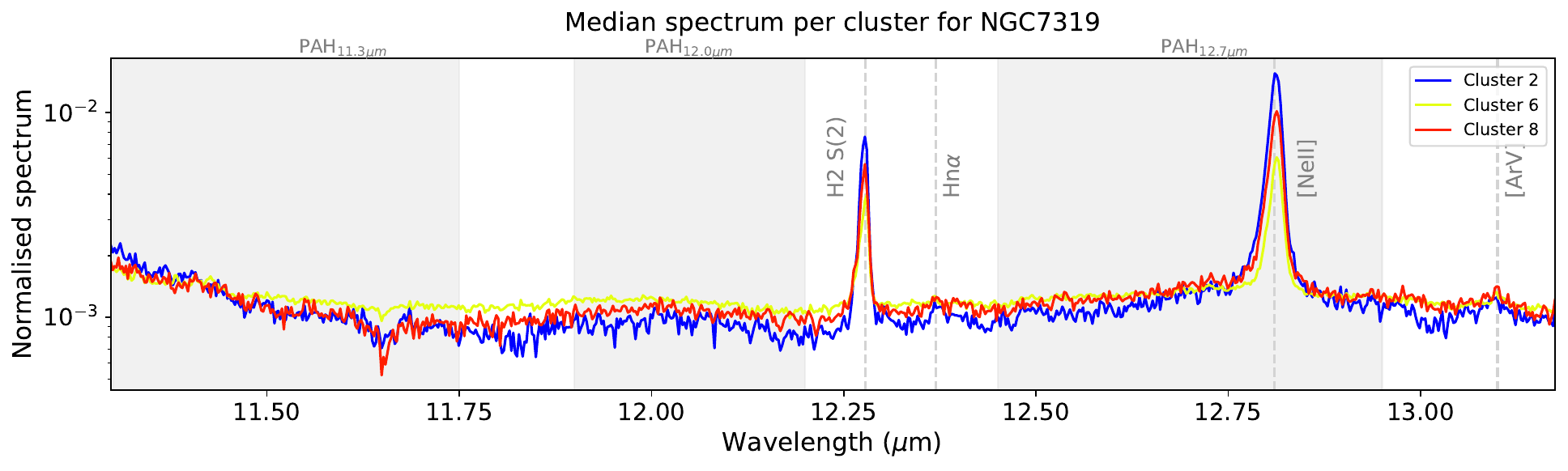}
    \includegraphics[width=.247\textwidth]{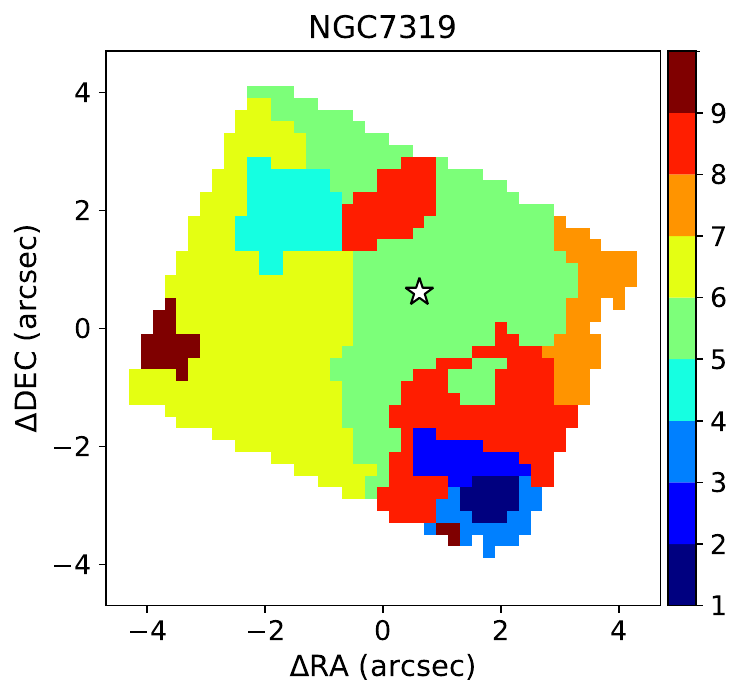}
	\includegraphics[width=.765\textwidth]{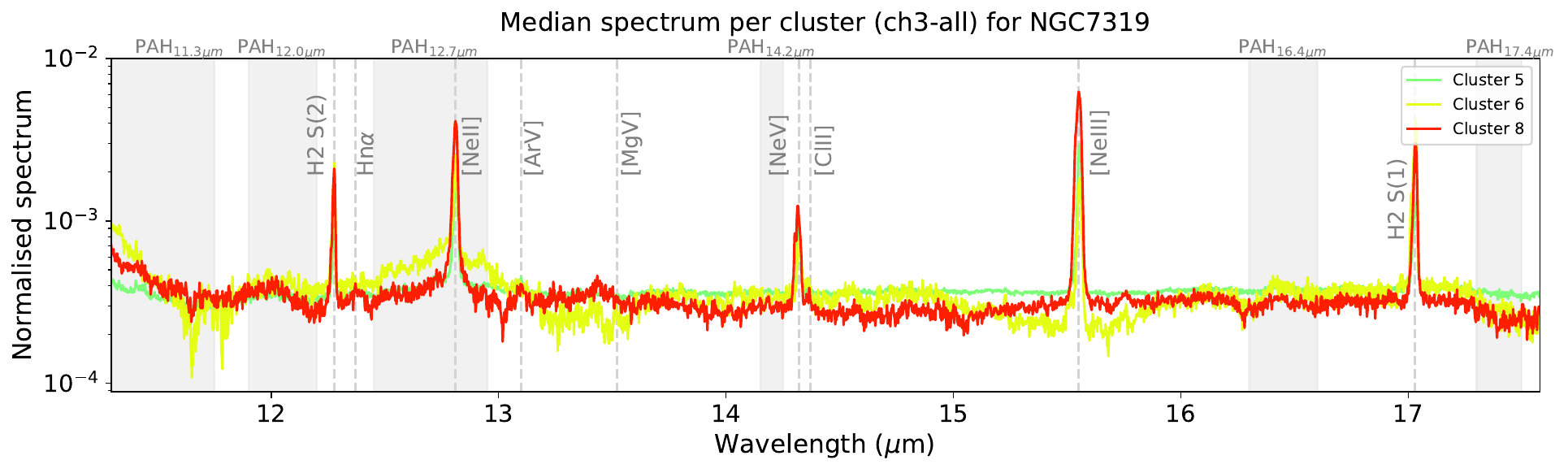}
	\caption{Same as Fig.~\ref{Fig:Cluster_NGC7172} but for NGC\,7319. We note that, for the top (bottom) panel, we do not show the spectrum for clusters 1, 3, 4, 5, 7, and 9 (1, 2, 3, 4, 7, and 9), as they are low S/N clusters.}
	\label{Fig:Cluster_NGC7319}
\end{figure*}

    \begin{figure*}
	\includegraphics[width=.254\textwidth]{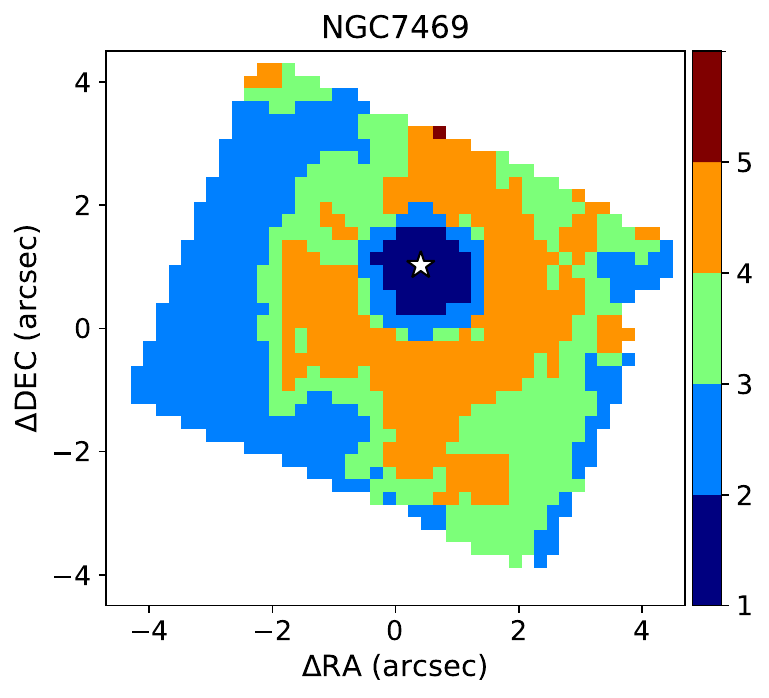}
	\includegraphics[width=.76\textwidth]{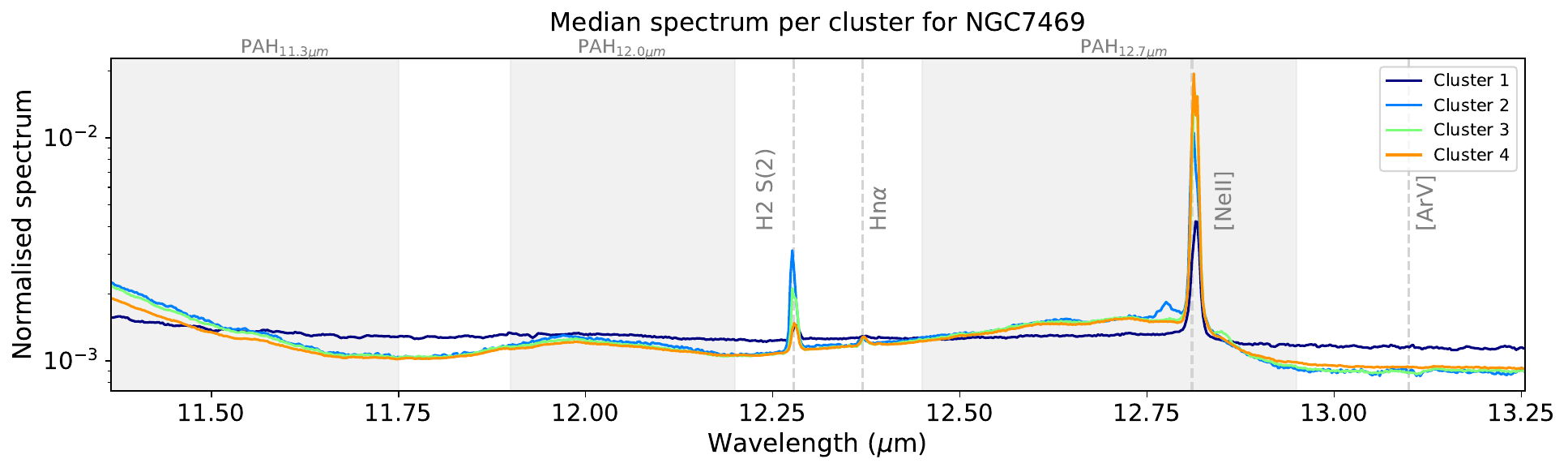}
    \includegraphics[width=.254\textwidth]{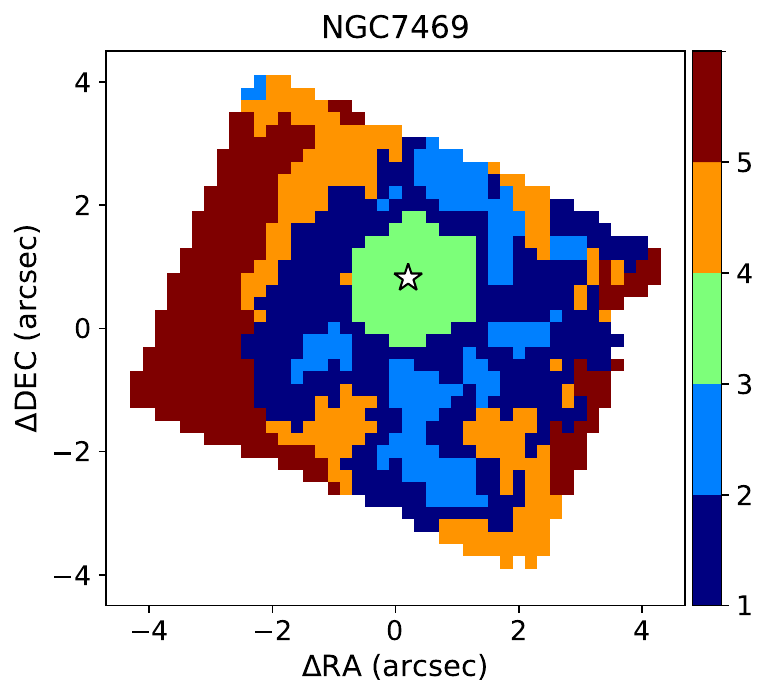}
    \includegraphics[width=.75\textwidth]{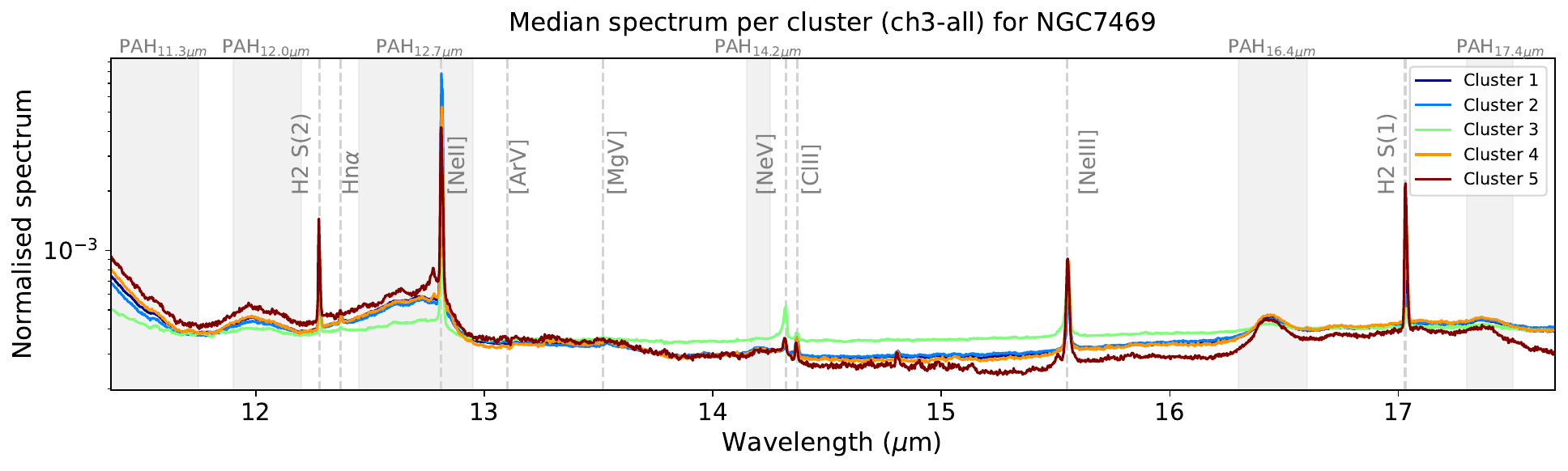}
	\caption{Same as Fig.~\ref{Fig:Cluster_NGC7172} but for NGC\,7469. We note that, for the top panel we do not show the spectrum for cluster 5, which is a low S/N cluster.}
\label{Fig:Cluster_NGC7469}
\end{figure*}

    \begin{figure*}
	\includegraphics[width=.26\textwidth]{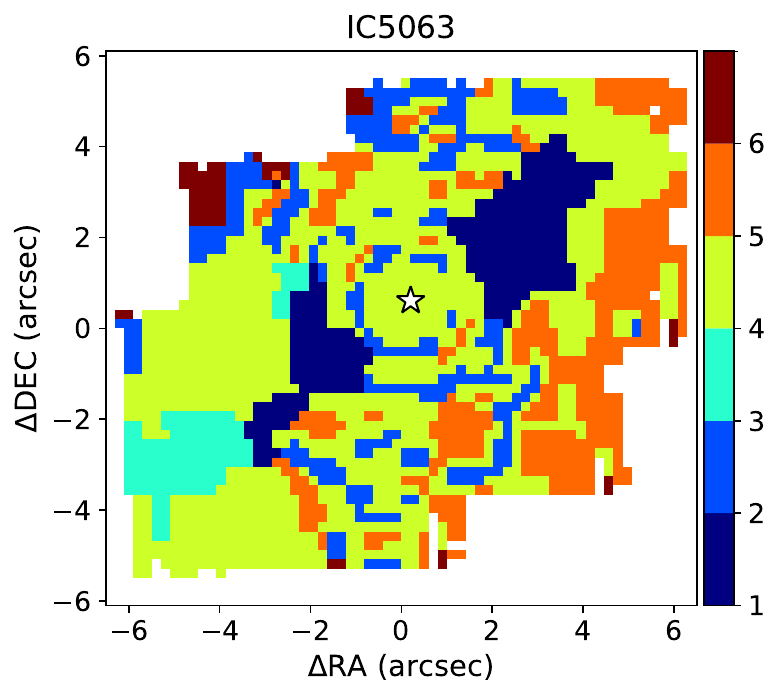}
	\includegraphics[width=.74\textwidth]{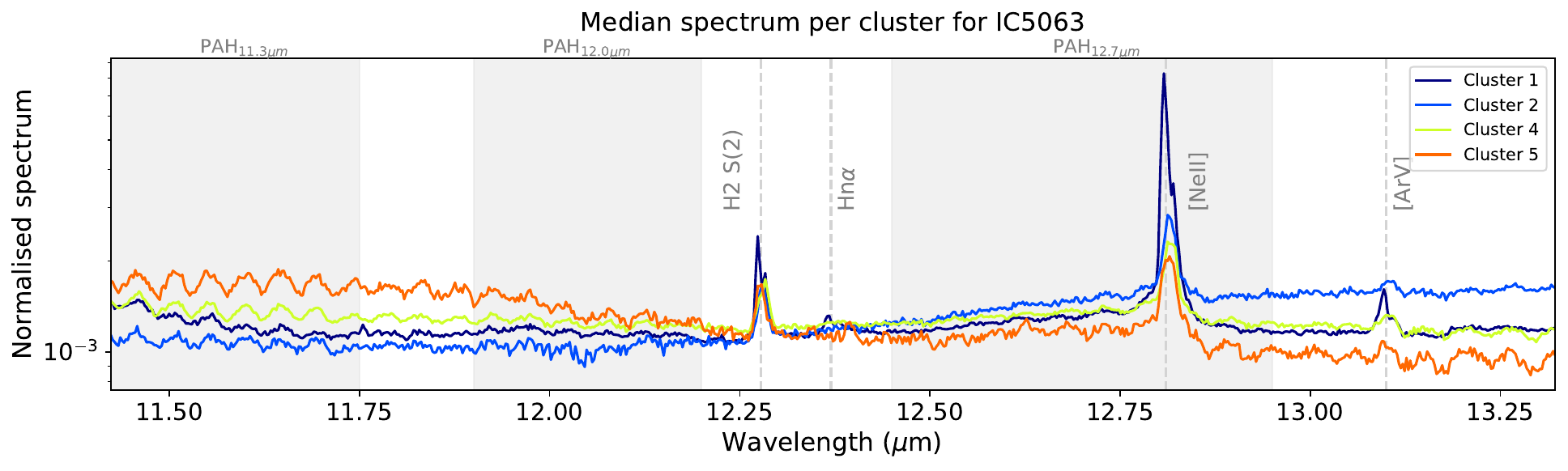}
    \includegraphics[width=.26\textwidth]{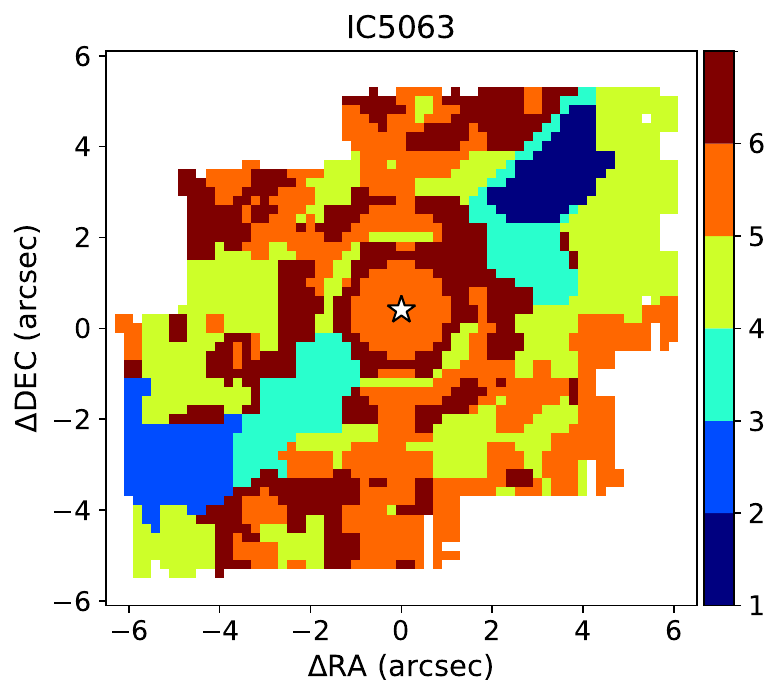}
	\includegraphics[width=.74\textwidth]{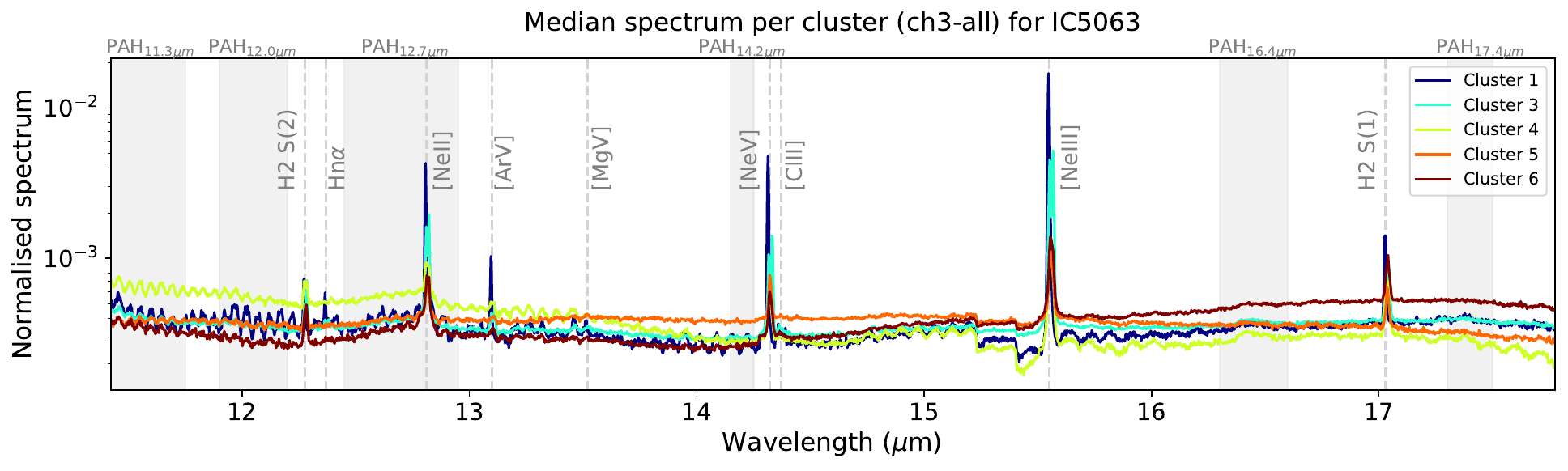}
	\caption{Same as Fig.~\ref{Fig:Cluster_NGC7172} but for IC\,5063. We note that, in the top (bottom) panel, we do not show the spectrum for cluster 3 and 6 (2), as they are low S/N clusters.}
	\label{Fig:Cluster_IC5063}
\end{figure*}

\begin{figure*}
	\includegraphics[width=.26\textwidth]{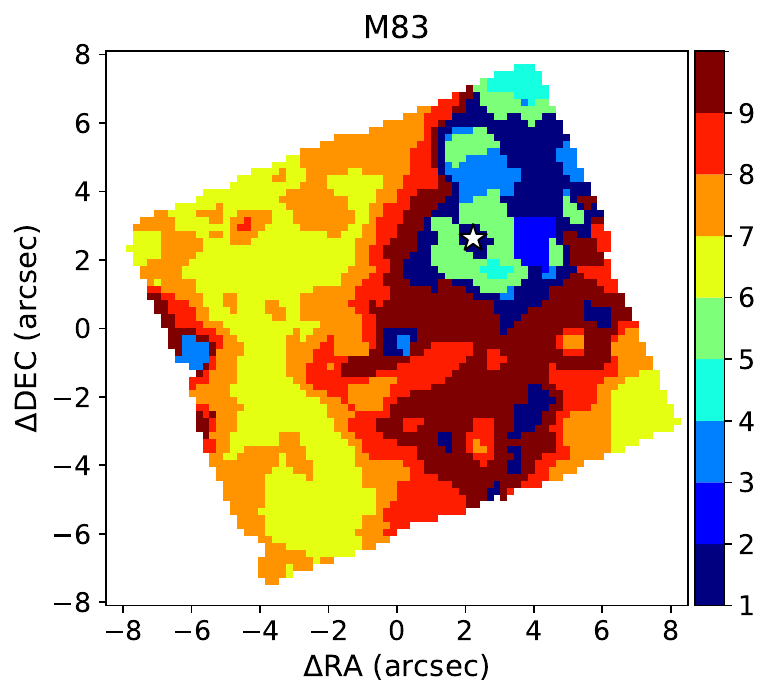}
	\includegraphics[width=.755\textwidth]{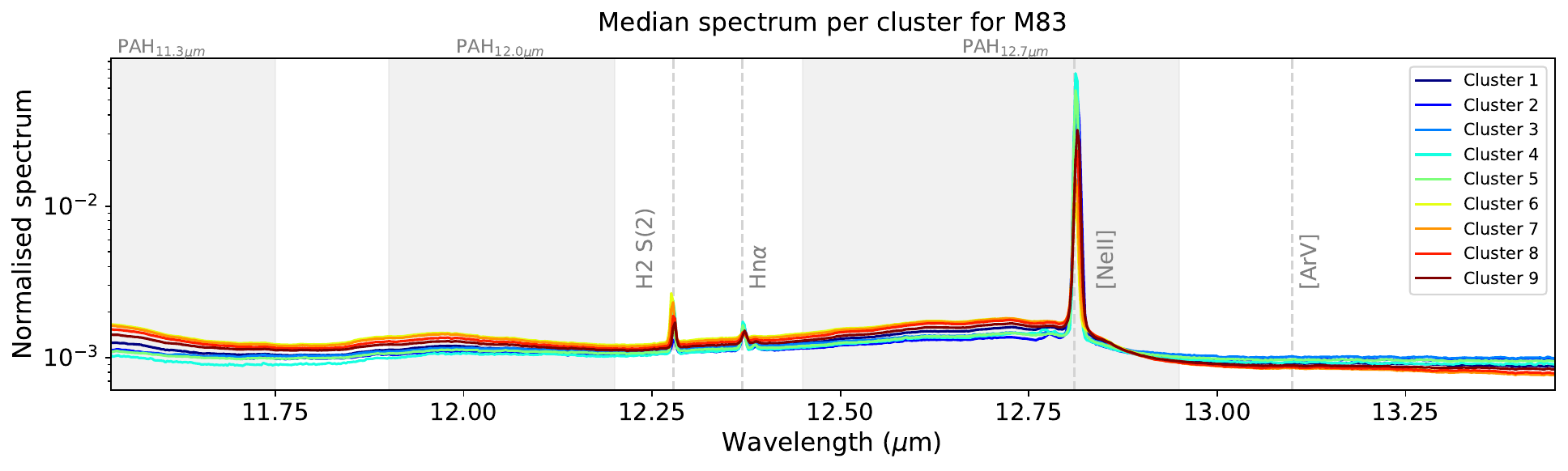}
    \includegraphics[width=.26\textwidth]{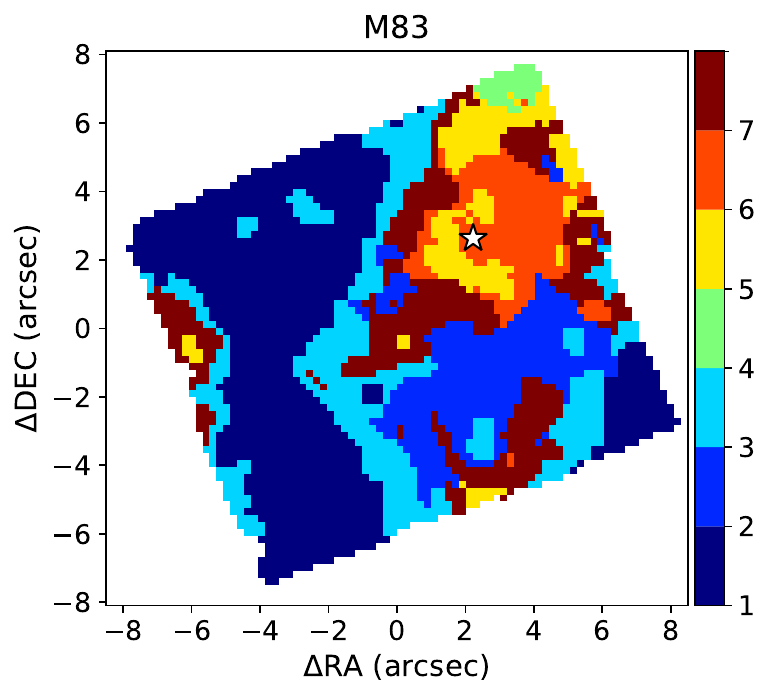}
	\includegraphics[width=.75\textwidth]{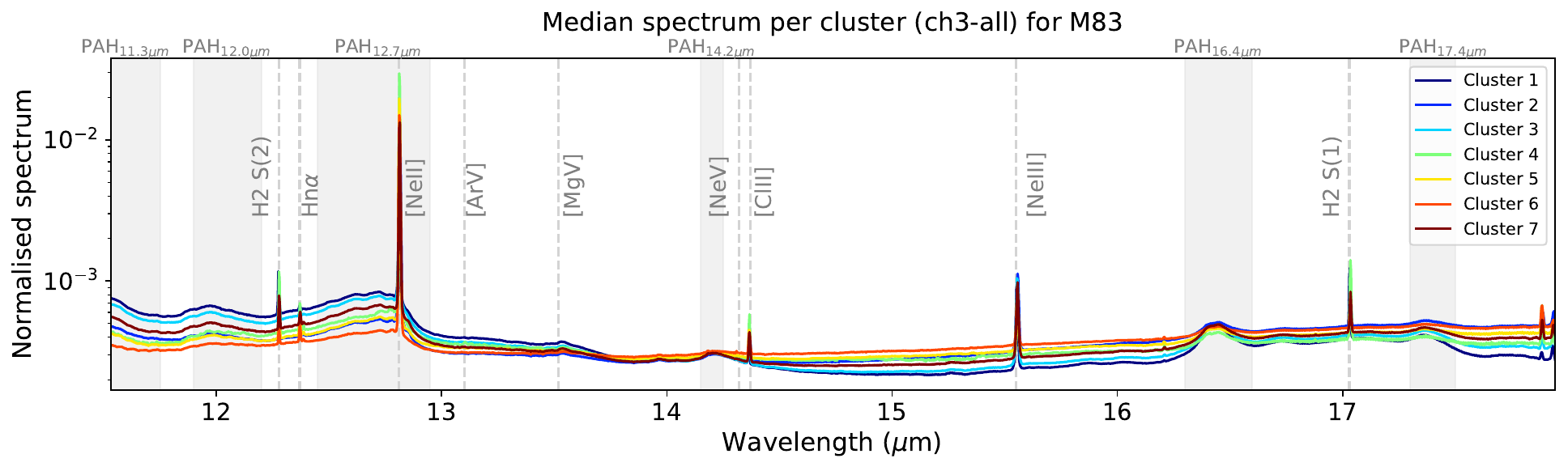}
	\caption{Same as Fig.~\ref{Fig:Cluster_NGC7172} but for M\,83. We note that the mid-IR photometric centre does not coincide with the optical one, but is close to the stellar kinematic centre \citep[see][and references therein]{Hernandez2025}.}
	\label{Fig:Cluster_M83}
\end{figure*}

\begin{figure*}
	\includegraphics[width=.23\textwidth]{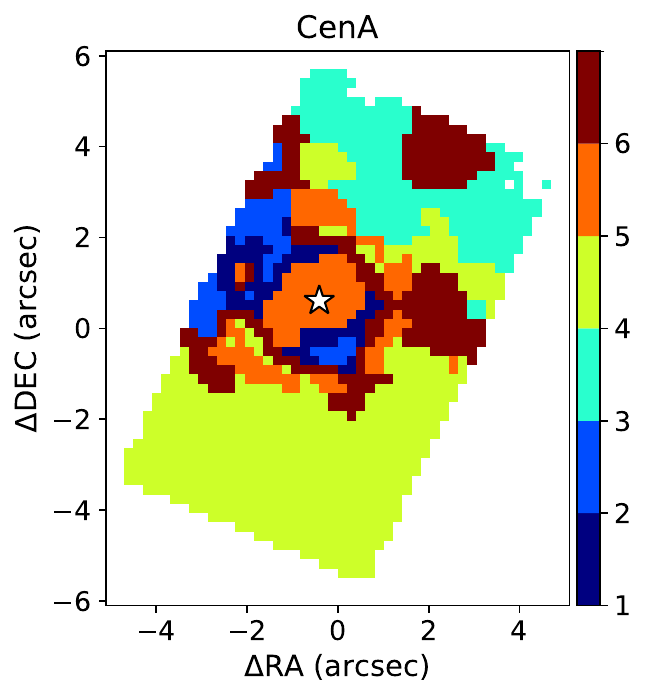}
	\includegraphics[width=.78\textwidth]{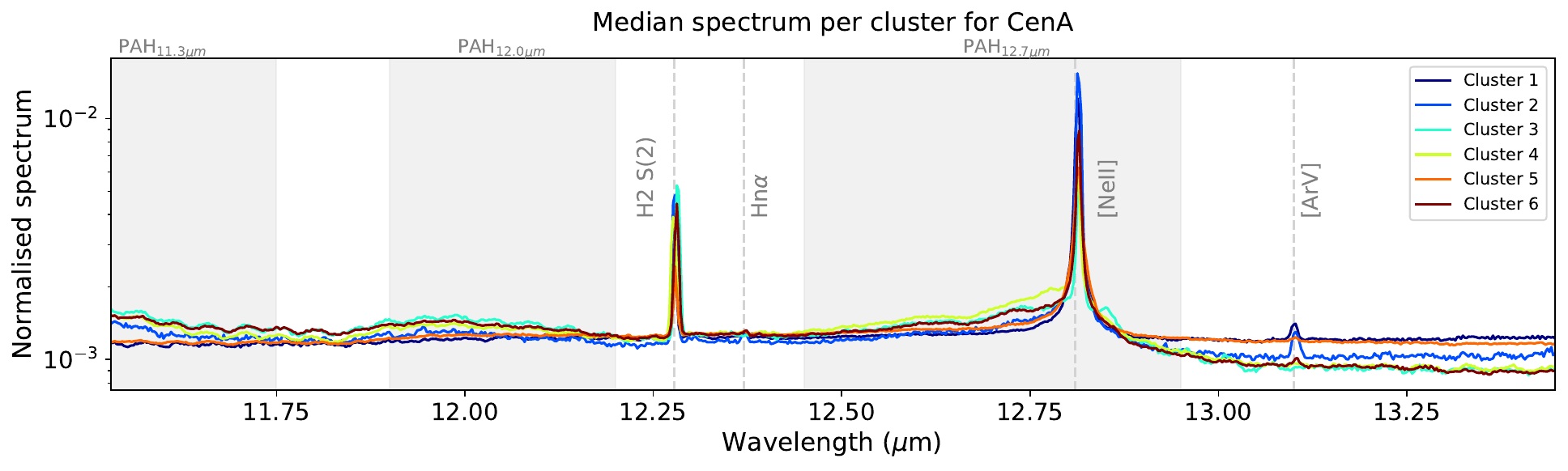}
        \includegraphics[width=.23\textwidth]{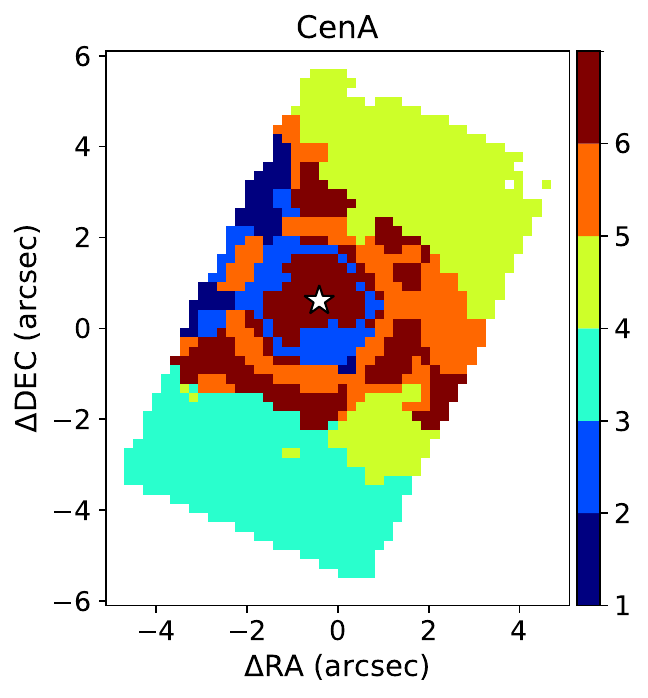}
	\includegraphics[width=.78\textwidth]{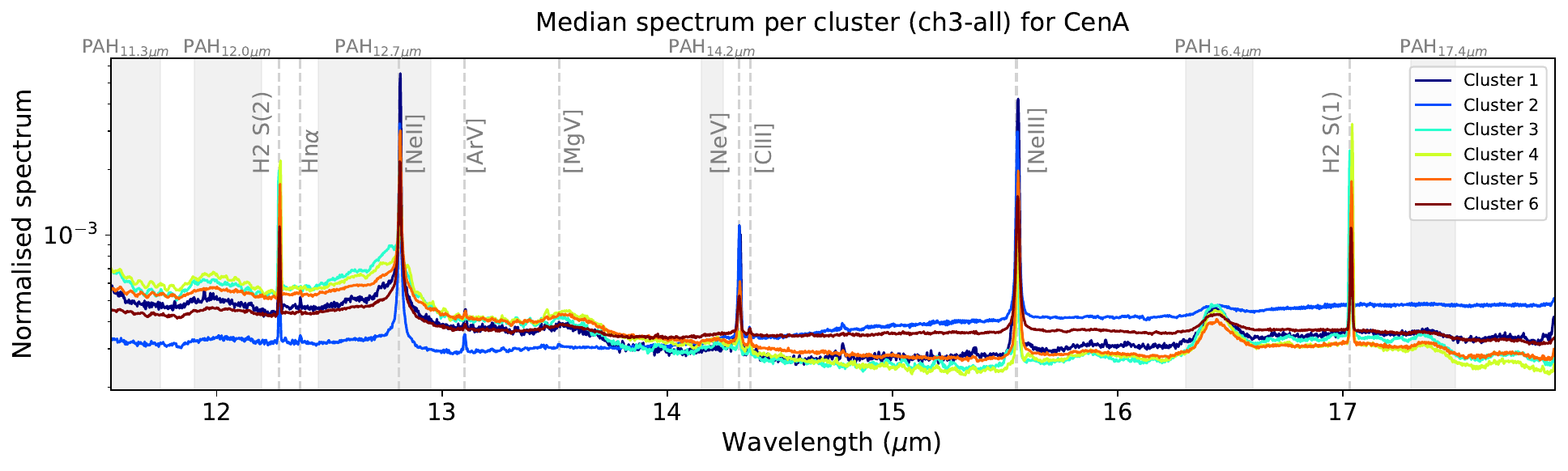}
	\caption{Same as Fig.~\ref{Fig:Cluster_NGC7172} but for Centaurus\,A.}
	\label{Fig:Cluster_CenA}
\end{figure*}

\begin{table*}
	\caption{Probabilistic classification of all the clusters from the ch3-all cubes of the galaxies used to train the RF models. }
	\label{Table:2}
	\centering
	\tiny 
\begin{tabular}{llccccc}
	\hline\hline
	Galaxy & Cluster & Init. label & RF label & AGN probability & SF probability & Other probability \\
	(a) & (b) & (c) & (d) & (e) & (f) & (g) \\
	\hline
CenA    & 1       & - & 0 & 0.64 $\pm$ 0.08 & 0.14 $\pm$ 0.05 & 0.21 $\pm$ 0.06 \\
	    & 2       & - & 0 & 0.50 $\pm$ 0.11 & 0.16 $\pm$ 0.08 & 0.34 $\pm$ 0.10 \\
	    & 3       & - & 1 & 0.23 $\pm$ 0.08 & 0.71 $\pm$ 0.08 & 0.06 $\pm$ 0.03 \\
	    & 4       & - & 1 & 0.26 $\pm$ 0.08 & 0.64 $\pm$ 0.08 & 0.10 $\pm$ 0.04 \\
	    & 5       & - & 0 & 0.61 $\pm$ 0.09 & 0.24 $\pm$ 0.08 & 0.15 $\pm$ 0.05 \\
	    & 6$^{*}$ & 0 & 0 & 0.79 $\pm$ 0.04 & 0.15 $\pm$ 0.04 & 0.06 $\pm$ 0.03 \\
IC5063  & 1       & 2 & 2 & 0.29 $\pm$ 0.05 & 0.06 $\pm$ 0.03 & 0.66 $\pm$ 0.04 \\
	    & 3       & - & 0 & 0.60 $\pm$ 0.12 & 0.06 $\pm$ 0.05 & 0.34 $\pm$ 0.11 \\
	    & 4       & - & 0 & 0.75 $\pm$ 0.10 & 0.09 $\pm$ 0.07 & 0.17 $\pm$ 0.07 \\
	    & 5$^{*}$ & 0 & 0 & 0.88 $\pm$ 0.04 & 0.04 $\pm$ 0.04 & 0.08 $\pm$ 0.04 \\
	    & 6       & 0 & 0 & 0.89 $\pm$ 0.04 & 0.02 $\pm$ 0.02 & 0.08 $\pm$ 0.03 \\
M83     & 1       & - & 1 & 0.03 $\pm$ 0.02 & 0.96 $\pm$ 0.03 & 0.01 $\pm$ 0.01 \\
        & 2       & 1 & 1 & 0.01 $\pm$ 0.01 & 0.99 $\pm$ 0.02 & 0.00 $\pm$ 0.01 \\
        & 3       & 1 & 1 & 0.00 $\pm$ 0.01 & 0.99 $\pm$ 0.01 & 0.00 $\pm$ 0.00 \\
        & 4       & 1 & 1 & 0.01 $\pm$ 0.01 & 0.97 $\pm$ 0.02 & 0.01 $\pm$ 0.01 \\
        & 5       & 1 & 1 & 0.01 $\pm$ 0.01 & 0.99 $\pm$ 0.02 & 0.01 $\pm$ 0.01 \\
        & 6$^{*}$ & 1 & 1 & 0.03 $\pm$ 0.02 & 0.96 $\pm$ 0.03 & 0.02 $\pm$ 0.01 \\
        & 7       & 1 & 1 & 0.01 $\pm$ 0.01 & 0.99 $\pm$ 0.01 & 0.00 $\pm$ 0.01 \\
NGC1052 & 1       & 2 & 2 & 0.31 $\pm$ 0.04 & 0.08 $\pm$ 0.03 & 0.61 $\pm$ 0.01 \\
	    & 2       & - & 0 & 0.50 $\pm$ 0.12 & 0.29 $\pm$ 0.12 & 0.20 $\pm$ 0.07 \\
	    & 4$^{*}$ & 0 & 0 & 0.83 $\pm$ 0.04 & 0.10 $\pm$ 0.04 & 0.07 $\pm$ 0.03 \\
	    & 5       & - & 0 & 0.51 $\pm$ 0.11 & 0.28 $\pm$ 0.10 & 0.21 $\pm$ 0.07 \\
NGC3081 & 1       & 0 & 0 & 0.87 $\pm$ 0.04 & 0.05 $\pm$ 0.03 & 0.08 $\pm$ 0.03 \\
	    & 2       & 0 & 0 & 0.84 $\pm$ 0.05 & 0.02 $\pm$ 0.02 & 0.14 $\pm$ 0.05 \\
	    & 3       & 1 & 1 & 0.05 $\pm$ 0.03 & 0.94 $\pm$ 0.03 & 0.01 $\pm$ 0.01 \\
	    & 4       & - & 0 & 0.55 $\pm$ 0.13 & 0.31 $\pm$ 0.13 & 0.14 $\pm$ 0.06 \\
	    & 5$^{*}$ & - & 0 & 0.75 $\pm$ 0.08 & 0.08 $\pm$ 0.05 & 0.16 $\pm$ 0.07 \\
	    & 6       & - & 0 & 0.59 $\pm$ 0.16 & 0.28 $\pm$ 0.18 & 0.14 $\pm$ 0.06 \\
NGC3256N & 1$^{*}$ & 1 & 1 & 0.02 $\pm$ 0.02 & 0.98 $\pm$ 0.02 & 0.01 $\pm$ 0.01 \\
	     & 2       & 1 & 1 & 0.00 $\pm$ 0.01 & 1.00 $\pm$ 0.01 & 0.00 $\pm$ 0.00 \\
	     & 3       & 1 & 1 & 0.07 $\pm$ 0.02 & 0.90 $\pm$ 0.02 & 0.03 $\pm$ 0.02 \\
	     & 4       & 1 & 1 & 0.01 $\pm$ 0.01 & 0.99 $\pm$ 0.01 & 0.00 $\pm$ 0.01 \\
	     & 5       & - & 1 & 0.02 $\pm$ 0.02 & 0.96 $\pm$ 0.03 & 0.02 $\pm$ 0.02 \\
	     & 6       & - & 1 & 0.00 $\pm$ 0.01 & 1.00 $\pm$ 0.01 & 0.00 $\pm$ 0.00 \\
	     & 7       & - & 1 & 0.00 $\pm$ 0.01 & 0.99 $\pm$ 0.01 & 0.00 $\pm$ 0.01 \\
	     & 8       & - & 1 & 0.02 $\pm$ 0.02 & 0.96 $\pm$ 0.03 & 0.02 $\pm$ 0.02 \\
NGC4594 & 1       & - & 0 & 0.51 $\pm$ 0.11 & 0.28 $\pm$ 0.12 & 0.21 $\pm$ 0.05 \\
	    & 2$^{*}$ & 0 & 0 & 0.84 $\pm$ 0.02 & 0.09 $\pm$ 0.02 & 0.06 $\pm$ 0.02 \\
	    & 3       & - & 0 & 0.58 $\pm$ 0.07 & 0.18 $\pm$ 0.05 & 0.24 $\pm$ 0.05 \\
	    & 6       & - & 0 & 0.45 $\pm$ 0.08 & 0.34 $\pm$ 0.07 & 0.21 $\pm$ 0.05 \\
	    & 7       & - & 0 & 0.49 $\pm$ 0.09 & 0.28 $\pm$ 0.08 & 0.23 $\pm$ 0.05 \\
NGC5506 & 1       & - & 0 & 0.77 $\pm$ 0.09 & 0.09 $\pm$ 0.06 & 0.14 $\pm$ 0.07 \\
	    & 3       & - & 0 & 0.78 $\pm$ 0.07 & 0.06 $\pm$ 0.04 & 0.16 $\pm$ 0.06 \\
	    & 4$^{*}$ & 0 & 0 & 0.84 $\pm$ 0.06 & 0.05 $\pm$ 0.03 & 0.11 $\pm$ 0.05 \\
	    & 5       & - & 0 & 0.63 $\pm$ 0.12 & 0.08 $\pm$ 0.05 & 0.29 $\pm$ 0.10 \\
	    & 6       & - & 0 & 0.61 $\pm$ 0.12 & 0.07 $\pm$ 0.05 & 0.32 $\pm$ 0.11 \\
NGC5728 & 1       & 1 & 1 & 0.11 $\pm$ 0.05 & 0.87 $\pm$ 0.06 & 0.02 $\pm$ 0.02 \\
	    & 2$^{*}$ & 0 & 0 & 0.81 $\pm$ 0.05 & 0.06 $\pm$ 0.03 & 0.14 $\pm$ 0.04 \\
	    & 3       & - & 0 & 0.54 $\pm$ 0.09 & 0.36 $\pm$ 0.09 & 0.10 $\pm$ 0.04 \\
	    & 4       & 2 & 2 & 0.17 $\pm$ 0.04 & 0.03 $\pm$ 0.02 & 0.80 $\pm$ 0.04 \\
	    & 5       & - & 0 & 0.70 $\pm$ 0.08 & 0.16 $\pm$ 0.06 & 0.14 $\pm$ 0.05 \\
	    & 7       & 2 & 2 & 0.24 $\pm$ 0.04 & 0.06 $\pm$ 0.03 & 0.70 $\pm$ 0.03 \\
	    & 8       & - & 0 & 0.64 $\pm$ 0.08 & 0.18 $\pm$ 0.06 & 0.19 $\pm$ 0.06 \\
NGC7172 & 1       & 1 & 1 & 0.04 $\pm$ 0.02 & 0.96 $\pm$ 0.03 & 0.01 $\pm$ 0.01 \\
	    & 2$^{*}$ & 0 & 0 & 0.86 $\pm$ 0.04 & 0.08 $\pm$ 0.04 & 0.06 $\pm$ 0.03 \\
	    & 3       & - & 1 & 0.34 $\pm$ 0.13 & 0.59 $\pm$ 0.15 & 0.07 $\pm$ 0.04 \\
	    & 4       & - & 0 & 0.65 $\pm$ 0.10 & 0.22 $\pm$ 0.08 & 0.13 $\pm$ 0.06 \\
NGC7319 & 5$^{*}$ & 0 & 0 & 0.88 $\pm$ 0.04 & 0.02 $\pm$ 0.02 & 0.10 $\pm$ 0.04 \\
	    & 6       & - & 0 & 0.73 $\pm$ 0.10 & 0.10 $\pm$ 0.10 & 0.16 $\pm$ 0.06 \\
	    & 8       & - & 0 & 0.76 $\pm$ 0.08 & 0.04 $\pm$ 0.03 & 0.19 $\pm$ 0.08 \\
NGC7469 & 1       & 1 & 1 & 0.01 $\pm$ 0.01 & 0.98 $\pm$ 0.02 & 0.00 $\pm$ 0.01 \\
	    & 2       & 1 & 1 & 0.01 $\pm$ 0.01 & 0.99 $\pm$ 0.01 & 0.00 $\pm$ 0.01 \\
	    & 3$^{*}$ & - & 1 & 0.34 $\pm$ 0.13 & 0.57 $\pm$ 0.16 & 0.08 $\pm$ 0.05 \\
	    & 4       & 1 & 1 & 0.03 $\pm$ 0.02 & 0.96 $\pm$ 0.03 & 0.01 $\pm$ 0.01 \\
	    & 5       & - & 1 & 0.16 $\pm$ 0.07 & 0.81 $\pm$ 0.07 & 0.03 $\pm$ 0.02 \\
	\hline        
\end{tabular}\\
\tablefoot{Columns indicate: (a) Galaxy name, (b) cluster number, (c) initial label assigned based on previous works (0 is AGN, 1 is SF, 2 is Other, see Sect.~\ref{SubSect2:RFtecnhique}), (d) final label assigned with the RF model, and (e), (f), and (g) probabilities and their corresponding standard deviation of being assigned to one of the available classes (AGN, SF, and Other, respectively). We note that clusters excluded due to S/N are not in this table (see Sect.~\ref{SubSect2:Clustering}). $^{*}$ indicates the cluster containing the nuclear region of the galaxy.}
\end{table*}

	
	\section{Additional figures}
	\label{Appendix}
	
\begin{figure*}
		\centering
		\includegraphics[width=.93\textwidth]{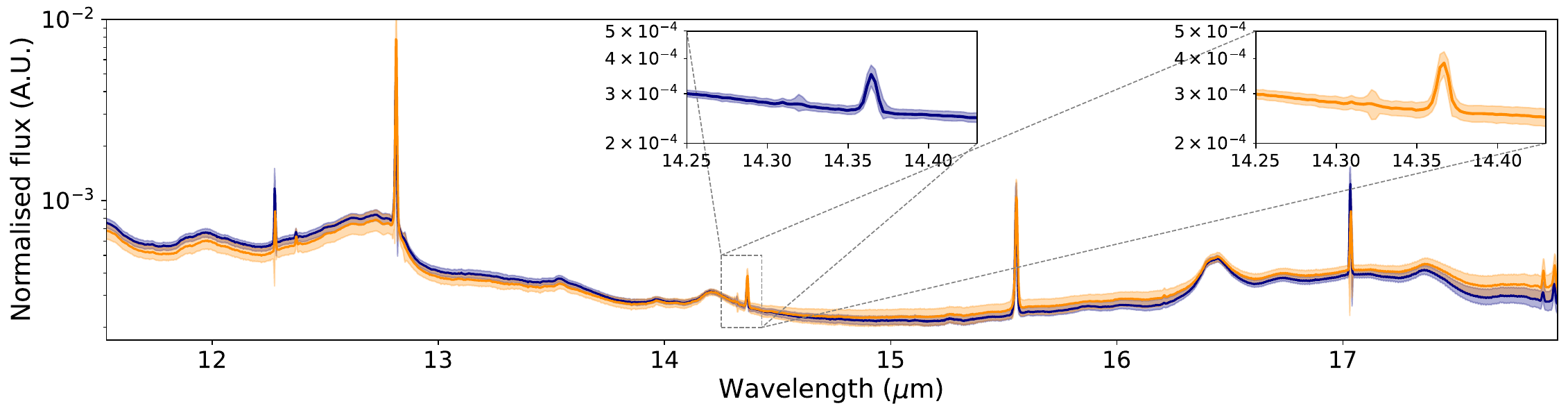}
		\caption{Total median spectra for clusters 1 and 3 (blue and orange, respectively) for M\,83 (ch3-all), with two insets showing the [Ne\,V] and [Cl\,II] lines. The shaded areas represent the uncertainty estimated as the standard deviation of all the spectra within each cluster (see Sect.~\ref{SubSect2:RFtecnhique}).}
		\label{FigAp:M83_NeV}
	\end{figure*}

\begin{figure*}
		\centering
		\includegraphics[width=.3\textwidth]{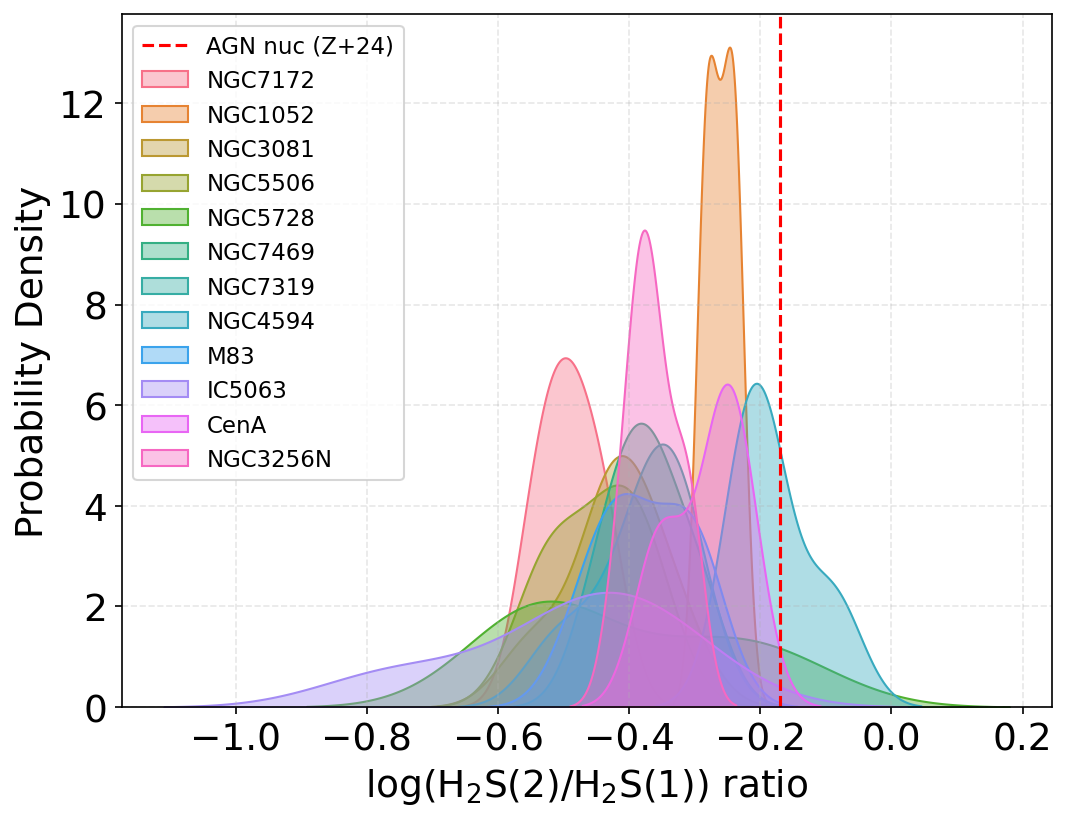}
		\includegraphics[width=.292\textwidth]{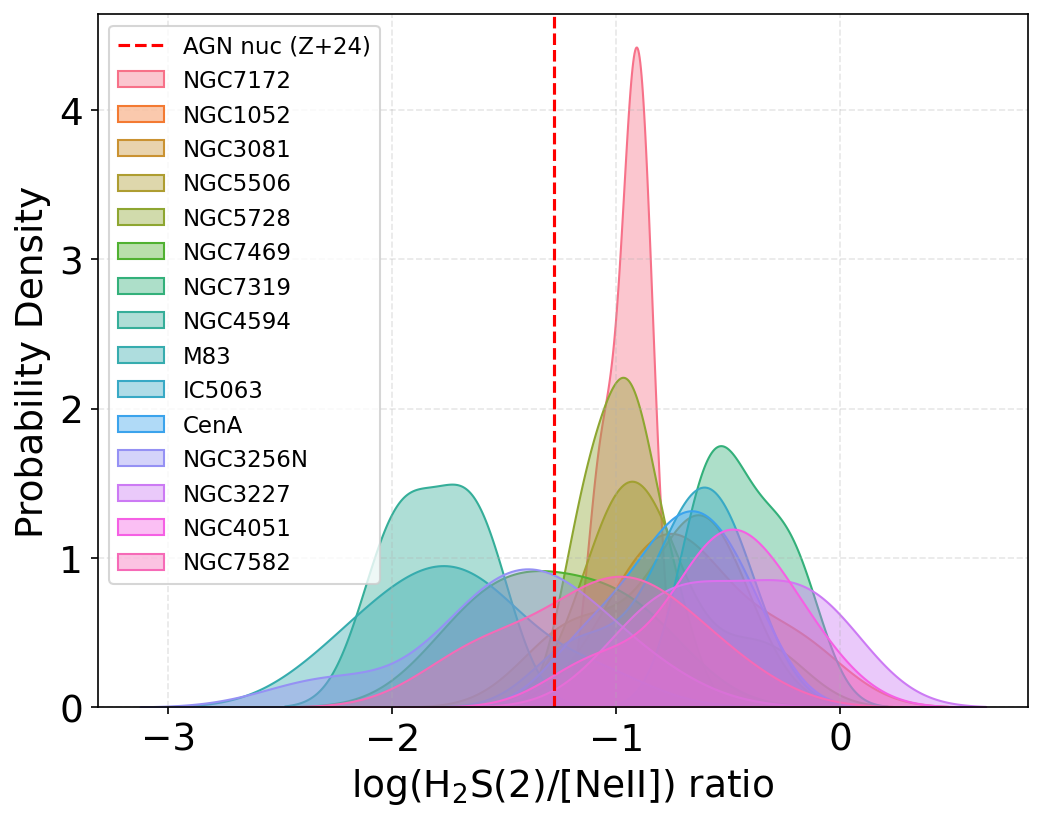}
		\includegraphics[width=.305\textwidth]{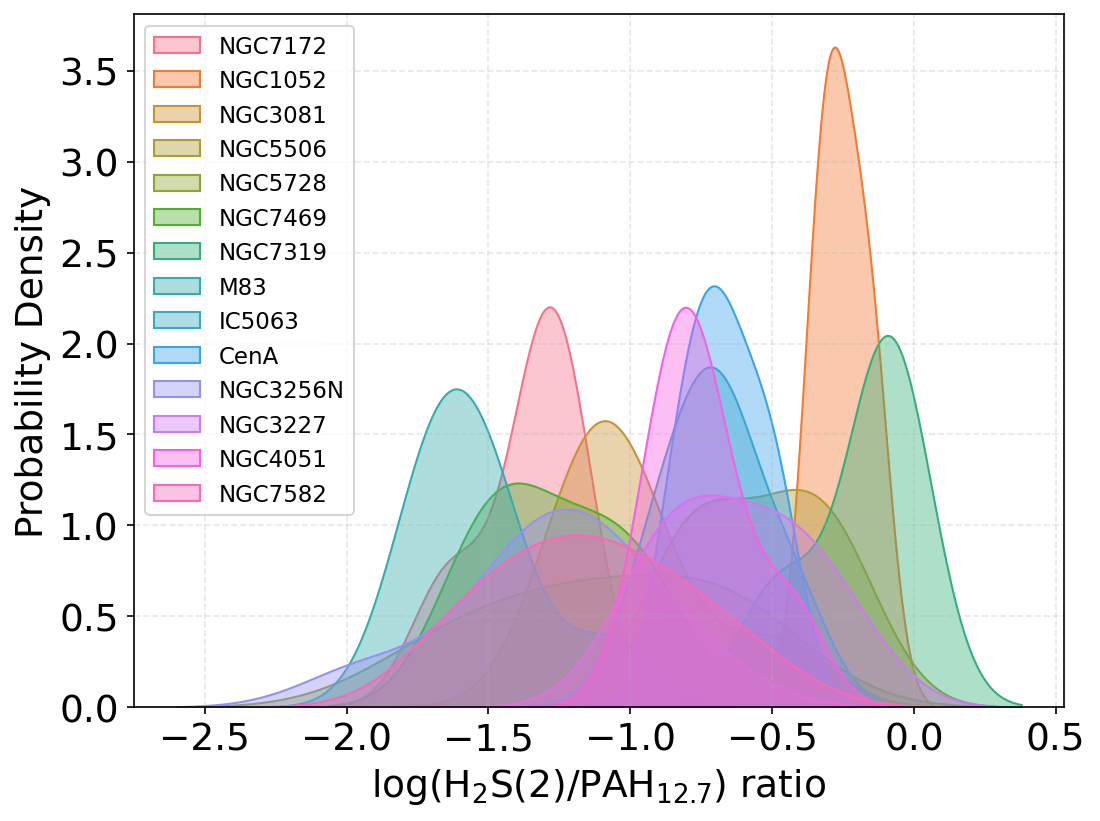}
		\includegraphics[width=.302\textwidth]{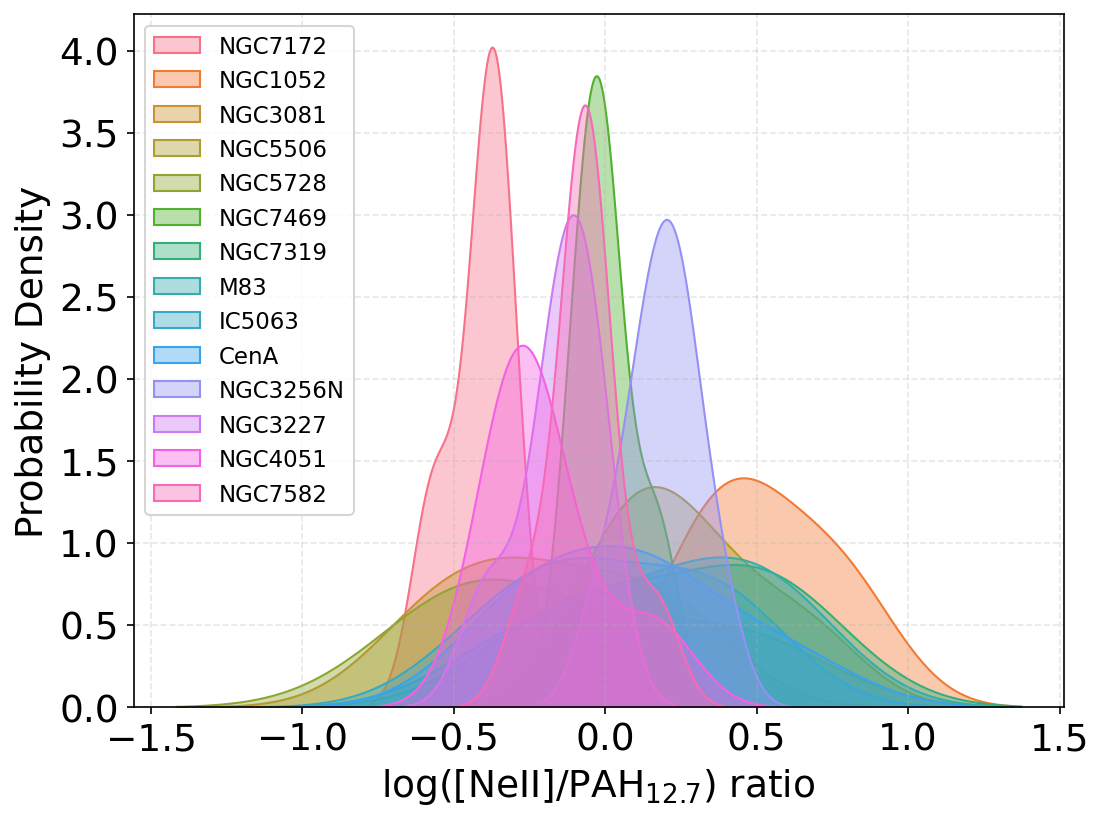}
		\includegraphics[width=.29\textwidth]{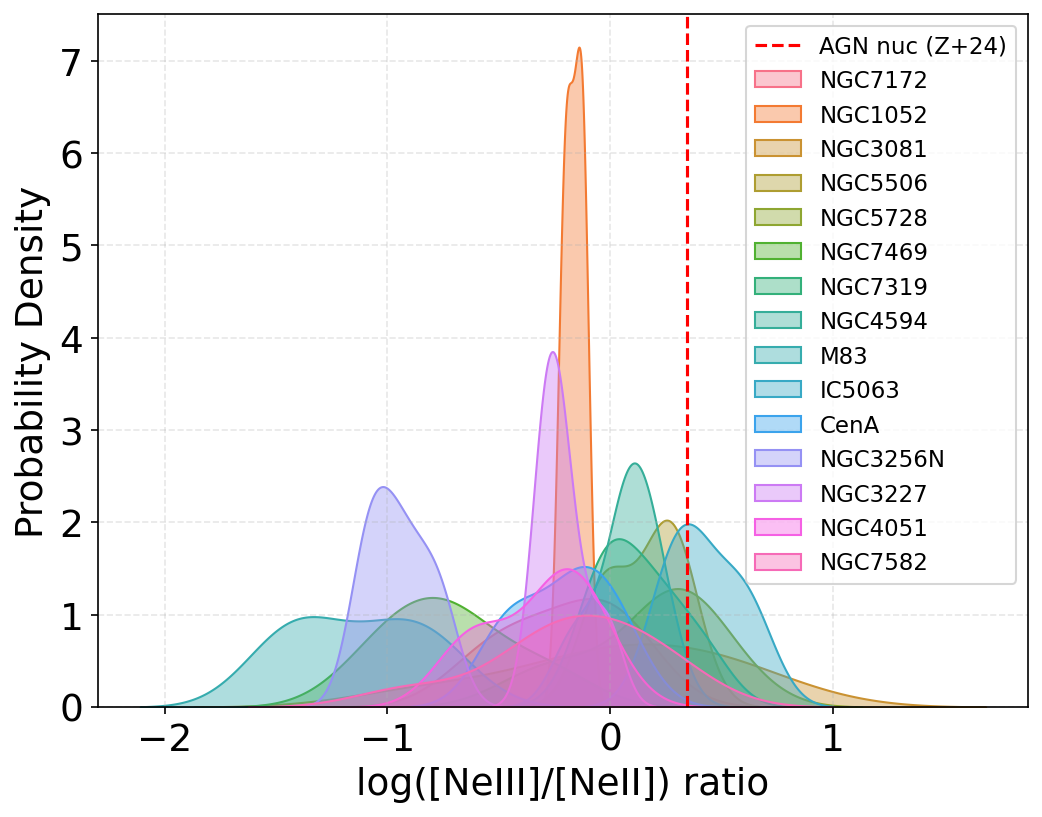}
		\includegraphics[width=.3\textwidth]{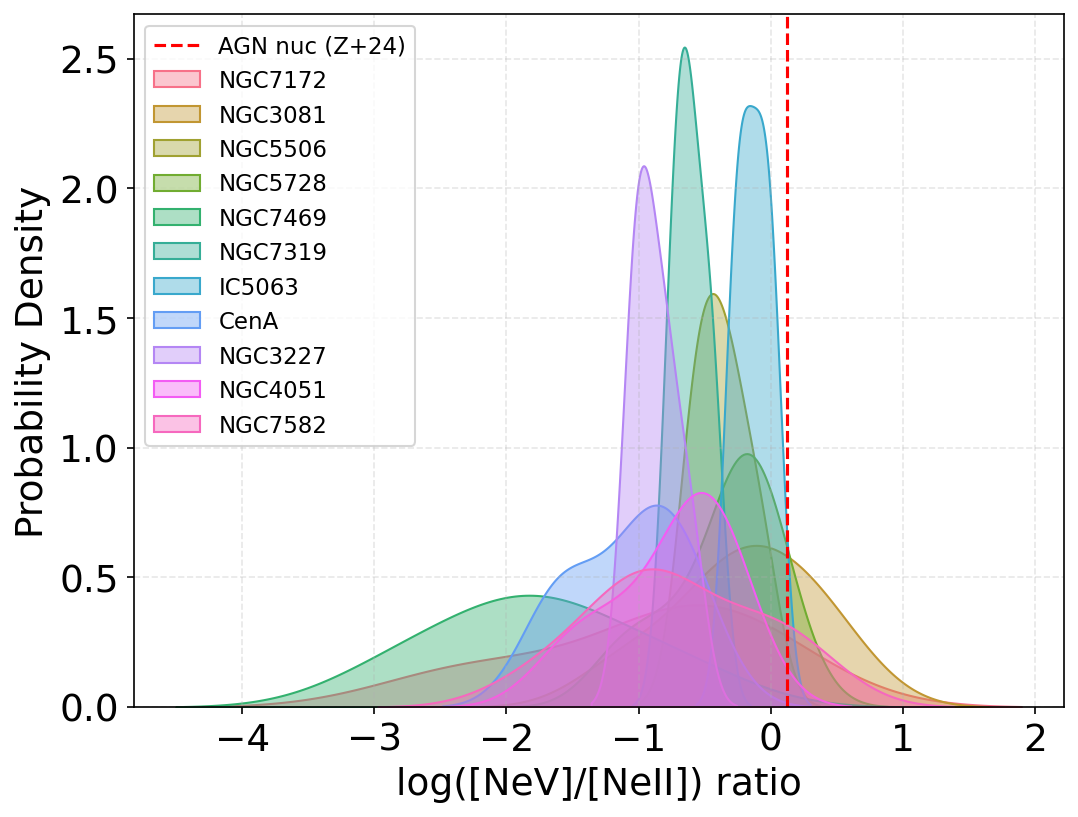}
		\caption{Histograms of the line ratios obtained for each cluster per galaxy using the ch3-all cubes. Instead of regular histograms, we use kernel density estimates (KDE) for visualisation purposes, that compute continuous probability density curves. The red, dashed line marks the median values of the line ratios measured in Sy galaxies with MIRI/MRS by \cite{Zhang2024} as a reference.}
		\label{FigAp:Histograms}
	\end{figure*}

\begin{figure*}
		\centering
		\includegraphics[width=.6\textwidth]{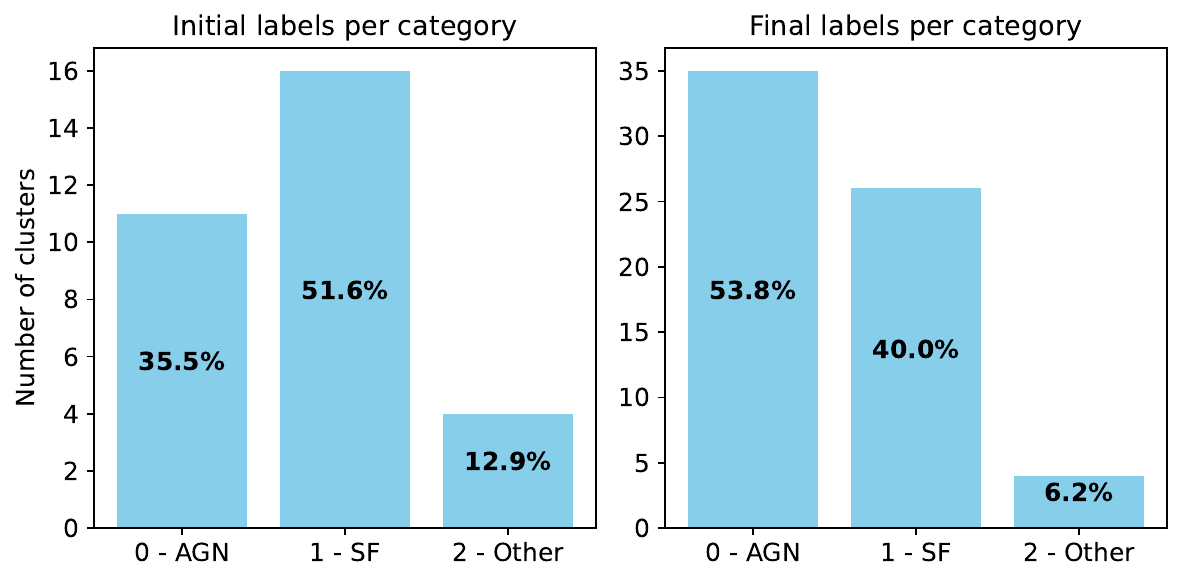}
		\caption{Percentage of clusters from the training sample labelled in each category for the inital classification, and after the prediction from the RF model (see also Table~\ref{Table:2}).}
		\label{FigAp:BalanceClasses}
	\end{figure*}

\begin{figure*}
		\centering
		\includegraphics[width=.54\textwidth]{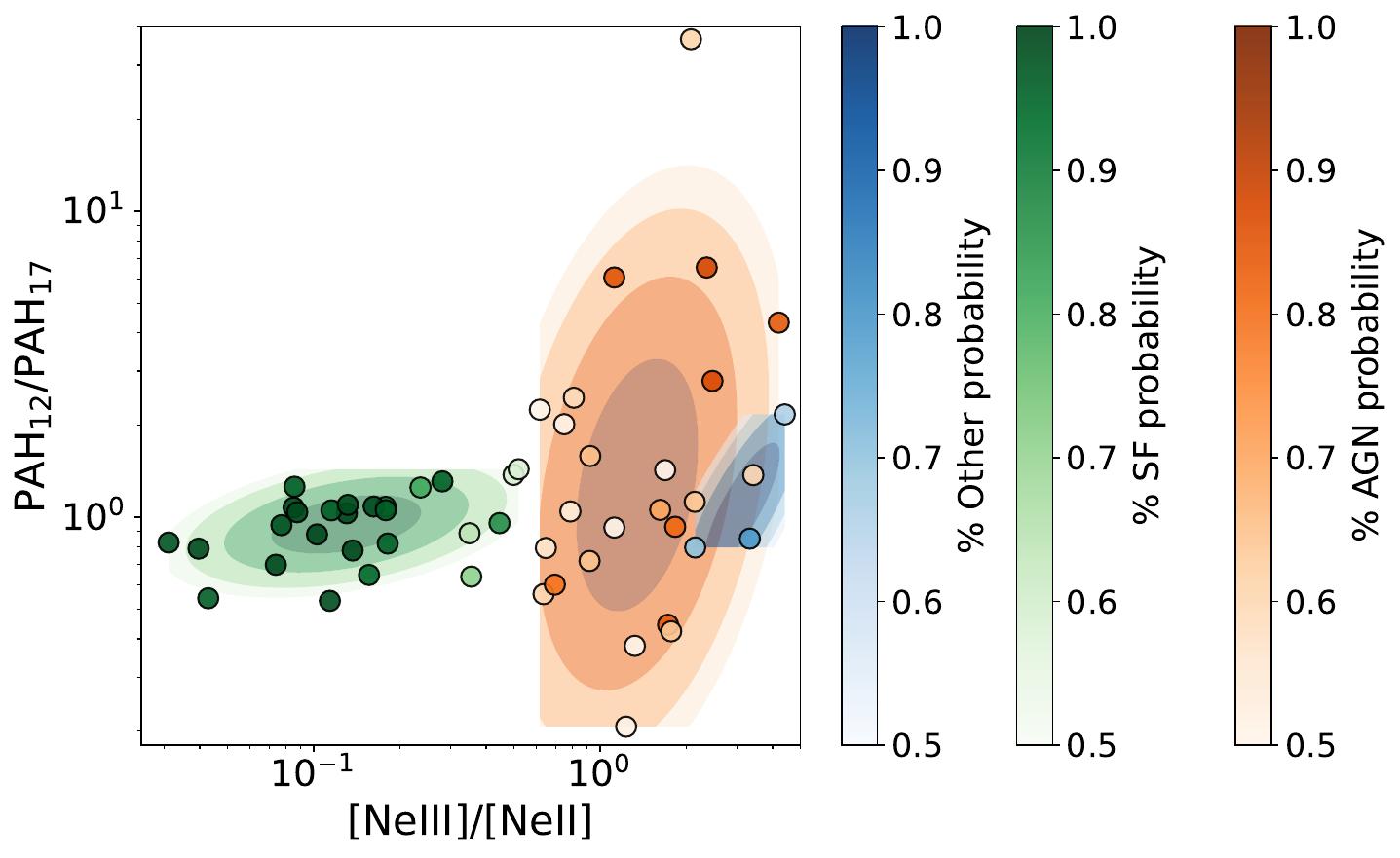}
		\includegraphics[width=.91\columnwidth]{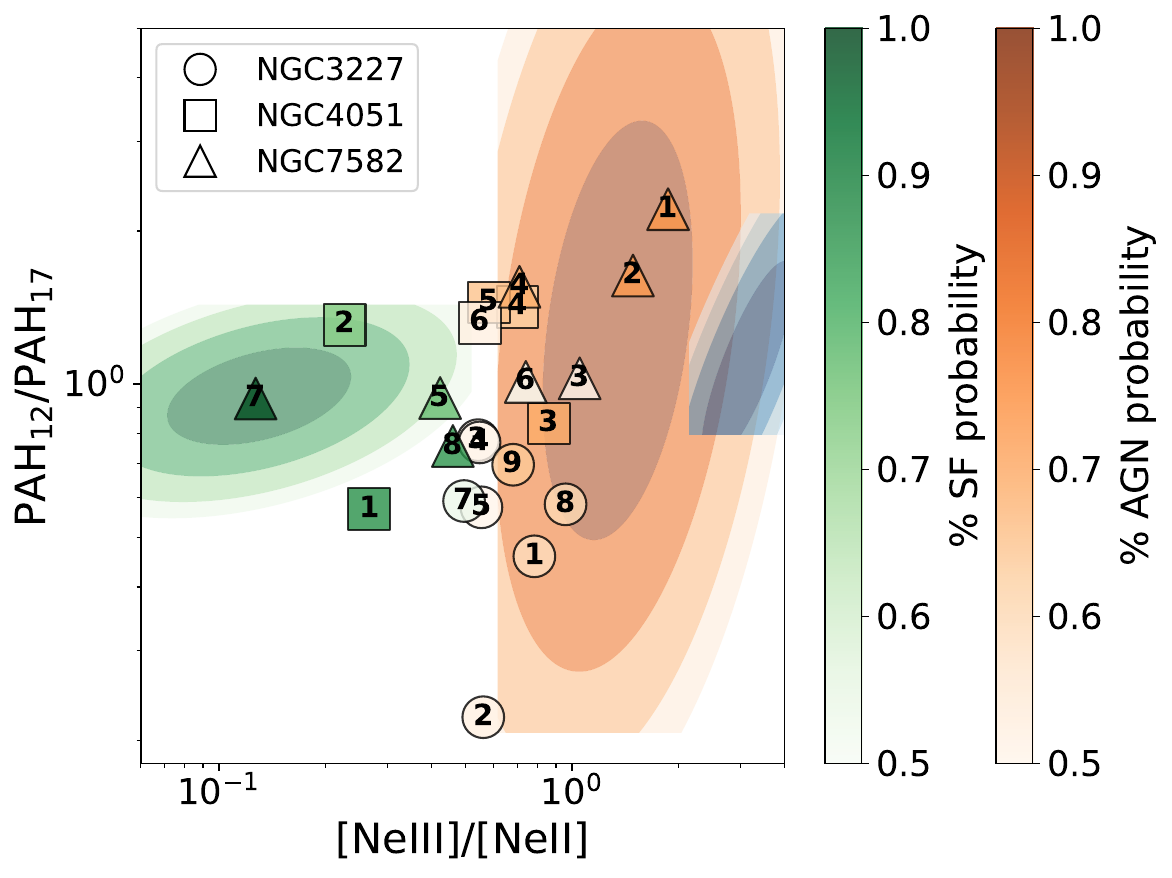}
		\caption{Diagnostic diagram in a logarithm scale similar to Figs.~\ref{Fig:LineRatios1} and~\ref{Fig:LineRatios_NGC7582}, using the PAH ratios derived from the ch3-all cubes for the training and testing samples (left and right, respectively, see Sect.~\ref{SubSect3:Results_Ionisation}).}
		\label{FigAp:DiagnosticsPAHs}
	\end{figure*}

	In this appendix we show the median spectra for clusters 1 and 3 of M\,83 (Fig.~\ref{FigAp:M83_NeV}). We then show several figures related to the line ratios and the random forest classifier. These include the probability distributions of the obtained line ratios (Fig.~\ref{FigAp:Histograms}). These are smoothed representations of the ratios from all clusters in a given galaxy, constructed using KDEs to provide a continuous visualisation of their overall distribution. In Fig.~\ref{FigAp:BalanceClasses}, we show the balance plots on the labels of the clusters from the training sample. Finally, we include the diagnostic diagrams created with the PAH$_{12}$/PAH$_{17}$ ratio (see Sect.~\ref{SubSect3:Results_Ionisation}), for the training and testing samples (see left and right panels in Fig.~\ref{FigAp:DiagnosticsPAHs}, respectively).
	
\end{appendix}

\end{document}